\documentclass{JHEP3}

\usepackage{latexsym}
\usepackage{graphicx}
\usepackage{cite}
\usepackage{bbm}

\def\AdSs5{$AdS_5$}
\def\AdS5s5{$AdS_5 \times S^5$}
\def\al{{\alpha^{\prime}}}
\def\gs{g_{st}}
\def\NSNS{{$NS\otimes NS$}}
\def\RR{{$R\otimes R$}}
\def\calE{{\cal E}}
\def\calB{{\cal B}}
\def\calZ{{\cal Z}}
\def\calP{{\cal P}}

\def\calN{{\cal N}}
\def\Tr{\mbox{Tr}}
\newcommand{\eg}{{\it e.g.~}}
\newcommand{\ie}{{\it i.e.~}}

\newcommand{\s}{\sigma}
\newcommand{\del}{\partial}
\newcommand{\aap}{{a^{\prime}}}
\newcommand{\be}{\begin{equation}}
\newcommand{\ee}{\end{equation}}
\newcommand{\ba}{\begin{eqnarray}}
\newcommand{\ea}{\end{eqnarray}}
\newcommand{\ra}{\rangle}
\newcommand{\la}{\langle}

\newbox\SlashedBox
\def\fs#1{\setbox\SlashedBox=\hbox{#1}
\hbox to
0pt{\hbox to 1\wd\SlashedBox{\hfil/\hfil}\hss}{#1}}
\def\hboxtosizeof#1#2{\setbox\SlashedBox=\hbox{#1}
\hbox to
1\wd\SlashedBox{#2}}

\def\ms#1{\setbox\SlashedBox=\hbox{$#1$}
\hbox to 0pt{\hbox to
1\wd\SlashedBox{\hfil/\hfil}\hss}#1}

\newcommand{\Dsm}{\,{\raisebox{1pt}{$/$} \hspace{-8pt} D}}

\newcommand{\one}{\mathbbm{1}}

\newcommand{\Z}{\mathbb{Z}}

\def\l{\lambda}
\def\t2{\tau_2}
\def\AdSS5{$AdS_5$}
\def\AdS5s5{$AdS_5\times S^5$}

\def\calC{{\cal C}}

\def\calR{{\cal R}}
\def\calK{{\cal K}}
\def\calZ{{\cal Z}}
\def\calP{{\cal P}}

\def\calF{{\cal F}}
\def\calO{{\cal O}}
\def\calH{{\cal H}}
\def\gy{g_{_{\rm YM}}}
\def\er{{\rm e}}

\def\IZ{\relax\ifmmode\mathchoice {\hbox{\cmss Z\kern-.4em Z}}
{\hbox{\cmss Z\kern-.4em Z}}
{\lower.9pt\hbox{\cmsss Z\kern-.4em Z}}
{\lower1.2pt\hbox{\cmsss Z\kern-.4em Z}}
\else{\cmss Z\kern-.4em Z}\fi}

\def\Tr{{\rm Tr}}
\def\tr{{\rm tr}}
\def\gs{g_{\rm s}}

\def\NSNS{{$NS\otimes NS$}}
\def\RR{{$R\otimes R$}}
\def\S{\Sigma}
\def\Sbar{{\bar\Sigma}}
\def\sbar{{\bar\sigma}}
\def\xib{{\bar \xi}}

\def\Mp{{{\cal M}'}}
\def\b{\beta}
\def\a{{\alpha}}
\def\g{\gamma}
\def\veps{\varepsilon}
\def\vt{\vartheta}
\def\bvt{{\bar\vartheta}}
\def\gdot{{\dot\gamma}}
\def\adot{{\dot\alpha}}
\def\bdot{{\dot\beta}}

\def\d{\delta}

\def\D{\Delta}
\def\Db{\bar \Delta}
\def\calA{{\cal A}}
\def\calH{{\cal M}}
\def\calQ{{\cal Q}}
\def\calX{{\cal X}}
\def\calT{{\cal T}}
\def\calJ{{\cal J}}

\def\c1{{\chi^1}}

\def\v{\varphi}

\def\tc{{\tau^c}}

\def\nub{{\bar \nu}}
\def\N4{{\cal N}=4}
\def\half{{1\over 2}}
\def\nn{\nonumber}

\def\nus{\begin{displaystyle}
    \bar\nu_{\rule{0pt}{2pt}}^{u[A}\nu^{B]}_u
    \end{displaystyle}}
\def\nut{\begin{displaystyle}
    \bar\nu_{\rule{0pt}{2pt}}^{u(A}\nu^{B)}_u
    \end{displaystyle}}
\def\nsix{(\bar\nu \nu)_{\bf 6}}
\def\nten{(\bar\nu \nu)_{\bf 10}}
\title{Instanton-induced Yang--Mills correlation functions at large $N$ and
their $AdS_5\times S^5$ duals}
\author{ Michael B. Green \\
Department of Applied Mathematics and Theoretical Physics \\
Wilberforce Road, Cambridge CB3 0WA, UK \\
E-mail: \email{M.B.Green@damtp.cam.ac.uk}}
\author{ Stefano Kovacs \\
Max-Planck-Institut f\"ur Gravitationsphysik \\
Albert-Einstein-Institut \\
Am M\"uhlenberg 1, D-14476 Golm, Germany \\
E-mail: \email{stefano.kovacs@aei.mpg.de}}

\abstract{Correlation functions of chiral primary operators and their
superconformal descendants in the $\calN=4$ supersymmetric $SU(N)$
Yang--Mills theory are studied in detail in a one-instanton background
and at large $N$. Whereas earlier calculations were restricted to
correlation functions that are saturated by the 16 exact
superconformal fermionic moduli, here the effect of the set
of additional fermionic moduli associated with the embeddings of the
$SU(2)$ instanton in $SU(N)$ is considered. The presence of the extra
fermionic modes is essential for matching Yang--Mills instanton
effects in various correlation functions with D-instanton effects in
type IIB string theory via the AdS/CFT conjecture.  The leading terms
of this kind on the string side contribute at order $\al^{-1}$ (where
the  Einstein--Hilbert terms are of order $\al^{-4}$), which is the
same order as the $\calR^4$ interaction.   For example, the instanton
contributions to correlation functions of higher dimensional chiral
primary operators are seen to match amplitudes involving Kaluza--Klein
excitations of the supergravity fields, as expected.  Another example
is the matching of certain multi-fermion correlation functions which
correspond to certain multi-fermion interactions required by
supersymmetry of the IIB string effective action. Careful analysis of
a variety of competing effects makes it possible to decipher
contributions corresponding to higher derivative interactions in the
IIB effective action.  In this manner it is possible to check  for the
presence of terms of order $\al$.  Comments are also made on the
structure of instanton contributions to near-extremal correlation
functions of chiral primary operators.}

\preprint{DAMTP-2002-159; AEI-2002-103; hep-th/0212332}

\begin{document}

\section{Introduction}
\label{intro}

The interconnections between gauge theory and gravity lie at the heart
of many of the most intriguing features of string theory.  The most
precise connection is that expressed by the AdS/CFT conjecture in the
maximally supersymmetric case, which relates superconformal $\calN=4$
$SU(N)$ Yang--Mills theory to type IIB superstring theory in an
\AdS5s5\ background \cite{maldacena,gkp,wittone}.   At present little
is known about the string theory in this background, even at tree
level.  What information exists on the string side comes from the
compactification of the leading terms in the derivative (or low
energy) expansion effective action of  ten-dimensional type IIB
supergravity in the \AdS5s5\ background.  This is supposed to be
equivalent to the Yang--Mills theory in the limit $N\to \infty$ and at
large values of the 't Hooft coupling, $\lambda = \gy^2 N$.  However,
this is the strong coupling limit, in which there is little
information concerning the Yang--Mills theory.  Since the domains in
which there is direct information on the two sides of the conjecture
are non-overlapping it has only been possible to make direct checks
for restricted classes of observables. Although the consistency of
these tests sometimes appears to be nontrivial it is often the case
that it follows purely from the very high degree of supersymmetry,
combined with other symmetries.  Nevertheless, there are instances in
which there has been unexpected agreement between the Yang--Mills and
string theory calculations in situations where no obvious symmetry
guarantees such agreement.  Such cases may serve to illuminate the
structure of the theory on one side of the conjecture or the other.

One situation in which the agreement is somewhat unexpected is in the
study of instanton contributions to correlation functions of composite
Yang--Mills operators. Here the comparison is with the effect of
D-instantons in the type IIB superstring theory
\cite{banksgreen,bgkr,doreya}.  Certain D-instanton contributions to
the low energy superstring effective action arise from interaction
terms at order $\al^{-1}$.  Such interactions lead to  very precise
expectations for a `minimal' set of correlation functions of chiral
primary operators and their superconformal descendants  in the
Yang--Mills instanton background.  Recall that a Yang--Mills instanton
breaks one half of the 32 superconformal symmetries, leaving 16
supermoduli (fermionic collective coordinates).  The minimal
correlation functions are those that are determined by saturating the
composite operators with the superconformal zero modes constructed out
of these moduli. The Yang--Mills calculation is semi-classical and
effectively sets $\lambda=0$, so the spectacular agreement with the
corresponding D-instanton effects in the string theory appears more
surprising and less trivial.  For example, the presence of terms
proportional to $\er^{-\gy^2 N}$  would have spoiled the agreement.
In the semi-classical Yang--Mills theory at fixed $N$ and $\gy \to 0 $
such terms have a finite limit whereas they vanish in the supergravity
limit, in which $\gy^2 N \to \infty$.

Given the success of the comparison of instanton effects for the
restricted class of minimal correlation functions it is natural to
enquire to what extent the comparison can be extended to more general
non-minimal correlation functions and to non-leading powers in the
$1/N$ expansion (always in the semi-classical approximation, \ie at
leading order in the Yang--Mills coupling constant).  The $1/N$
corrections arise from several sources in the Yang--Mills instanton
calculations.  Firstly, the one-instanton measure  can be expanded in
a power series in inverse powers of $1/N$ of the form, $\mu(N) =
N^{1/2}(a + b/N + \cdots)$, where $a$, $b$, $\ldots$, are simple
coefficients.  Further subleading terms arise from the expressions for
the composite gauge invariant operators in the background of an
instanton.  All of these subleading terms ought to be identified with
D-instanton contributions in IIB string theory at order $\al$
relative to the classical Einstein--Hilbert action.

The purpose of this paper is to study such one-instanton effects in
semi-classical approximation at large $N$ in $\calN=4$ Yang--Mills
theory and see to what extent it may be possible to use the
correspondence to restrict the terms of higher order in $\alpha'$ in
the effective action of type IIB string theory.

In section \ref{overii} we will review the derivative expansion (\ie
the $\alpha'$ expansion) of the type IIB effective action.  The
leading, classical, term is of order $(\alpha')^{-4}$.  A variety of
arguments \cite{greenvanhove,gs,pvw} have established the form of the
leading corrections, which are interaction terms of order
$(\alpha')^{-1}$ and, there is a certain amount of information about
the terms of order $(\alpha')$, although little is known about terms
at order $(\alpha')^0$ and less about higher powers of $\alpha'$.
These interactions are consistent with the $SL(2,\Z)$ duality of the
theory which requires the presence of D-instantons of arbitrary
integer charge.  The form of these interactions embodies detailed
information concerning the effects of D-instantons on supergravity
amplitudes.  The AdS/CFT correspondence relates this to detailed
information concerning the instanton contributions to $\N4$ $SU(N)$
Yang--Mills correlation functions in the large $N$ limit.  Conversely,
knowledge of the Yang--Mills instanton contributions can be used to
develop further understanding of the IIB effective action.  The
earlier comparisons of multi D-instanton contributions to   type IIB
supergravity  in \AdS5s5\ with multi instanton contributions in $\N4$
large-$N$ Yang--Mills   theory was restricted to the minimal
correlation functions.  In this paper we will extend this by including
the effects of other fermionic moduli that enter into both sides of
the correspondence.  The discussion in section \ref{overii} is
included in order to motivate the selection of Yang--Mills correlation
functions to be studied in subsequent sections.

In preparation for this discussion, section \ref{oneinst}  will
give an overview  of the one-instanton moduli space and measure
for the $\N4$ supersymmetric $SU(N)$ Yang--Mills theory and its
large-$N$ limit. In the  $SU(2)$ case there are five bosonic
moduli, $x_0^m$, $\rho$,  associated with broken translation and
scale symmetries plus three moduli corresponding to global gauge
rotations.  There are also sixteen  fermionic moduli,
$\eta^A_{\a}$, $\xib^A_\adot$ (where $A=1,2,3,4$ labels a ${\bf
4}$ of the $SU(4)$ R-symmetry group) associated with the eight
broken Poincar\'e supersymmetries and eight broken conformal
supersymmetries, respectively. Classically,  the $SU(N)$ case has
additional gauge-dependent bosonic and  fermionic moduli
associated with the embeddings of $SU(2)$ in $SU(N)$.   The
additional fermionic variables are denoted by $\nu^A_{u}$ and
$\nub^{A\,u}$, where $u=1,\ldots,N$ is a colour index in the
fundamental representation of $SU(N)$. These satisfy constraints
which leave a total of $8(N-2)$ independent fermionic $\nu$ and
$\bar \nu$ variables.   These  are not true moduli since the
Yukawa interactions induce couplings between them  so there is an
explicit dependence on these variables in the instanton action.
Following \cite{doreyn}, $\nu$ and $\bar\nu$ can be integrated out
of the measure by introducing six auxiliary bosonic variables,
$\chi^a$ ($a=1,\ldots,6$), that couple to a $\nus$ bilinear. This
reduces the integration over these variables to a gaussian one, so
that their contribution to the measure can be computed exactly for
any $N$. Our discussion will  extend \cite{doreyn} to allow for
the presence of factors of $\nu$ and $\bar\nu$ in operators inside
correlation functions. As we will discuss such insertions bring
additional factors of the coupling $\gy$. For this reason, in
subsequent sections it will be important to  include lowest order
perturbative corrections involving the scalar field propagator in
the instanton background. The structure of this propagator has
been explicitly derived in the case of $Sp(n)$ gauge groups and
for arbitrary tensor product representations in \cite{cgt}.  This
construction of the instanton propagator will be reviewed in
section \ref{oneinst}  and generalized to the $SU(N)$ groups
needed  in this paper.

In section \ref{oneinsttwo} and appendices \ref{compfields} and
\ref{appinsprofiles} we will derive the contributions of these extra
fermionic variables to composite gauge invariant operators.  We will
again consider the operators in the multiplet containing the
superconformal currents, which couple to the Kaluza--Klein ground
states of the type IIB supergravity  fields.  In addition, we will
also  briefly discuss  the higher-dimensional operators that
correspond to the Kaluza--Klein excitations of these supergravity
fields.

Armed with the expressions for these operators we can calculate
instanton contributions to more general classes of correlation
functions. This is the subject of the sections~\ref{nonmincorr},
\ref{kknonmincorr} and \ref{othercorr}.  The correlation functions
to be considered are of several types:
\begin{itemize}
\item[i)]
The most obvious of these consists of terms that arise from the
\AdS5s5\ point of view by expanding the
coefficients in the effective action, which are non trivial functions
of $\tau$ and $\bar\tau$.  This leads
to terms of the form $\hat\tau^r
\Lambda^{16}$, where $\hat\tau$ is the fluctuation of the dilaton
around its background value and $\Lambda$ is the dilatino
field\footnote{Throughout the paper we will denote by $\Lambda$ the
type IIB dilatino and use $\lambda$ for the Yang--Mills elementary
fermion. The fermionic operator in the supercurrent multiplet dual to
the dilatino is denoted by $\hat\Lambda$}  (a 16-component Weyl
fermion).  Likewise, the expansion of the $\sqrt{-\det g}$ factor in
any instanton induced interaction leads to terms of the form $(\tr
h^m)^n \Lambda^{16}$ where $h_{\mu\nu}$ is the metric fluctuation. These
examples, in which there is a $\Lambda^{16}$  factor, are particularly
simple to evaluate since each $\Lambda$ necessarily soaks up precisely
one superconformal fermion mode.  The generalisation to interactions
with factors of $\calR^4$, $G\bar G \calR^2$, and others that arise at
order $1/\alpha'$ is straightforward but more complicated.
Interactions such as these play an important r\^ole in the discussions
of sections~\ref{L16Q2sugra} and \ref{nminhalf}.
\item[ii)]
There are single-instanton  effects that contribute to the leading
$N$-dependence and are necessary from the \AdS5s5\ point of view in
order for various field strengths to transform supercovariantly.  For
example the complex supercovariant third-rank field strength is
defined to be
\be
\label{supcov}
\hat{G}_{\mu\nu\rho} = G_{\mu\nu\rho}-
3{\bar\psi}_{[\mu}\g_{\nu\rho]}\Lambda -
6i{\bar\psi}^*_{[\mu}\g_\nu\psi_{\rho]} \, ,
\ee
where $G=d B$ is a complex combination of the field strengths of the
antisymmetric tensor potentials in the \NSNS\ and \RR\ sectors,
$\psi_\mu$ is the complex gravitino and $\Lambda$ the complex
dilatino\footnote{Conjugation of any complex spinor $\theta$ is
defined by $\bar\theta^*\equiv \theta\gamma^0$.}. This means that  in
addition to the $G\Lambda^{14}$ interaction there is a sixteen-fermion
interaction of the form $\psi^2\Lambda^{14}$.  This  was discussed in
detail in \cite{gs} where the complete scalar field dependence of this
interaction, including the D-instanton contributions, was deduced from
the constraints of type IIB supersymmetry.  In $AdS_5\times S^5$ the
gravitini can have vector index in internal directions and are in this
case denoted by $\chi$. The Yang--Mills correlation function
corresponding to $\chi^2\Lambda^{14}$ is (symbolically) $\langle
\calX(x_1) \calX(x_2) \hat\Lambda(x_3) \ldots
\hat\Lambda(x_{16})\rangle$.  The composite operators $\calX^{[AB]C}$
dual to internal components of the gravitino  are in a ${\bf 20^*}$ of
the $SU(4)$ R-symmetry group and each of them soaks up three fermionic
moduli.  The correlation function therefore soaks up a total of twenty
fermionic moduli and therefore must involve two $\nu^A$'s and two
$\bar\nu^A$'s in addition to the sixteen superconformal moduli.  In
section~\ref{nminhalf} we will show explicitly how the interaction
arises from the Yang--Mills point of view by including these extra
fermionic variables.
\item[iii)]
The Kaluza--Klein excitations of the supergravity fields involve
higher harmonics on the five-sphere which are described, in the
Yang--Mills theory, by fermionic bilinears in the instanton
background.  The excitations of any field contributing to the
interactions of order $\al^{-1}$ correspond to correlation functions
in which each five-sphere excitation is represented by products of
such bilinears.  This will be discussed explicitly in
section~\ref{kknonmincorr}.  In section~\ref{kkcorrel} we will also
comment on the special example of a near-extremal correlation
function of chiral primary operators  involving composite operators
dual to Kaluza--Klein excited states   in order to compare with the
perturbative analysis in \cite{nearextremal}.  Specifically, we will
consider the correlation function $\langle \calQ_4(x_1) \calQ_2(x_2)
\calQ_2(x_3) \calQ_2(x_4) \calQ_2(x_5)\rangle$, where $\calQ_\ell$
denotes a chiral primary operator of dimension $\Delta=\ell$.
We will discuss the partial non-renormalisation of this correlation
function at the non-perturbative level, \ie  the possibility that the
factorisation observed in the leading order perturbative contribution
is still valid when instanton contributions are taken into account.
\item[iv)]
There are subtle one-instanton contributions to certain
Yang--Mills  correlation functions in which some of the  fields are
replaced by their instanton profiles and other (scalar) fields are
contracted using the propagator in the instanton background. Some of
these contributions appear at first sight to have no correspondence
with the \AdS5s5\ theory.  Terms of this kind will be discussed in
sections \ref{nonmincorr} and  \ref{othercorr}.  We will see that they
are  accounted for in the supergravity description by including tree
diagrams in which one vertex is an induced instanton vertex of the
kind described in i) above.
\item[v)] All these Yang--Mills correlation functions have $1/N$
corrections to the leading $N^{1/2}$ instanton effects.  These arise
from various sources in the instanton measure as well as in the
expressions for the composite operators.  We will argue that a subset
of these correspond to terms in  the AdS supergravity of the form
$\alpha' R^2 \Lambda^{16}$, $\alpha' F_5^4\Lambda^{16}$, $\alpha' G^4
\Lambda^{16}$ and many related terms of order $\alpha'$ that are
expected to arise from the type IIB superstring effective action
(where $F_5$ is the self-dual Ramond--Ramond five-form field strength
and $G$ is a complex combination of the Ramond--Ramond and
Neveu-Schwarz--Neveu-Schwarz three-form field strengths).
\end{itemize}

We will see, at least qualitatively, how all these results match with
expectations from the \AdS5s5\ bulk supergravity based on the AdS/CFT
correspondence. A major aim of this paper is to see how far higher
derivative terms in the string effective action are encoded in the
$1/N$ expansion of  Yang--Mills instanton-induced correlation
functions.  The arguments in sections~\ref{nonmincorr} and
\ref{othercorr} make some headway towards this end although there are
some remaining ambiguities.  The situation is more intricate than in
the earlier work for two main reasons noted above.  On the Yang--Mills
side it is essential to include contributions involving the scalar
propagator in a one-instanton background.  Correspondingly, on the
supergravity side  amplitudes get contributions  from tree diagrams in
which one vertex is a D-instanton induced interaction and one or more
of the vertices are interactions of the classical supergravity.

\section{An overview of the type IIB derivative expansion}
\label{overii}

The type IIB string theory has a low energy effective action that can
be expressed as a power series in $\alpha'$ of the form
\be
S = {1\over \al^4}(S^{(0)} + \al^3 S^{(3)} + \al^4 S^{(4)} +
\al^5 S^{(5)} + \cdots + \al^r S^{(r)}+ \cdots)
\label{effact}
\ee
The first term, $S^{(0)}$, defines the classical IIB supergravity.
A great deal of information has been obtained concerning $S^{(3)}$
based on duality symmetries as well as a direct implementation of
supersymmetry.  This is the leading term which contains D-instanton
contributions that are related, in the \AdS5s5\ background, to
Yang--Mills instanton effects in the boundary theory.  The
interactions that enter into $S^{(3)}$ have the schematic form, in
the string frame,
\ba
&& {1\over \al} \int d^{10}x\, \sqrt{-g} \,\er^{-\phi/2}\,
\left\{ f_1^{(0,0)}(\tau, \bar\tau)\, \left( \calR^4 + \hat G\bar {\hat G}
\hat G \bar {\hat G} + \cdots\right)\right.  \nn \\
&& \left. + \cdots +
f_1^{(8,-8)}(\tau,\bar\tau)\,  \left(\hat G^8 +\cdots\right)+ \cdots +
f_1^{(12,-12)}(\tau, \bar\tau)\, (\Lambda^{16})+ \cdots \right\}.
\label{ints}
\ea
The precise contractions of the terms in this equation are given
in \cite{greengut} and the $\cdots$ indicate a variety of other terms that
enter at the same order in $\al$ and whose structure is also known
precisely.  The coefficients are functions of the complex scalar,
defined by $\tau = \tau_1 + i\tau_2 = C^{(0)} + i \er^{-\phi}$, where
$\phi$ is the dilaton and $C^{(0)}$ is the pseudoscalar \RR\ field. The
action is invariant under $SL(2,\Z)$ transformations acting
projectively on $\tau$,
\be
\label{tautrans}
\tau\to {a\tau + b \over c\tau + d}
\ee
(where $ad-bc =1$ with integer
$a,b,c,d$).  Under these transformations any field, $\Phi$,  is
multiplied by a phase,
\be
\label{phitrans}
\Phi \to \left({c\tau+d\over c\bar\tau +d}\right)^{q_\Phi\over 2}\,
\Phi,
\ee
where $q_\Phi$ is the charge with respect to the local $U(1)$ symmetry
that rotates the two chiral supercharges.  The curvature and the
five-form self-dual field strength, $F^{(5)}$, have $q_\Phi =0$ and
are invariant.  The complex three-form field strength,
\be
\label{gdef}
G \equiv {1\over \sqrt \tau_2} \left(\tau\, dB_{\rm NSNS}
+ dB_{\rm RR} \right)
\ee
(where $B_{\rm NSNS}$ is the Neveu-Schwarz--Neveu-Schwarz two-form and
$B_{\rm RR}$ is the Ramond--Ramond two-form), has $q_G =1$ and transforms
with weight $(-1/2,1/2)$,  whereas the complex conjugate field
strength, $\bar G$, has $q_{\bar G}=-1$.  Similarly, the dilatino
field, $\Lambda$, is a Weyl spinor and has $ q_{\Lambda} =3/2$ while
$q_{\bar \Lambda} = -3/2$.   The gravitino, $\psi_\mu$, has weight
$q_{\psi} =1/2$ and  $\bar \psi_\mu$ has $q_{\bar\psi} =-1/2$.
Finally, the fluctuation of the complex scalar, $\delta \tau\equiv
\hat \tau$ has $q_{\tau} =-2$ while $q_{\bar \tau} =-2$.  In order for
the effective action to be invariant the coefficient functions,
$f_1^{(w,-w)}$, in (\ref{ints})   transform as modular forms where
$(w,-w)$ denotes the holomorphic and anti-holomorphic weights. This
means that $f_1^{(w,-w)}$ transforms with a phase,
\be
\label{ftrans}
f_1^{(w,-w)}(\tau,\bar\tau) \to \left({c\tau + d\over c\bar\tau +
d}\right)^w \, f_1^{(w,-w)}(\tau,\bar\tau).
\ee

These modular forms are simple non-holomorphic Eisenstein series.  For
example, $f_1^{(0,0)}$ is defined by
\be
\label{fodef}
f_1^{(0,0)}(\tau,\bar\tau) = \sum_{m,n}{\tau_2^{3/2} \over |m +
n\tau|^3}.
\ee
It is convenient to expand this function in Fourier modes by using the
expansion
\ba
&&  f_1^{(0,0)} (\tau, \bar \tau) =
\sum_{K=-\infty }^\infty \calF^1_K(\tau_2) \er^{2\pi i K \tau_1} \nn\\
&&= 2\zeta(3)\tau_2^{{3\over 2}} + {2\pi^2\over 3}\tau_2^{-{1\over
2}} + 4 \pi \sum_{K=1}^\infty |K|^{1/2} \mu(K,1) \nn\\
&& \times
\left(\er^{2\pi i K \tau} + \er^{-2\pi i K \bar \tau} \right) \left(1
+ \sum_{j=1}^\infty (4\pi K \tau_2)^{-j} {\Gamma( j -1/2)\over
\Gamma(- j -1/2) j!} \right)\, .
\label{newsum}
\ea
The non-zero Fourier  modes are D-instanton contributions with
instanton number $K$ ($K>0$ terms are D-instanton contributions while
$K<0$ terms are anti D-instanton contributions). The measure factor is
defined by $\mu(K,1) =\sum_{m|K} m^{-2}$, which is a sum over the
divisors of $K$.  The coefficients of the D-instanton terms in
(\ref{newsum}), include an infinite series of perturbative
fluctuations around any charge-$K$ D-instanton. The leading term in
this series is the one of relevance to the comparison with the
semi-classical Yang--Mills instanton calculations of
\cite{banksgreen,bgkr,doreya}.  Of significance is the fact that the  zero
D-instanton term, $\calF^1_0$,  contains  two power-behaved
contributions that arise in string perturbation theory as  tree-level
and one-loop contributions. No higher-loop terms arise.

The modular form $f_1^{(w,-w)}$ is obtained by acting with $w$ modular
covariant derivatives on $f^{(0,0)}_1$
\be
\label{covder}
f_1^{(w,-w)}(\tau,\bar\tau) = D_{w-1} D_{w-2} \dots D_0 \,
f_1^{(0,0)}(\tau,\bar\tau) \, ,
\ee
where $D_w = (\tau_2\, {\partial \over \partial\tau} - i{w\over 2})$ 
is the covariant derivative which maps $f_1^{(w,-w)}$ into
$f_1^{(w+1,-w-1)}$.  The D-instanton and anti D-instanton
contributions to these functions can be extracted by applying the
derivatives to the instanton terms in $f_1^{(0,0)}$ in (\ref{newsum}).
In general the anti D-instantons are suppressed by powers of the
string coupling relative to the D-instantons. For example, in  the
$\Lambda^{16}$ interaction with coefficient
$f^{(12,-12)}(\tau,\bar\tau)$ the anti D-instanton contribution is
suppressed by a power $\gs^{24}$.  We will see later that this
suppression is explained on the Yang--Mills side of the AdS/CFT
correspondence by the presence of a large number of fermionic moduli.

The D-instanton contributions to the expression (\ref{ints}) corresponds to
very specific contributions of Yang--Mills instantons to correlation
functions of gauge-invariant operators at leading order in the
large-$N$ limit.  The explicit Yang--Mills calculations that have been
performed are for correlation functions in which the sixteen superconformal
fermion moduli are soaked up, but none of the other fermionic moduli
are required.  This tests a subset of the `minimal' predictions that
only involve couplings at order $\al^{-1}$ with all the fields in
the lowest Kaluza--Klein level.

The set of couplings appearing at order $\al^{-1}$
include fluctuations of the complex scalar in the
interaction terms (\ref{ints}).  The factor of $\er^{2\pi iK \tau}$ in
the D-instanton induced terms of the expansion of the modular forms
$f_1^{(w,-w)}(\tau,\bar\tau)$ leads to the series
\be
\sum_r{1\over r!} (2\pi i K\hat \tau)^r \er^{2\pi i K\tau_0}
\label{expdil}
\ee
multiplying the factors of $\calR^4$ or $\Lambda^{16}$. Here
$\tau = \tau_0 + \hat \tau$ where $\tau_0 = C^{(0)}_0 + i \er^{-\phi_0}$
denotes the constant part of the complex scalar and
\be
\hat \tau = \hat C^{(0)} + i\left(\er^{-(\phi_0 + \hat \phi)} -
\er^{-\phi_0}\right) = \hat C^{(0)}  - i\tilde \phi + \cdots,
\label{flucdil}
\ee
where we have defined the fluctuating part of the dilaton by $\hat
\phi = \phi -  \phi_0$ and then rescaled it so that $\tilde \phi =
\er^{-\phi_0} \hat \phi$ has a canonically normalised kinetic term.  A
term with $r$ powers of $\hat\tau$ corresponds in the Yang--Mills
theory to the insertion of  $\prod_{k=1}^r \Tr_N(F^-(x_k))^2$ into
any of the minimal correlation functions.  There is the possibility of
additional dilaton fluctuation terms that come from expanding the
factor of
\be
\er^{-\phi/2} = \er^{-\phi_0/2}(1 - \half\hat \phi + {1\over
8} \hat \phi^2 + \cdots) = \er^{-\phi_0/2}(1 - {\er^{\phi_0/2}\over 2}
\tilde \phi + {\er^{\phi_0}\over 8} \tilde \phi^2)
\label{dilfluc}
\ee
in (\ref{ints}). However, the fluctuations $\tilde \phi$ are
suppressed by powers of $\gs = \er^{\phi_0}$ in this expression and we
shall ignore them since we are concerned with the leading power in the
string coupling constant.  Since $\hat \phi = (\hat \tau - {\bar{\hat
\tau}})/2i$, it follows that the higher order terms in (\ref{dilfluc})
involve powers of $\bar \tau$ as well as powers of $\tau$.  The field
$\bar\tau$ couples to $\Tr_N(F^{(+)})^2$ which gets contributions from
eight fermion moduli in the instanton background and would have to be
included in a more complete treatment.  Similarly from the $\sqrt{-g}$
factor in (\ref{ints}), expanding the metric in fluctuations around
the background, $g_{\mu\nu} = g^{(0)}_{\mu\nu} + h_{\mu\nu}$, leads to
vertices at order $\al^{-1}$ of the form $(\tr h^n)^m\,\calR^4$, $(\tr
h^n)^m \, \Lambda^{16}$ etc. These interactions give rise to
amplitudes that correspond to Yang--Mills  correlation functions that
require the insertion of the additional fermionic moduli. In the
following we shall discuss in detail some effects of these effective
couplings.

There has been very little discussion in the literature concerning the
term $S^{(4)}$ in (\ref{effact}), but there must be interactions at
this order $(\al)^0$ which correct the coefficient of the classical
action $S^{(0)}$.  Although no such terms have yet been derived
directly from string amplitudes in the flat ten-dimensional
background, the AdS/CFT correspondence suggests that they should give
non-zero effects in \AdS5s5.  Such terms would give the $1/N^2$
corrections to the leading Yang--Mills correlation functions that are
necessary since the corresponding Yang--Mills theory has gauge group
$SU(N)$ rather than $U(N)$.

Some of  the many interactions contained in $S^{(5)}$ have been
motivated by a variety of arguments \cite{berkovafa,greenvanhove}.
Among  these interactions are the following (in string frame)
\ba
S^{(5)} &=& \al \int d^{10}x\, \sqrt{-g}\,
\er^{\phi/2} \, \left\{f_2^{(0,0)}(\tau, \bar\tau)\, (D^4\calR^4  +
(G\bar G)^2 \calR^4 + \cdots)\right. \label{actfiv} \\
&&\left.  \qquad + f_2^{(2,-2)}(\tau, \bar\tau) G^4 \calR^4 +
f_2^{(12,-12)} (\tau, \bar\tau) \calR^2 \Lambda^{16}
+f_2^{(14,-14)}(\tau, \bar\tau) G^4\Lambda^{16} + \cdots \right\}
\, , \nn
\ea
where the notation is symbolic since it does not specify the detailed
contractions between fields and the coefficients are also not specified.
The modular functions $f_2^{(w,-w)}$, and more generally
$f_l^{(w,-w)}$, are generalisations of
$f_1^{(w,-w)}$ given by the double sum
\be
\label{fldef}
f^{(0,0)}_l (\tau,\bar \tau) =
{\sum_{(m,n)\ne (0,0)}} {\tau_2^{l+ {1\over 2}}\over |m+n
\tau|^{2l + 1 }}
\ee
and
\be
f^{(w,-w)}_l (\tau,\bar \tau) = D_{w-1} D_{w-2} \dots D_0 \,
f_l^{(0,0)}(\tau, \bar\tau) \, .
\label{flw-wdef}
\ee
The  instanton contributions to  $f_l^{(0,0)}$ may again
be extracted by considering the Fourier expansion,
\be
\label{asymfour}
f_l^{(0,0)} (\tau,\bar \tau)   = \sum_K \calF^l_K \er^{2\pi i K
\tau_1},
\ee
where
\be
\label{asymzero}
\calF^l_0  = 2\zeta(2l+1)
\tau_2^{l+{1\over 2}}   + 2 \tau_2^{{1\over 2}  -l} {\pi^{2l}
\Gamma({1\over 2} - l)\zeta(1-2l) \over \Gamma(l + {1\over 2})}
\ee
and
\be
\label{asymk}\calF^l_K
= {4\pi^{l+{1\over 2}}\over \Gamma(l+{1\over 2})} \, |K|^{l}
\,\mu(K,l)\, \tau_2^{{1\over 2}} \calK_l (2\pi|K|\tau_2)\,
\er^{-2\pi |K|\tau_2},
\ee
with
\be
\label{measuredef}
\mu(K,l) =\sum_{\hat m|K} \hat m^{-2l},
\ee
and where ${\cal K}_l$ is a modified Bessel function. Using the
asymptotic expansion of $\calK_l$ leads to the expansion
\ba
f_l^{(0,0)} (\tau,\bar \tau) &=& 2
\zeta(2l+1) \tau_2^{l+{1\over 2}}  + 2\pi^{2l} {\Gamma({1\over
2}-l)\zeta(1 -2l)\over \Gamma(l+{1\over 2})}
\tau_2^{{1\over 2} - l}
\label{asymterm} \\
&+& 2{\pi^{l+{1\over 2}}\over \Gamma(l+{1\over 2})}
\sum_{K\ne 0} \mu(K,l) \er^{-2\pi(|K|\tau_2 - i K \tau_1)}
|K|^{l-{1\over 2}}
\left(1+{(4l^2-1) \over 16\pi |K| \tau_2} +\cdots\right)
\, , \nn
\ea
of which (\ref{newsum}) is a special case. The zero Fourier mode
$\calF^l_0$ is the dominant contribution for large $\tau_2$ and
contains the tree-level and $l$-loop terms\footnote{Note that if the dilaton is
constant and  \RR\  fields are rescaled in the usual manner
the tree level terms have the correct $\gs^{-2}$ behaviour.}.
The nonzero modes $\calF_K^l$ ($K\ne 0$) are the charge-$K$ D-instanton
contributions when $K$ is positive and anti D-instanton contributions
when $K$ is negative.  The series of terms in the last parentheses in
(\ref{asymterm}) represents the infinite series of perturbative
fluctuations around each D-instanton.

Classes of terms of higher order in derivatives have also been
suggested  \cite{berkovafa}.  These generalise \cite{greenwrong} to
interactions of the form  (in string frame)
\ba
S_{\rm gen} &=& \sum_{l,\hat{l}=1}^\infty
\sum_{p=2-2\hat l}^{2\hat l-2} c_{l,\hat{l}}
\,(\al)^{2l+2\hat{l}-5} \int d^{10}x \sqrt{-g}\,
\er^{(5l+3\hat l-\frac{17}{2})\phi} \: F_5^{4l-4}
G^{2\hat{l}-2+p} \bar{G}^{2\hat{l}-2-p} \nn \\
&& \qquad \quad\left\{ f_{l+\hat{l}-1}^{(p,-p)}(\tau,\bar{\tau})
\calR^4 + \cdots + f_{l+\hat{l}-1}^{(p+12,-p-12)}(\tau,\bar{\tau})
\Lambda^{16} \right\} \, ,
\label{genterm}
\ea
where the constant coefficients $c_{l,\hat{l}}$ should, in principle, be
determined by supersymmetry.   It follows from the expansion of the
modular form  $f_{l+\hat{l}-1}^{(p,-p)}$ in powers of $\tau_2^{-1}$
that any term in (\ref{genterm}) of order $(\al)^{2l+2\hat{l}-5}$ has
perturbative contributions at tree-level  and $(l+\hat l -1)$ loops
only, together with an infinite number of D-instanton contributions.
Clearly at any order in the $\al$ expansion it is
possible to consider the interactions involving additional powers of
the fluctuation of the complex scalar $\hat\tau$ or of the metric $h$,
as in the case of the $(\al)^{-1}$ vertices.

In making a correspondence between these supergravity interactions
and Yang--Mills theory we will make use of the usual
relationship between the parameters of $\calN=4$ supersymmetric
$SU(N)$ Yang--Mills and type IIB string theory in $AdS_5\times
S^5$,
\be
\gy^2 = 4\pi \gs, \qquad   \gy^2 N =
\left({L^2\over \alpha'}\right)^2  \, ,
\label{paramrel}
\ee
where $L$ is the $AdS_5$ scale.  Recall that the leading term in the
derivative expansion of the effective action, the Einstein-Hilbert
action that describes classical supergravity, is proportional to $L^8
(\alpha')^{-4}\gs^{-2} \sim N^2$.  The higher derivative interactions
of (\ref{genterm}) are suppressed by powers of $\al$.  They contain
information about non-perturbative contributions to amplitudes that
arise from D-instanton induced multi-particle interactions which, in
turn, give information about the expected contributions of Yang--Mills
instantons in the boundary Yang--Mills field theory.  Rewriting
(\ref{genterm}) in terms of the Yang--Mills parameters in
(\ref{paramrel}) gives
\ba
S_{\rm gen} &=& \sum_{l,\hat{l}=1}^\infty \sum_{p=2-2\hat l}^{2\hat l-2}
c_{l,\hat{l}} \, \left({\gy^2\, N\over L^4}\right)^{5/2-l-\hat{l}}
\int d^{10}x \sqrt{-g} \:
\er^{(5l+3\hat{l}-17/2)\phi} F_5^{4l-4}
G^{2\hat{l}-2+p} \bar{G}^{2\hat{l}-2-p} \nn \\
&& \left\{ f_{l+\hat{l}-1}^{(p,-p)}(\tau,\bar{\tau}) \calR^4 +
\cdots + f_{l+\hat{l}-1}^{(p+12,-p-12)}(\tau,\bar{\tau}) \Lambda^{16}
\right\} \, .
\label{gentermstring}
\ea
Whereas the minimal set of interactions considered in
\cite{greengut,bgkr,doreya}, arises from the terms of order
$(\al)^{-1}$ (with $l=\hat l=1$) and corresponds to Yang--Mills
instanton contributions of order $N^{1/2}$, the D-instanton
contributions that arise from terms of higher order in $\al$ should
correspond to contributions to Yang--Mills correlation functions that
are of higher order in $1/N$, in particular the Yang--Mills
counterparts of processes induced by terms of order $\al$ are expected
to behave as $1/N^{1/2}$. However, the situation is more complicated
than this.  We will see that in order to decipher the various
contributions it is also necessary to consider certain special
\AdS5s5\ tree amplitudes in which one of the vertices is a
multi-particle instanton induced vertex.

\section{One instanton in the $SU(N)$ theory}
\label{oneinst}

The moduli space and integration measure of the $K$-instanton
configuration can be obtained by use of the  ADHM construction
\cite{adhm}, which has been generalised to a great number of
situations over the years, see \cite{dhkm-rep} for a recent review.
The connections between Yang--Mills theory and string theory have led
to major simplifications in the description of the instanton measure
for large classes of supersymmetric theories \cite{wittenmeas},
\cite{douglasmeas}.  A particularly  thorough discussion of the
supermoduli measure in the case of the $\calN=4$ $SU(N)$ Yang--Mills
has been given in \cite{doreya}. There it was shown that at large $N$
the measure may be evaluated by a saddle point method, leading to a
direct comparison with the AdS/CFT predictions of \cite{banksgreen}
that extends the analysis of the $SU(2)$ case carried out in
\cite{bgkr}.  A brief summary of the ADHM variables in the context of
the $\calN=4$ theory is given in appendix \ref{adhmsum}.  Here we are
interested only in the case of a single instanton.

\subsection{The one instanton measure}
\label{oneinstmeas}

The classical solution for a single $SU(2)$ instanton embedded in the
gauge group $SU(N)$ is well known to give rise to $4N$ independent
bosonic moduli.  In the $\calN=4$ supersymmetric theory there are also
$8N$ fermionic moduli.  With a particular choice of coordinates the
bosonic variables that enter into the construction are the position
$x_0^m$ and the gauge dependent coordinates, $w_{u\adot}$ and $\bar
w^{\adot u}$, in the $SU(N)/SU(N-2)$ coset that represent the
orientations of the $SU(2)$ inside $SU(N)$.  The $SO(4)$ index
$m=1,2,3,4$ labels a euclidean four-vector, $\a = 1,2$, $\adot = 1,2$
are $SO(4)$ spinor indices of opposite chiralities and  $u =
1,\dots,N$ is a $SU(N)$ colour index.  This is a redundant set of $4N
+8$ variables which are subject to constraints.  A priori, the
gauge-invariant bilinear, $W^\bdot{}_{\adot} \equiv \bar w^{\bdot u}\,
w_{u\adot}$ has four components
\be
W^\adot{}_\bdot \equiv \half W^0 \d^\adot{}_\bdot +
\half W^c \tau_c^\adot{}_\bdot
\, ,
\label{Wcompdef}
\ee
where $\tau_c$ are the Pauli matrices. However, the components
$W^c$ vanish in the one instanton sector because of the ADHM
constraints. There is thus only one gauge invariant degree of freedom
which is identified with the instanton scale $\rho$
\be
\label{gaginv}
W^\adot{}_{\bdot} \equiv \bar w^{\adot u} \, w_{u\bdot}
= \delta^\adot{}_\bdot \rho^2 \, .
\ee
In the case of a $SU(2)$ instanton ($N=2$) the eight parameters
consist of the position, $x_0^m$, the scale, $\rho$, and three global
$SU(2)$ gauge orientations.

The fermionic moduli consist of the gauge singlets, $\eta^A_{\a}$,
$\xib_{\adot}^A$,  which each have $8$ components. These correspond
respectively to Poincar\'e and special supersymmetries broken in the
instanton background. In addition there are $8N$ variables $\nu^{A}_u$
and ${\bar\nu}^{Au}$ which can be thought of as the superpartners of
the bosonic variables $\bar{w}$ and $w$ and  are subject to the $16$
constraints
\be
\label{wdefso}
\bar w^{\adot u}\, \nu^A_{u} = 0 \, , \qquad
\bar \nu^{Au}\, w_{u \adot} = 0 \, ,
\ee
so that $\nu^A_{u}$ and $\bar\nu^{Au}$ each have effectively $4(N-2)$
components.  In the special case with $N=2$ there are no $\nu$, $\bar
\nu$ moduli due to the constraints (\ref{wdefso}) and the only
fermionic moduli are those associated with the sixteen broken
supersymmetries. The extension to $K$-instanton configurations with
$K>1$ involves the relative orientations of the instantons and gives
matrix generalisations of the above variables. These may be described
by using the full power of the ADHM construction and its
supersymmetric generalisation, but they will not be considered in this
paper.

When interactions are taken into account only a small subset of these
variables are genuine moduli since the Yukawa couplings
lead to a nontrivial moduli space action.  The subset of true moduli
comprises those protected by $\N4$ supersymmetry, \ie eight bosonic
coordinates -- the overall position, scale and gauge $SU(2)$
orientation of the instanton -- and sixteen fermionic coordinates --
corresponding to the supersymmetries that are broken by the presence
of the instanton.  The remainder of the moduli are the bosonic variables
associated with the relative embeddings of the $SU(2)$ instantons in
$SU(N)$ and their fermionic partners.  As shown in \cite{doreyb} the
moduli-space action contains a four-fermi interaction involving the
$\nu$ and $\bar \nu$ variables. They can be integrated out by the
conventional trick (familiar from the Gross--Neveu model
\cite{grossneveu} and also referred to as Hubbard-Stratonovich
transformation in the context of condensed matter physics) of
introducing an auxiliary bosonic variable, $\chi^a$ (which is a
$SO(6)$ vector), coupling to a fermion bilinear, to rewrite the
four-fermi term as a Gaussian integral.

\subsection{Correlation functions of gauge invariant composite operators}
\label{ginvarcorr}

In order to compute correlation functions in the semiclassical
approximation one must integrate the classical expressions for the
composite operators (which will be discussed in the next section) with
the appropriate collective coordinate integration measure.  In the
fermionic sector we are interested in including the modes
${\bar\nu}^{Au}$ and $\nu^A_u$ in the integration measure, in addition
to the superconformal modes $\eta^A_\a$ and ${\bar\xi}^{\adot A}$. The
`physical' integration measure in the $K=1$ sector will be denoted by
\be
Z=\int d\mu_{\rm phys} \, \er^{-S_{4F}} \, ,
\ee
where $d\mu_{\rm phys}$ includes all the relevant moduli and $S_{4F}$
is quartic in the $\bar\nu$ and $\nu$ modes
\begin{equation}
S_{4F}=\frac{\pi^{2}}{2\gy^2\rho^{2}}\veps_{ABCD}F^{AB}F^{CD} \, ,
\label{S4F}
\end{equation}
where $F^{AB}$ is a fermion bilinear given by
\be
F^{AB} = \frac{1}{2\sqrt{2}}({\bar\nu}^{Au}\nu^B_u -
{\bar\nu}^{Bu}\nu^A_u) \; .
\label{defhatLamb}
\ee
The result, obtained in \cite{doreyb}, is
\ba
Z &=& \frac{c\pi^{-4N}\gy^{4N}\er^{2\pi i\tau}}
{(N-1)!(N-2)!} \int d\rho \, d^4x_0 \, d^6\chi \prod_{A=1}^4\,
d^2\eta^A \, d^2 {\bar\xi}^A\, d^{N-2}{\bar\nu}^A\, d^{N-2}\nu^A \nn \\
&& \rho^{4N-7} \exp\left[-2\rho^2 \chi^a\chi^a+
\frac{4\pi i}{\gy}\chi_{AB} {F}^{AB}\right] \, ,
\label{intmeas}
\ea
where $c$ is a numerical constant independent of $N$ and $g$.
In the following we will omit such numerical factors in
intermediate steps and only reinstate all the coefficients in the
final formulae.
In (\ref{intmeas}) we have used the relation (\ref{gaginv}), which
implies
\be
dW^0 (\det_2 W)^{N-2} = d\rho \, \rho \, (\rho^2)^{2N-4} \, .
\label{rhomeas}
\ee
We will be interested in the computation of correlation functions
of operators which explicitly depend on the fermionic modes $\bar\nu$
and $\nu$. Thus the dependence on these collective coordinates comes
both from the quartic action $S_{4F}$ in (\ref{intmeas}) and from the
operator insertions. In order to deal with the combinatorics involved
in the integrations over these additional fermionic modes we define a
generating function which will allow us to compute generic
contributions of the $\bar\nu$ and $\nu$ modes coming either from the
measure or from the insertions. To this end we introduce sources
${\bar\vt}_A^u$ and $\vt_{Au}$ coupled to $\nu^A_u$ and
${\bar\nu}^{Au}$ respectively and define
\ba
Z[\bar\vt,\vt] &=&\frac{\pi^{-4N}\gy^{4N}\er^{2\pi i\tau}}
{(N-1)!(N-2)!} \int d\rho \, d^4x_0 \, d^6\chi
\prod_{A=1}^4 d^2\eta^A \, d^2 {\bar\xi}^A\,
d^{N-2}{\bar\nu}^A\, d^{N-2}\nu^A \nn \\
&& \rho^{4N-7} \exp\left[-2\rho^2 \chi^a\chi^a+
\frac{\sqrt{8}\pi i}{\gy}{\bar\nu}^{Au}\chi_{AB}\nu^B_u +
{\bar\vt}^u_{A}\nu^A_u + \vt_{Au}{\bar\nu}^{Au} \right] \nn \\
&=& \frac{\gy^{8}\er^{2\pi i\tau}} {(N-1)!(N-2)!}
\int d\rho \, d^4x_0 \, d^6\chi \prod_{A=1}^4
d^2\eta^A \, d^2 {\bar\xi}^A \, \rho^{4N-7} \nn \\
&& \left[ \sum_{a=1}^6 \left( \chi^a \right)^2 \right]^{2(N-2)}
\exp\left[-2\rho^2 \chi^a\chi^a-\frac{i\gy}{\pi\sqrt{8}}
{\bar\vt}_A^u\left(\chi^{-1}\right)^{AB} \vt_{Bu} \right] \, ,
\label{genfunct2}
\ea
where the gaussian  ${\bar\nu}$ and $\nu$ integrals have been
performed and we have used the fact that in the one instanton sector
\be
\left(
\det_4\chi_{AB} \right) = \frac{1}{2^6} \left[ \sum_{a=1}^6 \left(
\chi^a \right)^2 \right]^2.
\label{detchione}
\ee
Finally introducing six-dimensional spherical coordinates,
\be
\chi^a ~ \longrightarrow ~ (r,\Omega) \: , \quad \sum_a (\chi^a)^2 =
r^2 \, ,
\ee
$Z[\vt,\bar\vt]$ can be expressed in the form
\ba
Z[\bar\vt,\vt] &=&
\frac{2^{-29}\pi^{-13}\,\gy^{8}\er^{2\pi i\tau}}{(N-1)!(N-2)!}
\int d\rho \, d^4x_0 \, d^5\Omega \prod_{A=1}^4 d^2\eta^A \,
d^2 {\bar\xi}^A \, \rho^{4N-7} \nn \\
&& \int_0^\infty dr \, r^{4N-3} \er^{-2\rho^2 r^2}
\calZ(\vt,\bar\vt;\Omega,r) \, ,
\label{genfunctfin}
\ea
where we have reinstated all the numerical coefficients and
we have introduced the density
\be
\calZ(\vt,\bar\vt;\Omega,r) = \exp\left[-\frac{i\gy}{\pi r}
{\bar\vt}_A^u \Omega^{AB} \vt_{Bu} \right] \, .
\label{density}
\ee
The simplectic form $\Omega^{AB}$ is related to the unit vector
on the five sphere by
\be
\Omega^{AB}= \frac{1}{\sqrt{8}}\bar\Sigma^{AB}_a\Omega^a
\label{defOmega}
\ee
and the symbols $\Sbar^{AB}_a$ are defined in appendix
\ref{conventions}. The contributions of insertions of $\bar\nu$ and
$\nu$ to correlation functions in the semiclassical approximation in
the instanton background are evaluated by taking suitable derivatives
of $\calZ(\vt,\bar\vt;\Omega,r)$ with respect to the sources $\vt$ and
$\bar\vt$ in the previous expression.  The resulting integrations over
the remaining exact collective coordinates are performed after setting
$\vt=\bar\vt=0$.

In the following sections we will be particularly  concerned with the
dependence on the zero modes of composite gauge-invariant operators in
the instanton background.  We will  restrict our considerations to the
semiclassical approximation which keeps terms that are leading order
in the coupling, $\gy$.  This means that for much of the following we
will be interested in  the `classical' zero modes induced by the
instanton although we will also need to consider certain effects that
involve propagators in the instanton background and occur at the same
leading order in $\gy$.  In this approximation there is a set of
`minimal' correlation functions of $M$ operators, $\langle \calO(x_1)
\ldots \calO(x_M)\rangle$, which can be specially chosen so as to soak
up only the sixteen exact fermion zero-modes.  In these cases the
correlation functions are given simply by replacing each operator by
its zero mode approximation and integrating over the fermionic and
bosonic moduli with the appropriate measure
\be
\langle \calO(x_1) \ldots \calO(x_M)\rangle =C_1(\gy,N) \int
\frac{d^4 x_{0} \,d\rho}{\rho^5} \, d^5\Omega^a \, d^8\eta^A_{\a}\,
d^8\xi^A_\adot \, \tilde\calO(x_1) \ldots \tilde\calO(x_M) \, ,
\label{corrinst}
\ee
where $\tilde\calO$ indicates the expression for $\calO$ in terms
of the zero modes in the instanton background and $C_1(\gy,N)$ includes
all the factors coming from the one-instanton measure.  The instanton
contributions to correlation functions of this type involving
composite operators in the supercurrent multiplet were computed in
\cite{bgkr} in the case of an $SU(2)$ gauge group and generalised to
$SU(N)$ and to arbitrary instanton number in the large $N$ limit in
\cite{doreyn,doreya}.

The correlation functions studied in those papers are related, via the
AdS/CFT correspondence, to terms of order $(\al)^{-1}$ in the type IIB
superstring effective action \cite{banksgreen}.  In sections
\ref{nonmincorr}, \ref{kknonmincorr} and \ref{othercorr}  we will
be  interested in analysing the effect of the additional `non-exact'
modes $\bar\nu$ and $\nu$ on more general correlation functions, that
we will refer to as ``non-minimal''.

We first observe that the $\nu^A_u$ and $\bar\nu^{uA}$ variables
always enter the expressions for gauge invariant composite operators
in colour singlet pairs of the form $\bar\nu^{uA} \nu^B_u$. This is
because the only other collective coordinates carrying a colour index
are the bosonic variables $w_{\dot\alpha u}$ and $\bar
w_u^{\dot\alpha}$, but the contractions $\bar w^{\dot\alpha u}\nu^A_u$
and $\bar\nu^{Au} w_{u\dot\alpha}$ vanish due to the constraints
(\ref{wdefso})\footnote{These constraints also imply that the colour
index $u$ effectively has only $N-2$ components.}. More precisely we
will find that the $\bar\nu^{Au} \nu^B_u$ pairs always arise in either
the colour singlet combination that is a ${\bf 6}$ of $SU(4)$,
\be
\label{colousix}
(\bar\nu^A \nu^B)_{\bf 6} \equiv
\nus = ({\bar\nu}^{Au}\nu^B_u - {\bar\nu}^{Bu}\nu^A_u) \, ,
\ee
or in a ${\bf 10}$ of $SU(4)$,
\be
\label{colouten}
(\bar\nu^A \nu^B)_{\bf 10} \equiv
\nut = ({\bar\nu}^{Au}\nu^B_u + {\bar\nu}^{Bu}\nu^A_u) \, .
\ee
It is notable that it is the combination $\nus$ that enters into the action
and whose expectation value is determined \cite{doreya} in terms of
the simplectic form $\Omega^{AB}$ (\ref{defOmega}).

In preparation for the calculations of the following sections and
their comparison with AdS supergravity we describe here some general
features of correlation functions involving insertions of $(\bar\nu \nu)$
bilinears. Of particular relevance in the context of the AdS/CFT
correspondence will be the dependence of the correlation functions on
$N$ and on the coupling $\gy$. Using the generating function
(\ref{genfunctfin}), (\ref{density}) we can determine the powers of
$\gy$ and $N$ induced by the presence of $\nub$ and $\nu$ modes in full
generality. Let us consider separately the contributions of $\nsix$ and
$\nten$ bilinears. From (\ref{density}) we find
\begin{eqnarray}
&& \left(\nub^{u_{1}[A_{1}} \nu^{B_{1}]}_{u_{1}} \right) \ldots
\left(\nub^{u_{n}[A_{n}} \nu^{B_{n}]}_{u_{n}} \right) =
\left.\frac{\d^{2n}\calZ[\vt,\bvt]}{\d \vt_{u_{1}[A_{1}}
\d\bvt^{u_{1}}_{B_{1}]} \ldots}\right|_{\vt=\bvt=0} \nn \\
&& \sim \left(\frac{\gy}{r}\right)^{n} \left[ (N-2)^{n} \,
\Omega^{A_{1}B_{1}}\Omega^{A_{2}B_{2}} \ldots
\Omega^{A_{n}B_{n}} + O\left((N-2)^{n-1}\right) \right]
\label{nnusix}
\end{eqnarray}
and
\begin{eqnarray}
&& \left(\nub^{u_{1}(A_{1}} \nu^{B_{1})}_{u_{1}} \right) \ldots
\left(\nub^{u_{m}(A_{m}} \nu^{B_{m})}_{u_{m}} \right) =
\left.\frac{\d^{2m}\calZ[\vt,\bvt]}{\d \vt_{u_{1}(A_{1}}
\d\bvt^{u_{1}}_{B_{1})} \ldots}\right|_{\vt=\bvt=0} \nn \\
&& \sim \left(\frac{\gy}{r}\right)^{m} \left[ (N-2)^{m/2} \,
\left(\Omega^{A_{1}B_{2}}\Omega^{A_{2}B_{1}} \ldots
\Omega^{A_{m}B_{m-1}}\Omega^{A_{m-1}B_{m}} + {\rm permutations}
\right) \right. \nn \\
&& \left. + \, O\left((N-2)^{m/2-1}\right) \right] \, .
\label{nnuten}
\end{eqnarray}
Let us then consider a correlation function in which the operator
insertions contain $n$ $\nus$ and $m$ $\nut$ bilinears. In the
semiclassical approximation this takes the
form~\footnote{We are again
omitting numerical factors which are irrelevant for this general
discussion of the $N$-dependence.}
\begin{eqnarray}
&& I = \frac{c \gy^{8}\er^{2\pi i \tau}}{(N-1)!(N-2)!} \int d\rho \,
d^{4}x_{0} \, d^{5}\Omega \, d^{8}\eta \, d^{8}\xib \,
\rho^{4N-7} \nn \\
&& \int dr \, r^{4N-3}\er^{-2\rho^{2}r^{2}} \left.\frac{\d^{2n+2m}
\calZ[\vt,\bvt]}{\d \vt_{u_{1}[A_{1}} \d\bvt^{u_{1}}_{B_{1}]}\ldots
\d \vt_{v_{1}(C_{1}} \d\bvt^{v_{1}}_{D_{1})}}\right|_{\vt=\bvt=0}
\tilde\calO(x_{i};x_{0},\rho,\eta,\xib) \nn \\
&& = \frac{c \gy^{8+n+m}N^{n+m/2} \er^{2\pi i \tau}}{(N-1)!(N-2)!}
\int d\rho \, d^{4}x_{0} \, d^{5}\Omega \, d^{8}\eta \, d^{8}\xib
\, \rho^{4N-7} f(\Omega) \, \tilde\calO(x_{i};x_{0},\rho,\eta,\xib) \nn \\
&& \int dr \, r^{4N-3-n-m} \er^{-2\rho^{2}r^{2}}
\label{genernm}
\end{eqnarray}
where $\tilde\calO(x_{i};x_{0},\rho,\eta,\xib)$ and $f(\Omega)$
contain the dependence on the exact moduli and on the angles $\Omega$,
respectively. Computing the last integral gives
\begin{equation}
I = c\gy^{8+n+m} \er^{2\pi i \tau} \a(N) \int \frac{d\rho \, d^{4}
x_{0}}{\rho^{5}} d^{5}\Omega \, d^{8}\eta \, d^{8}\xib \, \rho^{n+m}
f(\Omega) \, \tilde\calO(x_{i};x_{0},\rho,\eta,\xib) \, ,
\label{genNdep}
\end{equation}
where
\begin{equation}
\a(N) = \frac{2^{-2N}\Gamma\left(2N-1-\frac{n+m}{2}\right)}
{(N-1)!(N-2)!}\left(N^{n+\frac{m}{2}}+O(N^{n+\frac{m}{2}-1})\right)
\sim N^{\frac{1}{2}+\frac{n}{2}}(1+O(1/N)) \,.
\label{Ndepnu}
\end{equation}
In (\ref{Ndepnu}) the $1/N$ corrections come both from
subleading terms in (\ref{nnusix}) and (\ref{nnuten}) and from the
asymptotic series expansion of the factorials.
In conclusion, in the large $N$ limit each $\nus$ insertion brings a
factor of $\gy\sqrt{N}$ while a $\nut$ only contributes a power of $\gy$.
It is important to remark that although the previous expressions
contain additional powers of the coupling associated with the $\nub$
and $\nu$ modes they contribute at leading non-vanishing order to the
non-minimal correlation functions.

In the following we will show that there are other classes of
contributions of the same order that must be included for
consistency. These are obtained by taking into account the leading
effect of quantum fluctuations in the background of the instanton. In
order to compute these contributions we will need the expression for
the scalar propagator in the instanton background which is reviewed in
the next subsection.

We conclude this section with a comment on the string theory
interpretation of the $1/N$ corrections in
(\ref{genNdep})-(\ref{Ndepnu}).  In sections \ref{nonmincorr} and
\ref{othercorr} we will analyse the leading order contributions to
some non minimal correlation functions as well as the first $1/N$
corrections to the leading order result in the large $N$ limit. From
the above discussion it is clear, however, that the calculation of any
correlation function produces an infinite series in $1/N$. This is
true in particular for the minimal correlation functions, in which the
additional fermionic modes $\nub$ and $\nu$ do not appear in the
operator insertions. For these correlation functions the dominant
large $N$ contribution in field theory is of order $N^{1/2}$,
corresponding to $m=n=0$ in (\ref{Ndepnu}). This result was shown in
\cite{doreya} to be in remarkable agreement with string theory
calculations involving the D-instanton induced couplings of order
$(\al)^{-1}$ in (\ref{ints}). The explanation of the series of $1/N$
corrections to the leading order result on the string side involves a
subset of the higher derivative couplings in the effective action
discussed in section \ref{overii}. The relevant terms are those of the
form
\be
\sum_{l=1}^\infty (\al)^{2l-3} \int d^{10}x \sqrt{-g}\,
\er^{(5l-\frac{11}{2})\phi} \: F_5^{4l-4} \left[
f_{l}^{(0,0)}(\tau,\bar{\tau}) \calR^4 + \cdots +
f_{l}^{(12,-12)}(\tau,\bar{\tau}) \Lambda^{16} \right] \, ,
\label{1overN}
\ee
obtained by setting $\hat l=1$, $p=0$ in (\ref{genterm}). These terms can
contribute to the minimal correlation functions because $F_5$ can be
set to its non-vanishing background value in (\ref{1overN}) which then
becomes an infinite series in $(\al)^2/L^4$. This is exactly what is
needed to reproduce the field theory expansion in $1/N$. We will need
to take into account similar effects in our analysis of non-minimal
correlation functions.

\subsection{Scalar field propagator in a one-instanton background}
\label{scalarprop}

The issue of constructing the Green function for a scalar field in the
background of (anti) self-dual Yang--Mills configurations was first
addressed in \cite{bccl}. The propagators for scalar fields in the
fundamental and adjoint representations of the gauge group were
considered in the background of the special $SU(2)$ $K$-instanton
solution of t' Hooft. In the same paper it was also shown how to
express the Green functions for spinors and vectors in terms of the
scalar one. The generalisation of these results to the general $SU(2)$
(anti-)self-dual Yang--Mills solution was given in \cite{cws}. A very
elegant formalism for the construction of scalar propagators in the
most general (anti-)self-dual background, valid for arbitrary
(compact) groups, was then developed in \cite{cfgt} for the case of
the fundamental representation and extended to any representation in
\cite{cgt}.

These Green functions play a crucial r\^ole in the study of quantum
effects in the instanton background, \ie in computing perturbative
expansions in topologically non-trivial vacua. However, as we will
show later, the scalar propagator is of relevance in the computation
of the non-minimal correlation functions of composite operators we are
concerned with already at lowest non-vanishing order in the coupling.
We briefly review here the construction of \cite{cgt} in the specific
case of interest of the adjoint representation of $SU(N)$ and in the
particularly simple situation of the $K$=1 sector. Some important
formulae are given in appendix \ref{compfields}.

In \cite{cgt} the more general problem of determining the Green
function for a scalar transforming under a product group $G_1\times
G_2$ was considered. The case of the adjoint representation of a group
$G$ can be obtained by a suitable projection considering $G_1=G_2=G$
with the scalar transforming under ${\bf f}\otimes{\bf \bar{f}}$,
where ${\bf f}$ denotes the fundamental. In the case of $SU(N)$ we are
interested in we will consider ${\bf N}\otimes{\bf \bar{N}}$. In
\cite{cgt} the construction of the propagator for a scalar transforming
under a tensor product representation was given using the ADHM
formalism for simplectic groups, here we will give the results
relevant for the adjoint representation of $SU(N)$.

The Green function we are concerned with is, in matrix form,
\be
G^{ABCD}(x,y) = \langle \varphi^{AB}(x) \varphi^{CD}(y) \rangle \, ,
\label{defprop}
\ee
where $\varphi^{AB}$ are the six real scalars of the $\N4$ SYM
theory. $G^{ABCD}(x,y)$ satisfies the equation
\be
D^2 G^{ABCD}(x,y) = -2\veps^{ABCD} \d(x-y) \, ,
\label{propeq}
\ee
with the appropriate covariant derivative in the instanton
background.

Working in the ADHM formalism in the one-instanton sector of $SU(N)$
it is convenient to describe the elementary fields in terms of
$[N+2]\times[N+2]$ matrices as in appendix \ref{compfields}. We write
the scalars $\varphi^{AB}$ as
\be
\varphi_u^{AB\,v} = \bar{U}_{u}{}^{r\a}\tilde{\varphi}_{r\a}^{AB\,s\b}
U_{s\b}{}^{v} \, ,
\label{tildescal}
\ee
where $U$ is the ADHM matrix defined in appendix \ref{adhmsum},
and consider the Green function
\be
\tilde{G}[_{r\a,s\b}\hspace*{-21pt}^{u\g,v\d}](x,y) =
\la \tilde{\varphi}_{r\a}{}^{u\g}(x)
\tilde{\varphi}_{s\b}{}^{v\d}(y) \ra \, .
\label{tildeprop}
\ee

The propagator $\tilde{G}$ for a scalar field $\tilde{\varphi}$
transforming under the tensor product ${\bf N}\otimes{\bf \bar{N}}$
can be expressed in terms of the projection operator $\calP$ defined
in (\ref{proj}) as \cite{cgt}
\be
\tilde{G}[_{r\a,s\b}\hspace*{-21pt}^{u\g,v\d}](x,y)
= \frac{1}{4\pi^2 (x-y)^2} \left[\calP(x)
\calP(y)\right]_{r\a}{}^{v\d} \left(\left[\calP(x)
\calP(y)\right]^{*}\right)^{u\g}{}_{s\b} + \frac{1}{4\pi^2}
C[_{r\a,s\b}\hspace*{-21pt}^{u\g,v\d}](x,y) \, ,
\label{NNbarprop}
\ee
where the term $C(x,y)$ is non-singular as $x\to y$ and was
computed in \cite{cgt} in terms of the ADHM matrices of bosonic
collective coordinates. This term vanishes in the one-instanton sector
in the case of the adjoint representation of $Sp(n)$, which implies in
particular that it is absent in the case of $SU(2) \equiv Sp(1)$, in
agreement with the results of \cite{bccl,cws}. However, this is not the
case for the adjoint of $SU(N)$ when $N \ge 3$. The ADHM construction
for unitary groups does not distinguish between $SU(N)$ and $U(N)$.
In the $SU(N)$ case we are interested in we must suitably project the
${\bf N}\otimes{\bf \bar{N}}$ expression (\ref{NNbarprop}) in order to
obtain the adjoint propagator. Unlike the $Sp(n)$ case the projection
onto the adjoint of $SU(N)$, which makes the propagator traceless,
does not cancel the term $C(x,y)$ in (\ref{NNbarprop}). In conclusion
the expression we obtain for the adjoint scalar propagator in the
one-instanton sector reads
\ba
&& \tilde{G}^{ABCD}[_{r\a,s\b}\hspace*{-21pt}^{u\g,v\d}](x,y)
= \frac{1}{4\pi^2} {\tilde B}
[_{r\a,s\b}\hspace*{-21pt}^{u\g,v\d}](x,y)
+ \frac{1}{4\pi^2} {\tilde C}
[_{r\a,s\b}\hspace*{-21pt}^{u\g,v\d}](x,y) \nn \\
&& = \frac{\gy^2 \veps^{ABCD}}{2\pi^2 (x-y)^2}\left\{ \rule{0pt}{16pt}
\!\left[\calP(x) \calP(y)\right]_{r\a}{}^{\!v\d} \left[\calP(y)
\calP(x)\right]_{s\b}{}^{\!u\g} \right. \nn \\
&& -\frac{1}{N} \left(
\left[\calP(x)\right]_{r\a}{}^{\!u\g}
\left[\calP(y)\calP(x)\calP(y)\right]_{s\b}{}^{\!v\d} +
\left[\calP(y)\right]_{s\b}{}^{\!v\d}
\left[\calP(x)\calP(y)\calP(x)\right]_{r\a}{}^{\!u\g}
\right) \nn \\
&& \left.+\frac{1}{N^2} \left[\calP(x)\right]_{r\a}{}^{\!u\g}
\left[\calP(y)\right]_{s\b}{}^{\!v\d}
\, \Tr \left[\calP(x)\calP(y)\right] \right\} \nn \\
&& +\frac{\gy^2\veps^{ABCD}}{4\pi^2 \, \rho^2} \left\{ \rule{0pt}{16pt}
\!\left[ \calP(x)b\bar b \calP(x)\right]_{r\a}{}^{\!u\g} \, \left[
\calP(y)b\bar b \calP(y)\right]_{s\b}{}^{\!v\d} \right. \nn \\
&& - \frac{1}{N}\left(
\left[\calP(x)\right]_{r\a}{}^{\!u\g}\left[\calP(y)b\bar b\calP(y)
\right]_{s\b}{}^{\!v\d}\, \tr_2\left[\bar b\calP(x)b\right]
\right. \nn \\
&& \left. + \left[\calP(y)\right]_{s\b}{}^{\!v\d}\left[\calP(x)b\bar
b\calP(x)\right]_{r\a}{}^{\!u\g} \, \tr_2\left[\bar b\calP(y)b
\right] \right) \nn \\
&& \left. + \frac{1}{N^2} \left[\calP(x)\right]_{r\a}{}^{\!u\g}
\left[\calP(y)\right]_{s\b}{}^{\!v\d} \,
\tr_2\left[ \bar b\calP(x)b\right]\tr_2\left[\bar b\calP(y)b
\right] \right\} \, ,
\label{propfin}
\ea
where $b$ and $\bar b$ are the ADHM matrices defined in appendix
\ref{adhmsum} and we have used the hermiticity of the projector
$\calP$. We use the notation ${\tilde
B}[_{r\a,s\b}\hspace*{-21pt}^{u\g,v\d}](x,y)$ to denote the terms with
$1/(x-y)^2$ singularity as $y\to x$ and ${\tilde
C}[_{r\a,s\b}\hspace*{-21pt}^{u\g,v\d}](x,y)$ for the non-singular
terms corresponding to $C(x,y)$ in (\ref{NNbarprop}). In
(\ref{propfin}) we have reintroduced the $SU(4)$ indices and inserted
the factor of $\gy^2$ which follows from the normalisation we are
using in which the action is written with an overall $1/\gy^2$. The
above expression for the propagator is traceless as appropriate for
fields in the adjoint of $SU(N)$. Moreover the terms ${\tilde
C}[_{r\a,s\b}\hspace*{-21pt}^{u\g,v\d}](x,y)$, which are those in the
last four lines  of (\ref{propfin}), exactly cancel for $N=2$.

The expression for the propagator in (\ref{propfin}) is
rather complicated. The explicit form of its components can easily be
obtained using the expression for the projector given in
(\ref{projoneinst}) of appendix \ref{compfields}. However, we will
see that this is not  necessary in the computation of the correlation
functions of  gauge-invariant composite operators we are interested
in.  The contributions to correlation functions in the instanton
background produced by insertions of (\ref{propfin}) will be matched
with contributions to amplitudes computed in $AdS_5\times S^5$. In
particular the $1/N$ nd $1/N^2$ terms in (\ref{propfin}), which ensure
the tracelessness in colour space, will be important in reproducing
subleading effects in the AdS calculations.

\section{Composite operators in the one-instanton sector}
\label{oneinsttwo}

In this section we will discuss the construction of supersymmetry
multiplets of composite operators in $\N4$ SYM and their expressions
in the one instanton background. We will in particular analyse the
effect of the inclusion of the extra fermionic variables  $\bar\nu^{A u}$
and $\nu^A_u$.

The multiplets of operators we are interested in belong to the special
class of `ultra-short' multiplets of the superconformal group
$SU(2,2|4)$ characterised by the fact that their lowest component is a
scalar of conformal dimension $\ell$, transforming in the
representation $[0,\ell,0]$ of the $SU(4)$ R-symmetry group.
These operators are of the form
\be
\calO_\ell^{(a_1\dots a_\ell)} = \frac{N}{(\gy^2 N)^{\ell/2}}
\, \Tr \left(\v^{a_1}\dots\v^{a_\ell}\right)
\raisebox{-4pt}{$\Big|_{[0,\ell,0]}$} \: ,
\label{cpoell}
\ee
where the subscript $[0,\ell,0]$ on the right hand side means
projection on the corresponding $SU(4)$ representation, which requires
complete symmetrisation and subtraction of the traces in the $a_i$
indices. Notice in particular the normalisation we choose for the
Yang--Mills operators. This is the natural choice of normalisation to
be used in the AdS/CFT correspondence since it does not require any
further rescaling of the amplitudes on the gravity side to match the
dependence on $N$ in Yang--Mills correlation functions \cite{intri}.

The case of the supercurrent multiplet corresponds to $\ell=2$.  We
will also be interested in extending the list of operators to include
those that correspond to the Kaluza--Klein towers of states that arise
on the gravity side in the $AdS_5\times S^5$ background.   These are
multiplets in the same class with $\ell \ge 3$.  We will focus on the
multiplet corresponding to the first Kaluza--Klein excited level,
$\ell =3$. The cases $\ell=2,3$ that  we consider in more detail in
the following are characterised by a  further shortening \cite{gunmar}.

\subsection{Composite operators of the supercurrent multiplet}
\label{nusupercurr}

The $\N4$ supercurrent multiplet was first constructed in the abelian
case in \cite{dewit}. The construction is more involved in the non
abelian case because of additional terms in the supersymmetry
transformations of the elementary fields and the full multiplet has
not been given explicitly in the literature. We will describe here the
construction of the relevant terms (in a sense that will be explained
in the following) for the operators whose correlations functions we
will study later on.

The $\N4$ multiplet of elementary fields comprises six real scalars,
$\v^a$, four Weyl fermions, $\lambda^A_\a$, and a vector, $A_m$.
In the following it will often be convenient to represent the scalars
as
\be
\v^{AB} = \frac{1}{\sqrt{8}} \, \Sbar^{AB}_a \v^a \, ,
\label{antisymmscal}
\ee
satisfying the reality condition
$\bar\v_{AB}=\half\veps_{ABCD}\v^{CD}$.  The expressions for the
elementary fields in the instanton background can be generated by
acting on the BPST instanton solution for the gauge potential $A_m$
with the 16 broken Poincar\'e and special supersymmetries, namely
$Q_A^\a$ and $\bar S_A^\adot$.  The same procedure can be repeated to
generate the dependence on the superconformal modes in the
gauge-invariant composite operators in the supercurrent multiplet
starting from the top-component $\calC$.  We will instead  construct
the expressions of the composite operators in terms of elementary
fields and then obtain their profiles in the one instanton background
by substituting the classical expressions of the elementary fields
which are given in appendix \ref{compfields}. This will allow us to
derive the dependence on the additional modes $\nu^A_u$ and
$\bar\nu^{Au}$ as well. We will only consider explicitly the simple
cases of operators obtained by the first few supersymmetry variations
of the chiral primary, but the procedure can be extended to more
complicated cases.  An alternative way of constructing the multiplet
is to use the decomposition in terms of $\calN=1$ supermultiplets
along the lines of what was done in \cite{konishi} for the Konishi
multiplet. In this case a superfield formulation can be used.

For the multiplet of superconformal currents the starting point is the
chiral primary operator $\calQ_2$. This is a scalar of
dimension 2 transforming in the representation ${\bf 20^\prime}$ of
$SU(4)$ and can be written in the form
\be
\calQ_2^{ab} = \frac{1}{\gy^2} \Tr \left(\v^a \v^b
- \frac{1}{6}\d^{ab} \v^c\v_c\right) \, ,
\label{Q2defstand}
\ee
or equivalently in a form more suitable for instanton calculations
as~\footnote{Since this operator is not a conserved current its
normalization is arbitrary.}
\begin{equation}
\calQ_2^{[A_1B_1][A_2B_2]} = \frac{1}{\gy^2} \Tr \left(
2 \v^{A_1B_1}\v^{A_2B_2} + \v^{A_1A_2}\v^{B_1B_2} +
\v^{A_1B_2}\v^{A_2B_1}\right) \, .
\label{O2newdef}
\end{equation}
The other members of the multiplet are obtained by successive
applications of supersymmetry to this chiral primary operator.
Here we will consider only those operators generated by applying
the supersymmetries $Q_A$, of one particular $SU(4)$ chirality.
The first variation of $\calQ_2$ gives a spinor in the representation
${\bf 20}^*$ of the R-symmetry group,
\be
\calX^{A_1[A_2B_2]}_\a = \frac{1}{\gy^2} \Tr\left(
2 \lambda^{A_1}_\a \v^{A_2B_2} + \lambda^{A_2}_\a \v^{A_1B_2}
- \lambda^{B_2}_\a \v^{A_1A_2} \right) \, .
\label{chinewdef}
\ee
A further $Q_A$ supersymmetry generates two operators, a
scalar and an antisymmetric two-form, transforming  in the
representations {\bf 10} and {\bf 6} of $SU(4)$. Their complete
non abelian expressions are given by
\ba
&& \calE^{(A_1A_2)} = \frac{1}{\gy^2} \Tr\left(
- \lambda^{\a A_1}\lambda^{A_2}_\a
+ t^{(A_1A_2)}_{CDEFGH} \, \v^{CD}\v^{EF}\v^{GH} \right)
\label{Enewdef} \\
&& \calB^{[A_1A_2]}_{mn} = \frac{1}{\gy^2}\Tr \left( \lambda^{\a A_1}
\s_{mn\,\a}{}^\b \lambda^{A_2}_\b + 2i F_{mn} \v^{A_1A_2} \right) \, .
\label{Bnewdef}
\ea
where the tensor $t^{(A_1A_2)}_{CDEFGH}$ projects the product of three
{\bf 6}'s onto the {\bf 10} of $SU(4)$. The second term in
(\ref{Enewdef}) is not present in the abelian case. As shown in
\cite{dfs} its presence is crucial at the perturbative level in proving
non-renormalisation properties of correlation functions involving the
operator $\calE$. However, this term will not be relevant for our analysis of
non-perturbative effects  at leading order in the coupling. A
further supersymmetry transformation gives the spinor
$\hat\Lambda_\a^A$ in the representation {\bf 4} of $SU(4)$ which
will play an important r\^ole in our calculations of correlation
functions.  The form of the operator $\hat\Lambda^A_\a$ is
\be
\hat\Lambda^A_\a = \frac{1}{\gy^2} \Tr \left\{\s^{mn\,\b}_{\,\a}
F_{mn} \lambda^A_\b + \left[{\bar\v}_{BC},\v^{CA} \right] \lambda^B_\a
+ \left(\Dsm_{\a\adot} {\bar\lambda}^\adot_B + \left[\lambda^C_\a,
{\bar\v}_{BC}\right] \right) \v^{AB} \right\} \, .
\label{Lamnewdef}
\ee
Again the cubic terms are only present in the non abelian case and
will not be needed in the semiclassical approximation.
The last operator that we will consider is obtained by applying
another $Q_A$ supersymmetry transformation,
\be
\calC = \frac{1}{\gy^2} \Tr \left(F^-_{mn}F^{-\, mn}\right) + \cdots \, ,
\label{Cnewdef}
\ee
where the ellipsis stands for terms which will not be relevant in the
following.

The detailed expressions for these operators in the instanton
background are obtained in appendix \ref{appinsprofiles}   by
substituting the instanton profiles  of the elementary fields, $A_m$,
$\lambda^A_\alpha$ and $\varphi^{AB}$ that are given in appendix
\ref{compfields}.  These results can be  summarised as follows.
Starting from the component $\calC$, which does not contain any
superconformal zero-modes, the other components listed in
(\ref{Q2defstand})-(\ref{Lamnewdef}) are obtained by adding
superconformal modes and/or $\bar\nu\nu$ pairs in the combination
(\ref{colouten}), \ie, in a {\bf 10} of $SU(4)$.  The schematic
structure of the resulting expressions is
\ba
\calC_{\bf 1} &=& (\rho f)^4 \nn \\
\hat \Lambda_{\bf 4} &=& (\rho f)^4\, \zeta \nn \\
\calB_{\bf 6} &=& (\rho f)^4 \, (\zeta\zeta)
\label{opsnu} \\
\calE_{\bf 10} &=& (\rho f)^4 \, (\zeta\zeta) + \left(\rho^2
f^3\right) \, \nten \nn \\
\calX_{\bf 20^*} &=& (\rho f)^4 \, (\zeta\zeta\zeta) +
\left(\rho^2 f^3 \right)\, [\zeta\nten] \nn \\
\calQ_{\bf 20^\prime} &=& (\rho f)^4 \,
(\zeta\zeta\zeta\zeta) + \left(\rho^2 f^3\right) \, [\zeta\zeta\nten]
+ \left(f^2\right) \, [\nten\nten] \, ,  \nn
\ea
where
\be
f={1\over (x-x_0)^2 + \rho^2},
\label{ffdef}
\ee
and
\be
\zeta^A_\a = \frac{1}{\sqrt{\rho}} \left( \rho\eta^A_\a -
(x-x_0)_m\sigma^m_{\a\adot} {\bar\xi}^{\adot A} \right) \, .
\label{zetadef}
\ee
These expressions indicate the combinations of superconformal modes
and $\bar\nu \nu$ bilinears that enter into the various operators (the
precise index structure is given in appendix
\ref{appinsprofiles}). The operators in  (\ref{opsnu}) contain other
terms involving additional $\bar\nu\nu$ bilinears (but not $\zeta$'s)
which will not be relevant for our calculations. This is because, as
shown in section \ref{ginvarcorr}, the insertion of each
$(\bar\nu\nu)$ bilinear brings a power of the coupling
$\gy$. Therefore in the semiclassical approximation only the minimal
number of such insertions needed for a non-vanishing result must be
considered.  The operators in (\ref{opsnu}) are the ones which contain
terms with a minimum number of four or less fermions.  The
construction of the operators higher up the multiplet, which have a
larger number of fermionic factors, is more complicated.  For example,
the anti dilatino, $\bar {\hat\Lambda}_{\bf 4^*}$, and $\bar \calC$ 
can contain up to seven and eight superconformal modes respectively.

It is crucial for the comparison with $AdS$ calculations
that the  bulk-to-boundary propagator $K_\D(x;x_0,\rho)$ from the
point $(x_0,\rho)$ in $AdS_5$ to the point $(x,0)$ on the boundary for
a scalar field of dimension $\D$ is given by
\be
\left[ \rho
f(x;x_0,\rho) \right]^\D = K_\D(x;x_0,\rho) =
\frac{\rho^\D}{[(x-x_0)^2+\rho^2]^\D} \, .
\label{fKrel}
\ee
In (\ref{opsnu}) we have used the compact notation $\nten$ (defined in
(\ref{colouten}) to denote a $\bar\nu\nu$ pair in the {\bf 10}  of
$SU(4)$ and we will similarly denote a bilinear in the {\bf 6} by
$\nsix$ (see (\ref{colousix})). Note that the operator $\calC_{\bf 1}$
contains no  fermionic moduli at all while $\hat\Lambda^A_\a$ and
$\calB^{AB}_{mn}$ depend only on $\zeta^A_\a$ but not on
$\bar\nu^{Au}$ and $\nu^A_u$.

Although the detailed structure of the gauge invariant composite
operators in appendix \ref{appinsprofiles}) looks complicated the
general structure indicated in (\ref{opsnu}) is relatively simple due
to subtle cancellations.  The profiles of the elementary fields in the
instanton background contain terms with  different  numbers of
superconformal modes as well as $\bar\nu$ and $\nu$ modes. Moreover
elementary Yang--Mills fields depend on the superconformal zero modes
not only through the combination $\zeta_\alpha^A$ (\ref{zetadef}) but
$\eta^A_\a$   and $\bar\xi^A_\adot$ enter explicitly. Any of the
fundamental fields, $\Phi$,   has the structure
\be
\Phi = \hspace*{-0.3cm} \begin{array}[t]{c}
{\displaystyle \sum_{q=0}^4} \\
{\scriptstyle p+4q \le 16}
\end{array} \hspace*{-0.3cm}
\Phi^{(p+4q)} + {\rm terms ~ involving} ~ \bar\nu ~
{\rm and} ~ \nu \, ,
\label{eleminst}
\ee
where the superscript $(p+4q)$ refers to the number of superconformal
modes and $p=0$ for $A_m$, $p=1$ for $\lambda^A_\a$, $p=2$ for
$\v^{AB}$ and $p=3$ for $\bar\lambda_A^\adot$. When these are combined
to form the gauge invariant operators in the supercurrent multiplet
there are cancellations that lead to the expressions summarised in
(\ref{opsnu}). In particular, it is important that
 only the terms with the minimum number
of superconformal modes survive and these always appear in the
combination $\zeta_\alpha^A$ defined in (\ref{zetadef}).
These cancellations are crucial
for reproducing the correct form of the
supergravity amplitudes in $AdS_5\times S^5$ from
the Yang--Mills viewpoint. For instance,
 terms with five superconformal zero modes
cancel out of the complete expression for $\hat\Lambda^A_\a$,
\ie, in the notation of
(\ref{eleminst}) the combination
\ba
&& \Tr \left\{\s^{mn}_{\,\a}{}^\b \left(F^{(4)}_{mn} \lambda^{(1)\, A}_\b
+ F^{(0)}_{mn} \lambda^{(5)\, A}_\b \right) +
\left[{\bar\v}^{(2)}_{BC},\v^{(2)CA}\right]\lambda^{B(1)}_\a \right. \nn \\
&& \left. + \left(\Dsm_{\a\adot}
{\bar\lambda}^{(3)\adot}_B + \left[\lambda^{(1)C}_\a,
{\bar\v}^{(2)}_{BC}\right] \right) \v^{(2)AB} \right\} \,
\label{Lam5canc}
\ea
does not have a $\zeta^5$ contribution although the individual
terms in the sum do.

The operators listed in (\ref{opsnu}) are the simplest examples of
operators  in the superconformal current multiplet.  The complete
multiplet also contains the stress-energy tensor $\calT_{mn}$,
which is a $SU(4)$ singlet, the supersymmetry currents
$\Sigma_{m\,\a}^A$ and the R-symmetry currents $\calJ_{mA}{}^{B}$,
transforming respectively in the {\bf 4} and {\bf 15} of $SU(4)$. In
addition there are the operators conjugate to those in (\ref{opsnu}).
The dependence on the zero modes of these operators is
considerably more complicated to derive in detail.  In the following
we will restrict our considerations only to the components in
(\ref{opsnu}).

\subsection{Higher dimensional operators --- Kaluza--Klein excitations}
\label{nukaluza}

We will also consider operators corresponding to Kaluza--Klein modes on $S^5$
in ultra-short multiplets of the same type as the supercurrent multiplet.
The complete spectrum of type IIB supergravity in $AdS_5\times
S^5$ was constructed in \cite{kimromvan,gunmar}, where it was shown that the
Kaluza--Klein towers of states are organised in different branches.
The multiplets of operators  on the field theory side corresponding to
each level in the Kaluza--Klein tower are constructed from the lowest
component as with the supercurrent multiplet. For simplicity,
in the following we will mainly focus on operators dual to states in the
first Kaluza--Klein excited level. These belong to a multiplet that
descends from the chiral primary operator $\calQ_3$ in the {\bf 50} of
the $SU(4)$ R-symmetry group whose expression is
\ba &&
\calQ_3^{[A_1B_1][A_2B_2][A_3B_3]} = \frac{1}{(\gy^2N)^{3/2}}
{\rm Tr}\left(\varphi^{A_1B_1} \varphi^{A_2B_2} \varphi^{A_3B_3}  +
\varphi^{A_1B_1} \varphi^{A_3B_3} \varphi^{A_2B_2}  \right. \nn \\
&& + \varphi^{A_1B_1} \varphi^{A_2B_3} \varphi^{A_3B_2}  +
\varphi^{A_1B_1} \varphi^{A_3B_2} \varphi^{A_2B_3} + \varphi^{A_1B_3}
\varphi^{A_2B_2} \varphi^{A_3B_1} \nn \\
&& \left. + \varphi^{A_3B_1} \varphi^{A_2B_2}\varphi^{A_1B_3} +
\varphi^{A_1B_2} \varphi^{A_2B_1}
\varphi^{A_3B_3} + \varphi^{A_2B_1} \varphi^{A_1B_2}
\varphi^{A_3B_3}\right) \, .
\label{O3newdef}
\ea

The construction of the superconformal descendants is straightforward
but laborious. For definiteness, in the following we will illustrate
the way the correspondence works for correlation functions involving
the operator $\hat\Lambda^{A_1A_2A_3}_\a$ that is obtained from the
third supersymmetry variation of $\calQ_3$. This has the form
$\delta^3 \calQ_3 = \delta^3\varphi \varphi \varphi+ \delta^2\varphi
\delta\varphi \varphi + \delta\varphi\delta\varphi\delta\varphi$.
This is the operator corresponding to the first Kaluza--Klein excited
state of the dilatino.  An analogous analysis can be repeated  for all
the  operators in the multiplet as well as for multiplets
corresponding to higher Kaluza--Klein modes.  The expression for the
operator  $\hat\Lambda^{A_1A_2A_3}_\a$, which is a spinor of dimension
9/2 and transforms in the ${\bf 20^*}$ of $SU(4)$, reads
\ba
\hat\Lambda^{A_1(A_2A_3)}_\a &\!=\!&
\frac{1}{(\gy^2N)^{3/2}} \left\{ \Tr \left[ 2 \lambda^{A_1}_\a
\left( \lambda^{\b\,A_2}\lambda_\b^{A_3} + \lambda^{\b\,A_3}
\lambda_\b^{A_2} \right) + \lambda^{A_2}_\a \left(
\lambda^{\b\,A_1}\lambda_\b^{A_3} + \lambda^{\b\,A_3}
\lambda_\b^{A_1} \right) \right. \right. \nn \\
&& \left. + \lambda^{A_3}_\a \left(
\lambda^{\b\,A_1}\lambda_\b^{A_2} + \lambda^{\b\,A_2}
\lambda_\b^{A_1} \right) \right] \label{Lam20newdef} \\
&& \left. + \Tr \left[ F_{mn}
\s^{mn \, \b}_{\,\a} \left( \left\{ \lambda^{A_2}_\b, \v^{A_1A_3}
\right\} + \left\{ \lambda^{A_3}_\b, \v^{A_1A_2} \right\} \right)
\right] + \cdots \right\} \nn \, ,
\ea
where again the $\cdots$ refers to terms omitted since irrelevant for
calculations at leading order in the coupling.  These non-leading
terms have the form $\bar \lambda \varphi\varphi$ and
$[\varphi,\varphi]\lambda$, which give contributions of higher
order in $\gy$ because of the presence of extra fermionic
modes.

The precise form of $\hat\Lambda^{A_1A_2A_3}_\a$ in an instanton
background is given in terms of the bosonic and fermionic moduli in
appendix \ref{appinsprofiles}.  Since calculations of this type are rather
complicated  only one other operator in the $\ell=3$ Kaluza--Klein multiplet
is  considered in appendix \ref{appinsprofiles}, namely, the operator
conjugate to the first excited state of the dilaton, $\calC_{\bf 6}$.
However, from these examples it is clear that  the schematic structures of
the operators corresponding to the multiplet of first Kaluza--Klein
excited states (the $\ell=3$ supermultiplet) are given by
\ba
\calC_{\bf 6} &=&
\left(\rho^4 f^5\right) \nsix \nn \\ \hat \Lambda_{\bf 20^*} &=&
\left(\rho^4 f^5\right) [\zeta\nsix] \nn \\ \calB_{\bf 20^\prime} &=&
\left(\rho^4 f^5\right) [\zeta\zeta\nsix]
\label{kkopsnu} \\
\calE_{\bf 45} &=& \left(\rho^4 f^5 \right) [\zeta\zeta\nsix] +
\left(\rho^2 f^4\right) [\nten\nsix] \nn \\
\calX_{\bf 60^*} &=& \left(\rho^4 f^5\right)
[\zeta\zeta\zeta\nsix] + \left(\rho^2 f^4 \right) [\zeta\nten\nsix] \nn \\
\calQ_{\bf 50} &=& \left(\rho^4 f^5\right)
[\zeta\zeta\zeta\zeta\nsix] + \left(\rho^2 f^4\right)
[\zeta\zeta\nten\nsix] + \left(f^3\right) [\nten\nten\nsix] \nn \, ,
\ea
where only the analogues of the operators in (\ref{opsnu}) have been
considered.  These expressions are consistent with the dependence on
$N$ and $\gy$ that is expected on the basis of the AdS/CFT
correspondence, as we will see later.   As in the case of the
supercurrents described in the previous subsection these operators
also contain additional terms with more $\bar\nu\nu$ bilinears that we
do not need to consider at leading order in $\gy$.

We can now compare the form of the operators in this multiplet with
those in (\ref{opsnu}) and notice that in the instanton background the
profile of a given operator in the $\ell=3$ multiplet differs from the
corresponding operator in the $\ell=2$ multiplet by the prefactor
determined by the dimension of the field and by the presence of an
extra $\bar\nu\nu$ bilinear in the {\bf 6} of $SU(4)$. This structure
will prove essential in matching correlation functions involving these
higher dimensional operators with the corresponding amplitudes in AdS.
In particular the cancellation of terms with a higher number of
superconformal zero modes in (\ref{kkopsnu}) is crucial as will be
discussed in section \ref{kknonmincorr}. These cancellations are
again non trivial. For instance, in $\hat\Lambda^{A_1A_2A_3}_\a$
contributions cubic in the $\zeta$'s are cancelled between the two
traces in (\ref{Lam20newdef}).

The pattern that appears in (\ref{opsnu}) and (\ref{kkopsnu})
should generalise to higher values of $\ell$ and the systematics that
emerges is that operators at the same level in the multiplet with
increasing $\ell$ contain more zero modes of the $\nu$ and
$\bar\nu$ type always in the combination $\nsix$. For instance,
the operator corresponding to the second excited mode of the
dilaton is a scalar of dimension 6 in the ${\bf 20^\prime}$ of
$SU(4)$ would have the schematic form
\be
\calC_{\bf 20^\prime} = (\rho^4 f^6) [\nsix\nsix] \, .
\label{Cseconkk}
\ee

In section \ref{kknonmincorr} we will discuss how the general
structure  of the multiplets associated with Kaluza--Klein excited
modes fits with the  predictions of the AdS/CFT correspondence.  In
section \ref{othercorr} we will also briefly discuss  processes
involving the chiral primary operator $\calQ_4$ of dimension 4 in the
representation {\bf 105} of the R-symmetry group. The instanton
profile of $\calQ_4$ should be of the form
\ba
\calQ_{\bf 105} &=& (\rho^4 f^6) \,
[\zeta\zeta\zeta\zeta\nsix\nsix] + \left(\rho^2 f^5\right) \,
[\zeta\zeta\nten\nsix\nsix] \nn \\ && + \left(f^4\right) \,
[\nten\nten\nsix\nsix] \, .
\label{O4inst}
\ea
In particular we will comment on how the proof of partial
non-renormalisation properties of near extremal correlation functions can be
extended to include non-perturbative effects.

\section{Non-minimal correlation functions in the instanton background:
a detailed example}
\label{nonmincorr}

In this section we present calculations of non-minimal correlation
functions of $\N4$ SYM composite operators and of the corresponding
supergravity amplitudes in $AdS_{5}\times S^{5}$. We will focus on one
particular example which will allow us to highlight the r\^ole played
by the $\nub$ and $\nu$ modes and all the general features which apply
to other cases as well. The specific process we examine is one
obtained by additional insertions of chiral primary operators in a
minimal correlation function. The particular correlation function that
we will analyse in detail is
\be
\hat G_{\hat \Lambda^{16}\calQ_2^2}(x_{1},\dots,x_{16},y_1,y_2) =
\langle \hat\Lambda^{A_1}_{\a_1}(x_1) \ldots
\hat\Lambda^{A_{16}}_{\a_{16}}(x_{16})\calQ_2^{[B_1C_1][D_1E_1]}(y_1)
\calQ_2^{[B_2C_2][D_2E_2]}(y_2) \rangle \, .
\label{defLam16O22}
\ee
We have chosen a correlation function with  the maximal number of
$\hat\Lambda$ insertions and two additional lowest weight chiral
primaries, $\calQ_{2}$. This is the simplest possibility because each
$\hat\Lambda^{A}_{\a}$ soaks up exactly one fermion superconformal mode
and the dependence on the non-exact modes enters only in the
$\calQ_{2}$'s, so that the combinatorial analysis involved in the
computation of grassmannian integrals is minimised.

It might be natural to expect the correlation function
(\ref{defLam16O22}) to correspond to an amplitude on the AdS side
induced by a vertex of order $\al$ in the type IIB derivative
expansion of the form
\be
\al \int d^{10}x \, \sqrt{-g} \, \er^{\phi/2}
f_2^{(12,-12)}(\tau,\bar\tau) \calR^{2} \Lambda^{16} \, .
\label{R2Lam16vertex}
\ee
To match the amplitude involving such a vertex the Yang--Mills
correlation function should then be of order $N^{-1/2}$. We will show
that the situation is more complicated and the naive expectation based
on the generalisation of the analysis of minimal correlation functions
is incorrect.  In fact, we will find terms of order $N^{1/2}$ as well
as the expected contribution of order $N^{-1/2}$ in the Yang--Mills
instanton calculation. Moreover there is a contribution of order
$N^{5/2}$ corresponding to a disconnected diagram. We will show that
the combination of various different effects both on the Yang--Mills
and on the AdS side is required in order to reconcile the results with
the AdS/CFT correspondence.

The example that we discuss here illustrates all the different
phenomena encountered in the study of non-minimal correlation
functions.  On the Yang--Mills side consistency of the analysis in
semiclassical approximation requires to include all the non-vanishing
contributions at leading order in the coupling. The first type of
contributions are those in which the classical solutions for the
fields are used to saturate the integrations over the fermion
zero-modes. In non-minimal correlation functions the operator
insertions contain more than 16 fermionic modes and thus $\bar\nu\nu$
pairs must be included in addition to the 16 exact superconformal
modes. In the case of operators in the supercurrent multiplet these
bilinears are in the {\bf 10} of $SU(4)$. For higher dimensional
operators, which correspond to Kaluza--Klein excited modes in
supergravity, $\nsix$ bilinears will also appear. This is the
situation considered in section \ref{kknonmincorr}. At the same order
in $\gy$ there are contributions coming from the leading quantum
fluctuations around the instanton which are obtained contracting pairs
of scalars and involve the propagator discussed in section
\ref{scalarprop}. In general there is the possibility of having
fermion and vector contractions as well, but these are more
complicated and present subtleties related to infrared divergences
\cite{bccl}. We will restrict our analysis to examples in which only
scalar contractions can contribute. This is the case for
(\ref{defLam16O22}) since all the fermions involved have the same
chirality and cannot be contracted and similarly there is no $\la F^-
F^-\ra$ propagator. The common feature of all the non-minimal
correlation functions considered in this paper is the presence of a
leading contribution of order $N^{1/2}$ plus corrections of order
$N^{-1/2}$.  Therefore we will need to take into account the $1/N$
corrections to the measure on the instanton moduli space as well.

The interpretation of the results of the Yang--Mills instanton
calculation in the \AdS5s5\ string theory also requires the sum of
different types of amplitudes.  As already observed there are
processes in \AdS5s5\ involving vertices at order $\al$ in the type
IIB effective action which can produce amplitudes with the right
structure to match the expectation values computed on the Yang--Mills
side. The amplitudes involving these vertices  naturally correspond to
the subleading terms of order $N^{-1/2}$ in the Yang--Mills result. In
the case of the correlator (\ref{defLam16O22}) for instance the
relevant vertex at order $\al$ is the one in
(\ref{R2Lam16vertex}). Much less obvious is the interpretation of the
leading $N^{1/2}$ terms.  The explanation of these unexpected
contributions is based on new effects that come from the
well-known $O({\al}^{-1})$ terms, $\sqrt{-g} \, \er^{-\phi/2}\,
f_1^{(0,0)}(\tau,\bar\tau) \calR^4$, $\sqrt{-g}\,\er^{-\phi/2}
f_1^{(12,-12)}(\tau,\bar\tau) \Lambda^{16}$ and related ones.

As described in section \ref{overii} the expansion of the $\er^{2\pi i
\tau}$ factor in the modular forms $f_1^{(w,-w)}(\tau,\bar\tau)$ gives
rise to `effective' couplings of the type $\hat\tau^k \calR^4$,
$\hat\tau^k \Lambda^{16}$ etc. at order $\al^{-1}$. Similar terms come
from the expansion of the $\er^{-\phi/2}$ factor that also gives rise
to vertices with powers of $\hat{\bar\tau}$, which however we will not
need to consider since they produce contributions which are suppressed
by powers of the coupling.  These vertices contribute to Witten
diagrams describing amplitudes with additional  multiple insertions of
the dilaton, $\tau$, together with its conjugate, $\bar \tau$.  Each
factor of $\tau$ in its Kaluza--Klein ground state corresponds to an
insertion of the operator $\calC$ in the supercurrent multiplet in the
Yang--Mills correlation function. More generally $\tau$ may be in an excited
Kaluza--Klein state, in which case the dual operator soaks up some
$\nu$ and $\bar\nu$ moduli. In this case the contributions are
restricted by $SU(4)$ invariance.  For example, the correlation
function $\la\hat \Lambda^{16}\, \calC^2_{\bf 6} \ra$ is non-zero and
corresponds to the effect of soaking up four $\nu$'s and $\bar\nu$'s
in the Yang--Mills theory.  A $\bar\tau$ insertion in the supergravity
amplitude corresponds to the insertion of a $\bar\calC$ operator which
soaks up eight fermionic zero modes.  Similarly, expanding the
$\sqrt{-g}=\sqrt{-\det(g^{(0)}+h)}$  factor in powers of the
fluctuation part of the metric  produces vertices with additional
factors of $(\tr\,h^p)^q$, which can give rise to amplitudes that
correspond to multiple insertions of chiral primaries. These also soak
up extra fermionic moduli.   There are therefore contributions to
specific non minimal correlation functions that are expected to arise
at order $(\al)^{-1}$, \ie $N^{1/2}$, via the processes just described.

\FIGURE[!h]{ \hspace*{0.2cm}
\includegraphics[width=0.4\textwidth]{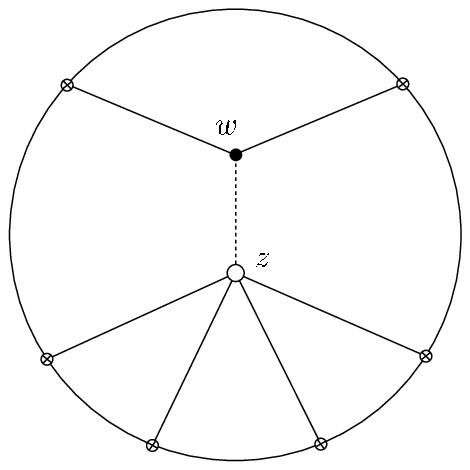}
\caption{A tree process with a D-instanton vertex at $z$
and a classical supergravity vertex at $w$. Solid lines denote
bulk-to-boundary propagators, the dotted one a bulk-to-bulk propagator.}
\label{adstree}
}
When considering the non minimal correlation functions we are
interested in there are other effects which need to be taken into
account. These correspond to processes in which one of the vertices
is a multiparticle vertex induced by a D-instanton and there are
additional tree level perturbative interactions in the bulk. An
example of such a diagram is illustrated in figure \ref{adstree}. Here
the point $z$ represents a D-instanton induced vertex. The amplitude
involves another interaction vertex, at the point $w$, which is a
standard supergravity interaction. The dashed line corresponds to a
bulk-to-bulk propagator for one of the fields entering the D-instanton
induced interaction. In view of the above discussion this can be
either one of the fields in $\calR^4$, $\Lambda^{16}$ etc. or a
dilaton or graviton in the induced $\hat\tau^k \calR^4$, $(\tr
h^k)^l\calR^4$, $\hat\tau^k\Lambda^{16}$, $(\tr h^k)^l\Lambda^{16}$,
\dots interactions. Moreover the intermediate state can also be any
Kaluza--Klein excited state allowed by the symmetries. The plain lines
in the figure represent usual bulk-to-boundary propagators. We will
discuss various similar processes in the following sections. It
is important to notice that all the tree level AdS diagrams with fixed
external states are of the same order when all the appropriate factors
of the coupling associated with vertices and bulk propagators are
included~\footnote{When expressed in terms of
Yang--Mills parameters the classical supergravity action in the string
frame has an overall factor of $N^2$. This means that all the vertices
are proportional to $N^2$ and all bulk-to-bulk propagators have a
factor of $N^{-2}$. There are no powers of $N$ associated with
bulk-to-boundary propagators since the overall $N^2$ drops out of
the field equations.}.
The contribution of amplitudes of this type is needed in
particular to match the Yang--Mills result in the case of the
correlation function (\ref{defLam16O22}).

The general analysis of such processes would be very complicated.  We
will focus on those relevant for the interpretation of the Yang--Mills
calculation of some simple non-minimal correlators.  In these cases
very specific diagrams are selected by the conditions imposed by the
$SU(4)$ symmetry and by the $U(1)$ symmetry of classical supergravity.

In the next subsection we present the Yang--Mills calculation of
(\ref{defLam16O22}) in the one-instanton sector, then in the following
subsection we describe the supergravity calculation and show how the
agreement with field theory is recovered. In section
\ref{kknonmincorr} we discuss another example of non minimal
correlation function with insertions of higher dimensional operators
corresponding to Kaluza--Klein excited states on the gravity side.
Section \ref{othercorr} contains an overview of other correlation
functions which also involve the various processes described here.

\subsection{One-instanton contribution to ${\hat
\Lambda^{16}\calQ_2^2}$ in $\N4$ SYM}
\label{1instcorr}

The operator insertions in (\ref{defLam16O22}) contain 24 fermion
modes at lowest order in the coupling, thus $\nub\nu$ pairs must be
included beyond the 16 exact superconformal modes. Each
$\hat\Lambda^A_\a$ operator soaks up one superconformal mode and does
not depend on the $\nu$ and $\nub$ modes, so that the latter only
enter into the $\calQ_2$ insertions.  A non-vanishing result in the
semiclassical limit is obtained by replacing each $\calQ_2$ by its
expression in terms of $\nten$ bilinears given in appendix
\ref{appinsprofiles}. This contribution contains four $\nten$ pairs,
which according to the general analysis of section \ref{ginvarcorr}
bring four powers of $\gy$. The other  relevant contributions are
obtained by contracting pairs of $\v$ fields in the two $\calQ_2$'s
with a propagator: one can either contract two $\v$'s and saturate the
remaining two with $\nub$ and $\nu$ modes or contract both pairs of
scalar fields.

Taking into account the powers of $\gy$ associated with the
propagators discussed in section \ref{scalarprop} and those associated
with $\nub\nu$ insertions, (\ref{nnusix}) and (\ref{nnuten}), we
see that all three types of contributions occur at the same {\em
leading non-vanishing order} in the coupling and must be included in
the semiclassical approximation. Notice that the only fields whose
fluctuations must be considered are the scalars. We will denote the
above three types of processes as follows
\be
\hat G_{\hat\Lambda^{16}\calQ_2^2} =
\hat G_{\hat\Lambda^{16}\calQ_2^2}^{\rm (a)} +
\hat G_{\hat\Lambda^{16}\calQ_2^2}^{\rm (b)} +
\hat G_{\hat\Lambda^{16}\calQ_2^2}^{\rm (c)}
\label{L16O22exp}
\ee
They are represented diagrammatically in figure \ref{Lam16Q2figYM}.

\FIGURE[!h]{
{\includegraphics[width=0.3\textwidth]{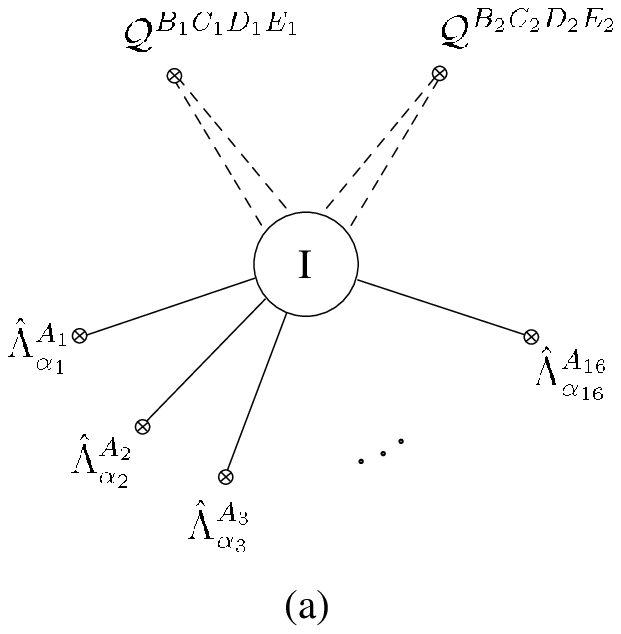}
\hspace*{0.3cm}
\includegraphics[width=0.3\textwidth]{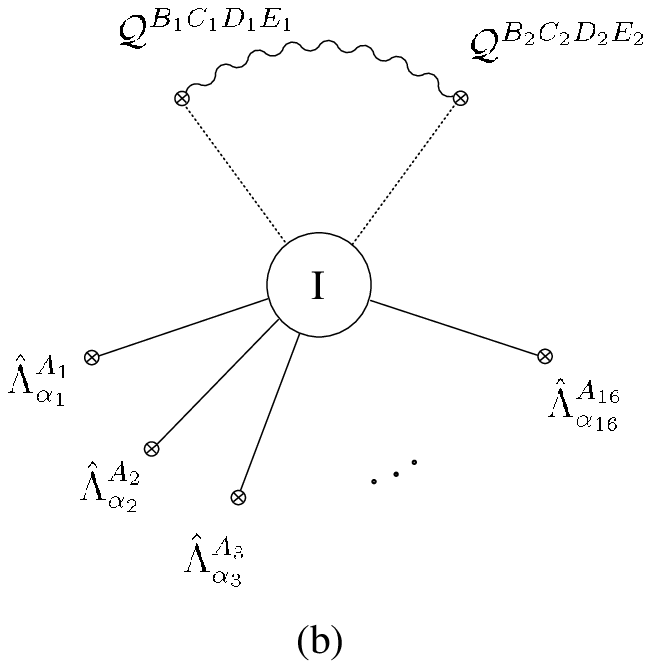}
\hspace*{0.3cm}
\includegraphics[width=0.3\textwidth]{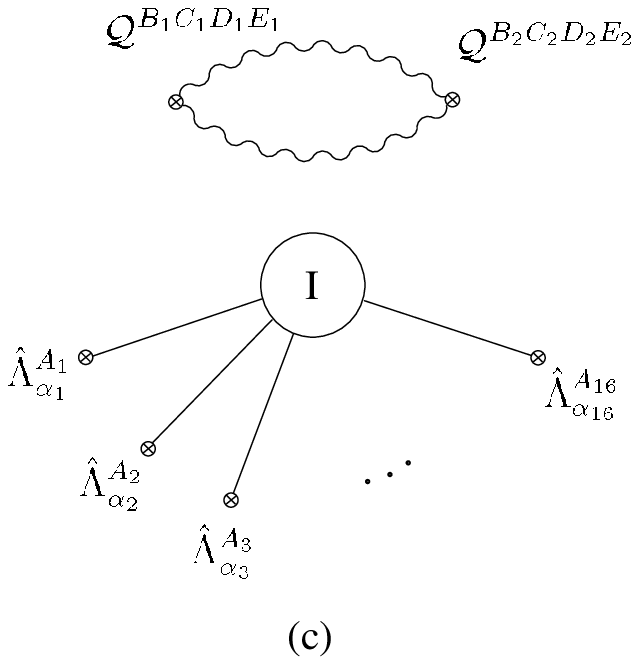}}
\caption{The three types of contributions to the correlation
function $\hat G_{\hat\Lambda^{16}\calQ_{2}^{2}}$: wiggly lines
denote scalar propagators, a plain line corresponds to the insertion of
a superconformal mode, a $\bar\nu \nu$ pair is indicated by a dashed
line when in the {\bf 10} and by a dotted line when in the {\bf 6}.}
\label{Lam16Q2figYM}
}
We first consider the `purely instantonic' process depicted in figure
\ref{Lam16Q2figYM}(a). In the semiclassical approximation we must
evaluate~\footnote{In all the following
calculations in this section we will omit numerical coefficients in
intermediate steps and reinstate them in the final formulae.}
\ba
\hat G_{\hat\Lambda^{16}\calQ_2^2}^{\rm (a)} &\!=\!&
\int d\mu_{\rm phys}\, \er^{-S_{4F}} \left[
\hat\Lambda^{A_1}_{\a_1}(x_1) \ldots
\hat\Lambda^{A_{16}}_{\a_{16}}(x_{16})
\calQ_2^{[B_1C_1][D_1E_1]}(y_1) \calQ_2^{[B_2C_2][D_2E_2]}(y_2)
\right] \nn  \\
&\!=\!& \frac{1}{\gy^{36}}\int d\mu_{\rm phys}\,
\er^{-S_{4F}} \, \left[ \prod_{i=1}^{16} \frac{\rho^4
\zeta^{A_i}_{\a_i}}{[(x_i-x_0)^2+\rho^2]^4} \right. \nn \\
&& \left. \prod_{j=1}^2 \frac{
\begin{displaystyle}
({\bar\nu}^{u(B_j}\nu^{D_j)}_{u})
({\bar\nu}^{v(C_j}\nu^{E_j}_{v})-({\bar\nu}^{u(B_j}\nu^{E_j)}_{u})
({\bar\nu}^{v(C_j}\nu^{D_j)}_{v})
\end{displaystyle}}
{[(y_j-x_0)^2+\rho^2]^2} \right] \, ,
\label{Lam16Q2-a1}
\ea
where we have substituted the classical expressions for the
operators $\hat\Lambda$ and $\calQ_2$ which are given by
(\ref{explLamb}) and (\ref{explO2}), respectively.
The combinatorics necessary to evaluate the $8(N-2)$ fermionic
integrations over
$\bar\nu^A_u$ and $\nu^A_u$ is performed with the aid of the generating
function (\ref{genfunctfin}) by rewriting the above expression as
\ba
\hat G_{\hat\Lambda^{16}\calQ_2^2}^{\rm (a)} &\!=\!&
\frac{\gy^{-28}\er^{2\pi i\tau}} {(N-1)!(N-2)!}
\int d\rho \, d^4x_0 \, d^5\Omega \prod_{A=1}^4 d^2\eta^A \, d^2
{\bar\xi}^A \, \rho^{4N-7}\int_0^\infty dr \, r^{4N-3}
\er^{-2\rho^2 r^2} \nn \\
&& \left\{ \prod_{i=1}^{16}
\frac{\rho^4}{[(x_i-x_0)^2+\rho^2]^4}\zeta^{A_i}_{\a_i} \prod_{j=1}^2
\frac{1}{[(y_j-x_0)^2+\rho^2]^2} \right. \label{Lam16Q2a2} \\
&& \left. \left.  \left[ \frac{\d^8\calZ(\vt,\bvt;\Omega,r)}
{\d\vt_{u_1(B_1}\d\bvt_{C_1)}^{u_1}\d\vt_{u_2(D_1}\d\bvt_{E_1)}^{u_2}
\d\vt_{u_3(B_2}\d\bvt_{C_2)}^{u_3}\d\vt_{u_4(D_2}\d\bvt_{E_4)}^{u_4}}
- (D_j \leftrightarrow E_j) \right] \right\vert_{\vt=\bvt=0}
\right\} \, ,  \nn
\ea
where the measure has been explicitly written in terms of
integrations over the collective coordinates.

The $\nten$ insertions are replaced by $\vt$ derivatives of
$\calZ(\vt,\bvt;\Omega,r)$. The resulting integration over angular
variables, $\Omega$, selects the combination of $\nub$ and $\nu$ modes
forming a singlet of $SU(4)$. After performing the $\vt$ derivatives
and symmetrising the indices as in (\ref{Lam16Q2a2}) we obtain angular
integrals which can be performed using
\ba
&& \int \! d^5\Omega \,
\Omega^{A_1B_1}\Omega^{A_2B_2}\Omega^{A_3B_3} \Omega^{A_4B_4} =
\frac{1}{2^{6}}\!\int_{\sum_{e=1}^6 \! \Omega_e^2}
\prod_{e=1}^6 d\Omega_e \,
\Omega_a\Omega_b \Omega_c \Omega_d \Sbar^{A_1B_1}_a
\Sbar^{A_2B_2}_b \Sbar^{A_3B_3}_c \Sbar^{A_4B_4}_d \nn \\
&& \rule{0pt}{16pt} =\frac{1}{2^6} \left(\d_{ab}\d_{cd} +
\d_{ac}\d_{bd} + \d_{ad}\d_{bc} \right) \Sbar^{A_1B_1}_a
\Sbar^{A_2B_2}_b \Sbar^{A_3B_3}_c \Sbar^{A_4B_4}_d \nn \\
&& \rule{0pt}{20pt} = \frac{1}{2^4}
\left(\veps^{A_1B_1A_2B_2}\veps^{A_3B_3A_4B_4} +
\veps^{A_1B_1A_3B_3}\veps^{A_2B_2A_4B_4} + \veps^{A_1B_1A_4B_4}
\veps^{A_2B_2A_3B_3} \right) \, .
\label{spherint}
\ea
The $SU(4)$ singlet tensor resulting from the five-sphere integration
is
\begin{eqnarray*}
&& t^{B_1C_1D_1E_1B_2C_2D_2E_2} =
- \varepsilon^{B_1E_1C_2E_2} \: \varepsilon^{C_1D_1B_2D_2}
+ \varepsilon^{B_1E_1C_2D_2} \: \varepsilon^{C_1D_1B_2E_2} \nn \\
&& + \varepsilon^{B_1E_1B_2E_2} \: \varepsilon^{C_1D_1C_2D_2}
- \varepsilon^{B_1E_1B_2D_2} \: \varepsilon^{C_1D_1C_2E_2}
+ \varepsilon^{B_1D_1C_2E_2} \: \varepsilon^{C_1E_1B_2D_2} \nn \\
&& - \varepsilon^{B_1D_1C_2D_2} \: \varepsilon^{C_1E_1B_2E_2}
- \varepsilon^{B_1D_1B_2E_2} \: \varepsilon^{C_1E_1C_2D_2}
+ \varepsilon^{B_1D_1B_2D_2} \: \varepsilon^{C_1E_1C_2E_2} \nn \\
&& + 2\left(\varepsilon^{B_1E_1D_2E_2} \: \varepsilon^{C_1D_1C_2B_2}
+ \varepsilon^{B_1E_1C_2B_2} \: \varepsilon^{C_1D_1D_2E_2}
+ \varepsilon^{B_1D_1D_2E_2} \: \varepsilon^{C_1E_1B_2C_2} \right. \nn \\
&& + \varepsilon^{B_1D_1B_2C_2} \: \varepsilon^{C_1E_1D_2E_2}
+ \varepsilon^{B_1C_1C_2E_2} \: \varepsilon^{D_1E_1B_2D_2}
+ \varepsilon^{B_1C_1D_2C_2} \: \varepsilon^{D_1E_1B_2E_2} \nn \\
&& \left. + \varepsilon^{B_1C_1E_2B_2} \: \varepsilon^{D_1E_1C_2D_2}
+ \varepsilon^{B_1C_1B_2D_2} \: \varepsilon^{D_1E_1C_2E_2}\right)
+ 4\left(\varepsilon^{B_1C_1D_2E_2} \: \varepsilon^{D_1E_1B_2C_2}
\right. \nn \\
&& \left. + \varepsilon^{B_1C_1B_2C_2} \: \varepsilon^{D_1E_1D_2E_2}
\right) \, .
\label{tensor1}
\end{eqnarray*}
The expression for $\hat G_{\hat\Lambda^{16}\calQ_2^2}^{\rm (a)}$
becomes
\ba
\hat G_{\hat\Lambda^{16}\calQ_2^2}^{\rm (a)} &\!=\!&
t^{B_1C_1D_1E_1B_2C_2D_2E_2}
\gy^{-24}\er^{2\pi i\tau}
\int d\rho \, d^4x_0 \prod_{A=1}^4
d^2\eta^A \, d^2 {\bar\xi}^A \, \rho^{4N-7} \nn \\
&& \left[ \prod_{i=1}^{16}
\frac{\rho^4}{[(x_i-x_0)^2+\rho^2]^4}\zeta^{A_i}_{\a_i} \prod_{j=1}^2
\frac{1}{[(y_j-x_0)^2+\rho^2]^2} \right] \nn \\
&& \frac{(N^2-5N+6)}{(N-1)!(N-2)!}\int_0^\infty dr \, r^{4N-7}
\er^{-2\rho^2 r^2} \, ,
\label{Lam16Q2a3}
\ea
where the numerator $(N^2-5N+6)=(N-2)^2-(N-2)$ in the last line is the
result of the colour contractions in the $\nten$ insertions.
All factors of $N$ are now isolated by performing the $r$ integral,
giving
\be
\frac{N^2-5N+6}{(N-1)!(N-2)!} \int_0^\infty dr \, r^{4N-7}
\er^{-2\rho^2r^2} =  \frac{(N^2-5N+6)\Gamma(2N-3)}{(N-1)!(N-2)!}
2^{2-2N}\rho^{6-4N} \, .
\label{radint}
\ee
Including all the numerical coefficients the final result is
\ba
\hat G_{\hat\Lambda^{16}\calQ_2^2}^{\rm (a)} &=&
C_{\rm a}(N) \, t^{B_1C_1D_1E_1B_2C_2D_2E_2}
\frac{2^{39}3^{19}\er^{2\pi i\tau}}{\pi^{35/2}\gy^{24}}
\int \frac{d\rho \, d^4x_0}{\rho^5} \prod_{A=1}^4 d^2\eta^A \, d^2
{\bar\xi}^A \nn \\
&& \left[ \prod_{i=1}^{16}
\frac{\rho^4}{[(x_i-x_0)^2+\rho^2]^4}\zeta^{A_i}_{\a_i} \prod_{j=1}^2
\frac{\rho^2}{[(y_j-x_0)^2+\rho^2]^2} \right] \, ,
\label{Lam16Q2afin}
\ea
where
\be
C_{\rm a}(N) =
N^{1/2}\left(1-\frac{25}{8N}-\frac{47}{128N^2}+O(1/N^3)\right) \, .
\label{NdepL16Q2a}
\ee
We will discuss the interpretation of this result in the next
subsection, let us only note here that the  integration over
$(\Omega,r)$ determines the $SU(4)$ tensorial structure and  produces
the powers of $N$ that lead to the above overall $N$-dependence as
well as the factors of $\rho$ that reconstruct the functions
$K_2(y_j;x_0,\rho)$ for each $\calQ_2$ insertion, see (\ref{fKrel}).

The $\hat G^{\rm (b)}_{\hat\Lambda^{16}\calQ_2^2}$ term in
(\ref{L16O22exp}) corresponds to contributions in which one pair of
scalar fields is contracted via a propagator, see figure
\ref{Lam16Q2figYM}(b),
\ba
\hat G^{\rm (b)}_{\hat\Lambda^{16}\calQ_2^2} &\!=\!&
\la \Tr\!\left(F_{m_1n_1}\sigma^{m_1n_1}_{\a_1}{}^{\b_1}
\lambda^{A_1}_{\b_1} \right)\!(x_1) \ldots
\Tr\!\left(F_{m_{16}n_{16}}
\sigma^{m_{16}n_{16}}_{\a_{16}}{}^{\b_{16}}
\lambda^{A_{16}}_{\b_{16}} \right)\!(x_{16}) \nn \\
&& \left[2\Tr\left(\v^{B_1C_1}\v^{D_1E_1}\right)(y_1) \,
2 \Tr\left(\v^{B_2C_2}\v^{D_2E_2}\right)(y_2)
\raisebox{-10pt}{\hspace*{-6.79cm}
\rule{0.4pt}{4pt}\rule{3.8cm}{0.4pt}\rule{0.4pt}{4pt}}
\hspace*{2.85cm} + \cdots \right] \ra \, ,
\label{lam16Q2b1}
\ea
where the ellipsis refers to 35 other terms. The 36 terms in
(\ref{lam16Q2b1}) come from the expansion of the product of the two
$\calQ$ insertions using (\ref{O2newdef}) which gives nine terms  and
from the application of Wick's theorem which gives four contractions
for each term.  In this expression the sixteen factors of $\hat\Lambda$
soak up the superconformal modes as in the case of  $\hat G^{\rm
(a)}_{\hat\Lambda^{16}\calQ_2^2}$ and the remaining $\v$  fields which
are not contracted into a propagator should be saturated by $\bar\nu$
and $\nu$ modes.

The calculation of the contribution of a contraction between a pair
of scalar fields in two $\calQ$ operators is reported in appendix
\ref{scalcontr}. All the contractions in (\ref{lam16Q2b1}) give rise
to the same spatial structure and combining them together determines
the $SU(4)$ tensorial form of $\hat G^{\rm
(b)}_{\hat\Lambda^{16}\calQ_2^2}$. In computing a single contraction
only the terms in the first and fourth lines in the expression for the
propagator (\ref{propfin}) contribute because the other terms produce
traces over single $\v$ fields which vanish since the fields are in
$SU(N)$. This simplification would not occur in the case of a similar
contraction between operators formed by the product of three or more
elementary fields as we will see in the example discussed in the next
section.

Substituting the result (\ref{onecontr2}) in the appendix into
(\ref{lam16Q2b1}) and replacing the remaining fields with their
expressions in the instanton background gives
\ba
\hat G^{\rm (b)}_{\hat\Lambda^{16}\calQ_2^2} &\!=\!&
\gy^{-34} \veps^{B_{1}C_{1}B_{2}C_{2}}
\int d\mu_{\rm phys} \er^{-S_{4F}} \,
\left[ \prod_{i=1}^{16}\frac{\rho^4}{[(x_i-x_0)^2+\rho^2]^4}
\zeta^{A_i}_{\a_i} \right] \nn \\
&& \left\{ \frac{1}{32 \pi^{2}(y_{1}-y_{2})^{2}
[(y_{1}-x_{0})^{2}+\rho^{2}] [(y_{2}-x_{0})^{2}+\rho^{2}]}
\left[ (\bar\nu^{[D_1}\nu^{D_2]})(\bar\nu^{[E_2}\nu^{E_1]})
\right. \right. \nn \\
&& \rule{0pt}{18pt} \!
-(\bar\nu^{[D_1}\nu^{E_2]})(\bar\nu^{[D_2}\nu^{E_1]})
-(\bar\nu^{[D_1}\nu^{E_1]})(\bar\nu^{[D_2}\nu^{E_2]})
+(\bar\nu^{(D_1}\nu^{D_2)})(\bar\nu^{(E_1}\nu^{E_2)}) \nn \\
&& \rule{0pt}{18pt} \! \left. \left. - (\bar\nu^{(D_1}\nu^{E_2)})
(\bar\nu^{(D_2}\nu^{E_1)}) \right] + \cdots \right\}
\label{lam16Q2b2} \, ,
\ea
where the dots stand for the contributions of the other contractions.
Notice in particular that combining the two types of terms induced by
the scalar propagator in (\ref{onecontr})-(\ref{onecontr2}) leads to
the cancellation of a contact contribution, \ie a term non singular in
the limit $y_2\to y_1$, so that only one spatial structure appears in
(\ref{lam16Q2b2}).  To evaluate this expression we follow the same
steps as in the case of $\hat G^{\rm
(a)}_{\hat\Lambda^{16}\calQ_2^2}$,  we first rewrite the $\nub\nu$
bilinears as derivatives of the density $\calZ(\vt,\bvt;\Omega,r)$ and
then compute the resulting six-dimensional integral over
$(\Omega,r)$. In particular the angular integrals over $\Omega$ select
an $SU(4)$ singlet. This is a common feature of all the non-minimal
correlation functions, the $\nub\nu$ pairs in the operator insertions
must always be combined to form a singlet of the $SU(4)$ R-symmetry
group. This implies that the last two terms in (\ref{lam16Q2b2})
vanish when integrated over the five-sphere because the product ${\bf
10}\otimes{\bf 10}$ does not contain the singlet. After rewriting
(\ref{lam16Q2b2}) in terms of the generating function
$\calZ(\vt,\bvt;\Omega,r)$ and performing the $\vt$ derivatives, all
the angular integrals are computed using
\be
\int d^{5}\Omega \,\Omega^{AB} \Omega^{BC} = \frac{1}{2^2}
å\, \veps^{ABCD} \, .
\label{spherint2}
\ee
We then have to combine the terms in (\ref{lam16Q2b2}) corresponding
to all the possible Wick contractions. In conclusion we find
\ba
\hat G^{\rm (b)}_{\hat\Lambda^{16}\calQ_2^2} \!&=\!&
\gy^{-24}\er^{2\pi i\tau}t^{B_1C_1D_1E_1B_2C_2D_2E_2}
\int d\rho \, d^4x_0 \prod_{A=1}^4 d^2\eta^A \, d^2 {\bar\xi}^A \nn \\
&& \rho^{4N-7} \left[ \prod_{i=1}^{16}
\frac{\rho^4}{[(x_i-x_0)^2+\rho^2]^4}\zeta^{A_i}_{\a_i} \right]
\frac{1}{(y_1-y_2)^2} \left[ \prod_{j=1}^2
\frac{1}{[(y_j-x_0)^2+\rho^2]} \right] \nn \\
&& \frac{(N^2-5N+6)}{(N-1)!(N-2)!} \int_0^\infty dr \, r^{4N-5}
\er^{-2\rho^2 r^2} \, , \label{lam16Q2b3}
\ea
where $t^{B_1C_1D_1E_1B_2C_2D_2E_2}$ is the same tensor defined in
(\ref{tensor1}) which appeared in $\hat G^{\rm
(a)}_{\hat\Lambda^{16}\calQ_2^2}$. Note, however, that this same
$SU(4)$ tensorial structure is obtained here in a completely different
way. In $\hat G^{\rm (a)}_{\hat\Lambda^{16}\calQ_2^2}$ it was the
result of five dimensional integrals of the form (\ref{spherint})
whereas here it is obtained combining the 36 terms associated with all
the possible contractions with the appropriate weights.  The
$N$-dependence is determined by the integral over the radial variable
$r$. Performing this integral and reintroducing all the numerical
coefficients we finally obtain
\ba
\hat G_{\hat\Lambda^{16}\calQ_2^2}^{\rm (b)} &\!=\!&
C_{\rm b}(N) t^{B_1C_1D_1E_1B_2C_2D_2E_2}
\frac{2^{46}3^{16}\er^{2\pi i\tau}}{\pi^{35/2} \gy^{24}}
\int \frac{d\rho \, d^4x_0}{\rho^5} \prod_{A=1}^4 d^2\eta^A \,
d^2 {\bar\xi}^A  \nn \\
&&\left[ \prod_{i=1}^{16}
\frac{\rho^4}{[(x_i-x_0)^2+\rho^2]^4}\zeta^{A_i}_{\a_i} \right]
\frac{1}{(y_1-y_2)^2} \left[ \prod_{j=1}^2
\frac{\rho}{[(y_j-x_0)^2+\rho^2]} \right] \, ,
\label{Lam16Q2bfin}
\ea
where
\be
C_{\rm b}(N) = N^{3/2} \left( 1 - \frac{37}{8N}
+ \frac{553}{128N^2} + O(1/N^3) \right) \, .
\label{NdepL16Q2b}
\ee
Notice that the leading term in $\hat
G_{\hat\Lambda^{16}\calQ_2^2}^{\rm (b)}$ is of order $N^{3/2}$. Such a
contribution could not have an AdS counterpart, since D-instanton
effects appear in the type IIB effective action at order $(\al)^{-1}$,
\ie $N^{1/2}$. As we will show the leading term in (\ref{Lam16Q2bfin})
is cancelled by an equal and opposite term from the other contribution
to $\hat G_{\hat\Lambda^{16}\calQ_2^2}$ which we will now discuss.

The third type of instanton correction to the correlation function
$\la\hat\Lambda^{16}\calQ^2\ra$ is obtained by contracting both pairs
of scalars in the two $\calQ_{2}$ insertions by propagators, see
figure \ref{Lam16Q2figYM}(c),
\ba
\hat G_{\hat\Lambda^{16}\calQ_2^2}^{\rm (c)} &\!=\!&
\la \Tr\!\left(F_{m_1n_1}\sigma^{m_1n_1}_{\a_1}{}^{\b_1}
\lambda^{A_1}_{\b_1} \right)\!(x_1) \ldots
\Tr\!\left(F_{m_{16}n_{16}}
\sigma^{m_{16}n_{16}}_{\a_{16}}{}^{\b_{16}}
\lambda^{A_{16}}_{\b_{16}} \right)\!(x_{16}) \nn \\
&& \left[ 4 \Tr\left(\v^{B_1C_1}\v^{D_1E_1}\right)(y_1) \,
\Tr\left(\v^{B_2C_2}\v^{D_2E_2}\right)(y_2)
\raisebox{-15pt}{\hspace*{-6.6cm}
\rule{0.4pt}{10pt}\rule{4.6cm}{0.4pt}\rule{0.4pt}{10pt}}
\raisebox{-10pt}{\hspace*{-3.75cm}
\rule{0.4pt}{4pt}\rule{2.6cm}{0.4pt}\rule{0.4pt}{4pt}}
\hspace*{2.8cm} + \cdots \right] \ra \, ,
\label{Lam16Q2c1}
\ea
where again the ellipsis is for 17 terms corresponding to the other
Wick contractions. In (\ref{Lam16Q2c1}) all the scalars are
contracted, the $\hat\Lambda$ insertions saturate the sixteen
superconformal modes and no $\nu$ and $\nub$ modes are involved. The
double contraction is significantly more complicated to evaluate
than the single contraction entering $\hat
G_{\hat\Lambda^{16}\calQ_2^2}^{\rm (b)}$ since now all the terms in
the propagator (\ref{propfin}) contribute. The result is given in
(\ref{doublecontr2}) in appendix \ref{scalcontr}.
Since there is no dependence on the additional fermion modes $\nub$
and $\nu$, substituting into (\ref{Lam16Q2c1}) leads to the same type
of integrals encountered in the computation of minimal correlation
functions. Collecting terms of the same order in $N$ from the double
contraction we have
\ba
\hat G_{\hat\Lambda^{16}\calQ_2^2}^{\rm (c)} &\!=\!&
\frac{\veps^{B_{1}C_{1}D_{2}E_{2}}\veps^{D_{1}E_{1}B_{2}C_{2}}
\, \er^{2\pi i\tau}} {\gy^{24}(N-1)!(N-2)!}
\int d\rho \, d^4x_0 \, d^5\Omega \prod_{A=1}^4 d^2\eta^A \,
d^2{\bar\xi}^A \, \rho^{4N-7} \nn \\
&& \int_0^\infty dr \, r^{4N-3}\er^{-2\rho^2 r^2}
\left[\prod_{i=1}^{16}
\frac{\rho^4}{[(x_i-x_0)^2+\rho^2]^4}\zeta^{A_i}_{\a_i} \right]
\left\{\rule{0pt}{20pt}\frac{(N^2-1)}{(y_1-y_2)^4} \right. \nn \\
&& - N \frac{1}{(y_1-y_2)^2}
\prod_{j=1}^2 \frac{\rho}{[(y_j-x_0)^2+\rho^2]} + \left[
5\prod_{j=1}^2 \frac{\rho^2}{[(y_j-x_0)^2+\rho^2]^2} \right. \nn \\
&& \left. + \frac{1}{(y_1-y_2)^2} \prod_{j=1}^2
\frac{\rho}{[(y_j-x_0)^2+\rho^2]} \right] + \frac{1}{N} \left[
2 \prod_{j=1}^2 \frac{\rho^2}{[(y_j-x_0)^2+\rho^2]^2} \right. \nn \\
&& \left.\left. -\half \frac{1}{(y_1-y_2)^2} \prod_{j=1}^2
\frac{\rho}{[(y_j-x_0)^2+\rho^2]} \right] \right\} + \cdots \, .
\label{lam16Q2c2}
\ea
In (\ref{lam16Q2c2}) we have not included terms of order $1/N^2$ from
(\ref{doublecontr2}) since they give rise to subleading effects
beyond the order we are interested in for the comparison with string
theory.

Combining the contributions of the various contractions and computing
the integrals over $(\Omega,r)$ gives
\ba
&&\!\! \hat G_{\hat\Lambda^{16}\calQ_2^2}^{\rm (c)} =
\hat G_{\hat\Lambda^{16}\calQ_2^2}^{\rm (c1)} +
\hat G_{\hat\Lambda^{16}\calQ_2^2}^{\rm (c2)} +
\hat G_{\hat\Lambda^{16}\calQ_2^2}^{\rm (c3)} \nn \\
&& \!\! = t^{B_1C_1D_1E_1B_2C_2D_2E_2}
\frac{2^{46}3^{16}\er^{2\pi i\tau}}{\pi^{35/2}\gy^{24}}
\int \frac{d\rho \, d^4x_0}{\rho^5} \prod_{A=1}^4 d^2\eta^A \,
d^2 {\bar\xi}^A \left[ \prod_{i=1}^{16}
\frac{\rho^4}{[(x_i-x_0)^2+\rho^2]^4}\zeta^{A_i}_{\a_i} \right] \nn \\
&& \!\! \left\{ C_{\rm c1}(N) \frac{1}{(y_1-y_2)^4} +
C_{\rm c2}(N) \left[ \frac{1}{(y_1-y_2)^2}
\prod_{j=1}^2 \frac{\rho}{[(y_j-x_0)^2+\rho^2]} \right] \right. \nn \\
&& + \left. C_{\rm c3}(N) \left[\prod_{j=1}^2
\frac{\rho^2}{[(y_j-x_0)^2+\rho^2]^2} \right] \right\} \, ,
\label{Lam16Q2cfin}
\ea
where the tensor $t^{B_1C_1D_1E_1B_2C_2D_2E_2}$  is the same which
appears in the previous contributions and we have reinstated all the
numerical coefficients. There three different spatial structures that
appear in (\ref{Lam16Q2cfin}) and  the $N$-dependence is encoded in
the coefficients $C_{\rm c1}(N)$, $C_{\rm c2}(N)$ and $C_{\rm c3}(N)$
for which we obtain
\ba
&& C_{\rm c1}(N) = \frac{1}{4}(N^2-1)N^{1/2} \left( 1 - \frac{5}{8N}
- \frac{23}{128N^2} + O(1/N^3) \right)
\label{Ndepc1} \\
&&  C_{\rm c2}(N) = N^{3/2} \left( -1 + \frac{1}{6N}
- \frac{139}{288N^2} + O(1/N^3) \right)
\label{Ndepc2} \\
&&  C_{\rm c3}(N) = N^{1/2} \left( 1 + \frac{5}{4N}
+ \frac{101}{48N^2} + O(1/N^3) \right) \, .
\label{Ndepc3}
\ea
In (\ref{Lam16Q2cfin}) we have separated the contribution $\hat
G_{\Lambda^{16}\calQ_2^2}^{\rm (c1)}$ which starts at order $N^{5/2}$
which, as will be discussed in the next subsection, corresponds to a
disconnected AdS diagram. The other two terms, $\hat
G_{\Lambda^{16}\calQ_2^2}^{\rm (c2)}$ and $\hat
G_{\Lambda^{16}\calQ_2^2}^{\rm (c3)}$,  must be combined respectively
with $\hat G_{\Lambda^{16}\calQ_2^2}^{\rm (b)}$ and $\hat
G_{\Lambda^{16}\calQ_2^2}^{\rm (a)}$.

In conclusion we get
\be
\hat G_{\hat\Lambda^{16}\calQ_2^2} =
\hat G_{\hat\Lambda^{16}\calQ_2^2}^{\rm (disc)} +
\hat G_{\hat\Lambda^{16}\calQ_2^2}^{\rm (conn)} \, ,
\label{lam16Q2YMsum}
\ee
where
\be
\hat G_{\hat\Lambda^{16}\calQ_2^2}^{\rm (disc)} = \hat
G_{\Lambda^{16}\calQ_2^2}^{\rm (c1)}
\label{Lam16Q2disc}
\ee
and
\ba
\hat G_{\hat\Lambda^{16}\calQ_2^2}^{\rm (conn)} &\!=\!&
t^{B_1C_1D_1E_1B_2C_2D_2E_2}
\frac{2^{46}3^{16}\er^{2\pi i\tau}}{\pi^{35/2}\gy^{24}}
\int \frac{d\rho \, d^4x_0}{\rho^5} \prod_{A=1}^4 d^2\eta^A \,
d^2 {\bar\xi}^A \nn \\
&& \left[ \prod_{i=1}^{16}
\frac{\rho^4}{[(x_i-x_0)^2+\rho^2]^4}\zeta^{A_i}_{\a_i} \right]
\left\{ C_1(N) \frac{1}{(y_1-y_2)^2} \left[ \prod_{j=1}^2
\frac{\rho}{[(y_j-x_0)^2+\rho^2]} \right] \right. \nn \\
&& + \left. C_2(N) \left[\prod_{j=1}^2
\frac{\rho^2}{[(y_j-x_0)^2+\rho^2]^2} \right] \right\}
\label{Lam16Q2conn}
\ea
\ba
&& C_1(N) = -\frac{107}{24}N^\half + \frac{4421}{1152} N^{-\half}
+O\left(N^{-\frac{3}{2}}\right)
\label{Ndepconn1} \\
&& C_2(N) = \frac{155}{128}N^\half + \frac{605}{1024} N^{-\half}
+O\left(N^{-\frac{3}{2}}\right) \, .
\label{Ndepconn2}
\ea
As already anticipated in the introductory discussion we obtain, apart
from the disconnected term $\hat G_{\hat\Lambda^{16}\calQ_2^2}^{\rm
(disc)}$, two different spatial structures contributing to the
correlation function $\hat G_{\hat\Lambda^{16}\calQ_2^2}$ and both
have a leading term of order $N^{1/2}$, whereas one would have naively
expected a result of order $N^{-1/2}$. In the next subsection we
compare the result with the supergravity calculation of the dual
amplitude and we discuss the interpretation of this result.

\subsection{The ${\hat \Lambda^{16}\calQ_2^2}$ correlator in supergravity}
\label{L16Q2sugra}

The AdS computation of the correlation function (\ref{defLam16O22}) at
leading non-vanishing order in the coupling constant involves various
different processes of the type described in the introduction to this
section. Let us first list all the relevant contributions at each
order in $N$, then we will proceed with the description of the
individual diagrams. The AdS amplitudes we have to consider are
represented in figure \ref{Lam16Q2figAdS}.
\FIGURE[!h]{
{\includegraphics[width=0.3\textwidth]{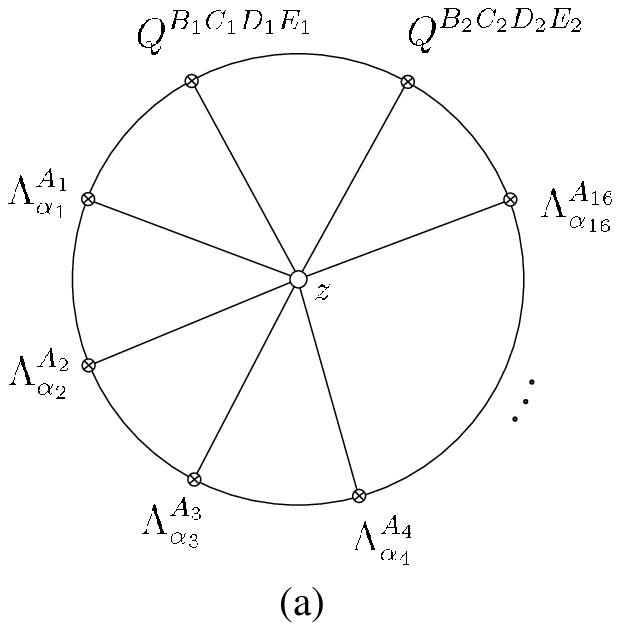}
\hspace*{0.4cm}
\includegraphics[width=0.3\textwidth]{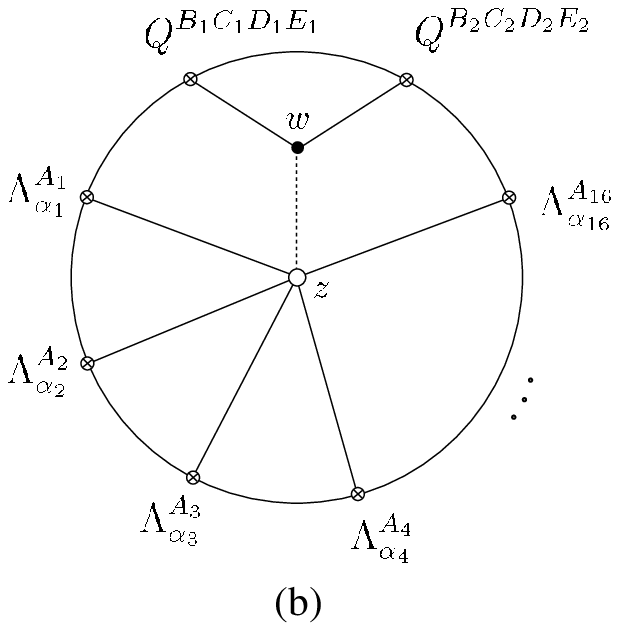}
\hspace*{0.4cm}
\includegraphics[width=0.3\textwidth]{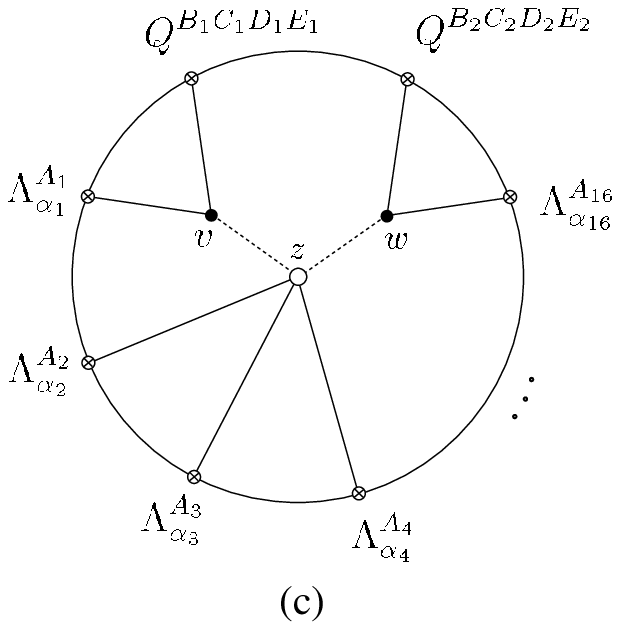}
} \\
\rule{0pt}{10pt} \\
{\includegraphics[width=0.3\textwidth]{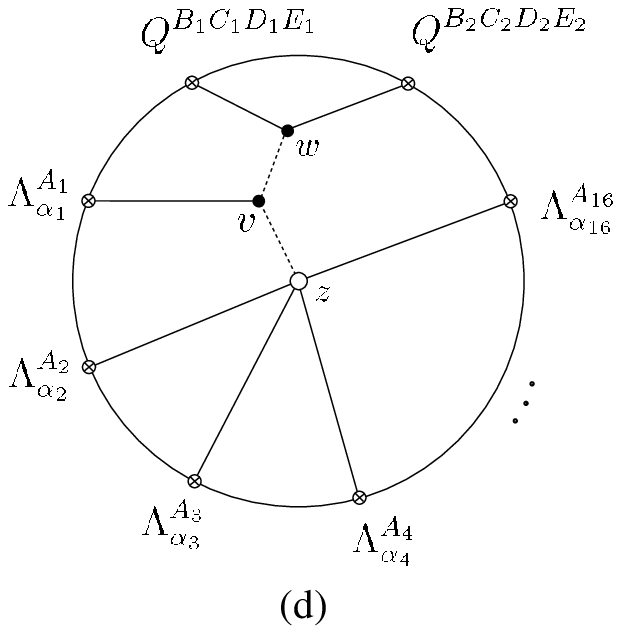}
\hspace*{0.4cm}
\includegraphics[width=0.3\textwidth]{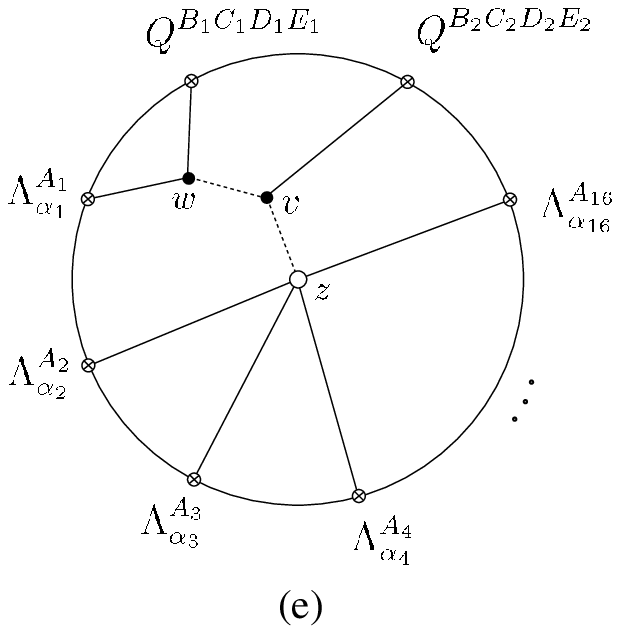}
\hspace*{0.4cm}
\includegraphics[width=0.3\textwidth]{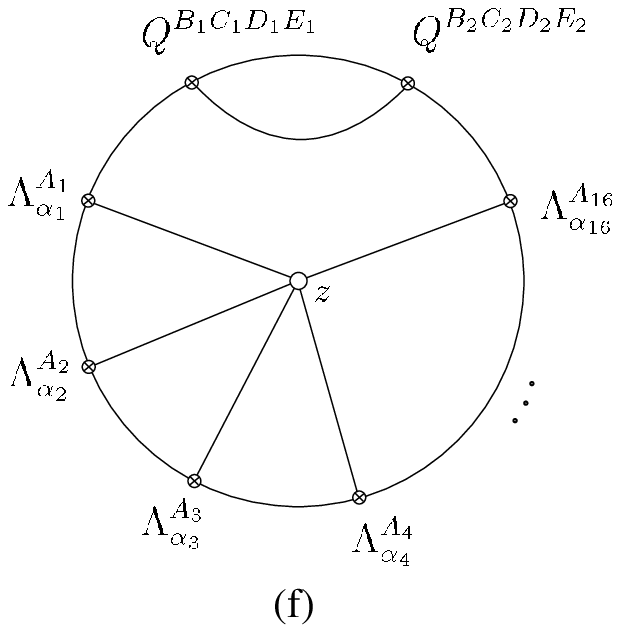}}
\caption{AdS diagrams contributing to the correlation function
$G_{\Lambda^{16}Q_{2}^{2}}$.}
\label{Lam16Q2figAdS}
}

Diagram (a) corresponds to a structure which appears at order
$N^{1/2}$ as well as $N^{-1/2}$. The $N^{1/2}$ contribution comes from
a $\Lambda^{16}$ vertex at order $\al^{-1}$  with two $Q_2$ insertions
coming from the expansion of $\sqrt{-g}$. Here and in the following we
are using the symbol $Q_\ell$ as a shortcut to denote the bulk
field corresponding to the operator $\calQ_\ell$ in the boundary field
theory, \ie a linear combination of the trace part of the metric and
the RR four-form with indices on the five-sphere. The order $N^{-1/2}$
contribution with the structure in (a) comes from the $\calR^2
\Lambda^{16}$ vertex at order $\al$ as well as from subleading
corrections to the leading $N^{1/2}$ term. Diagram (b) represents an
amplitude starting at order $N^{1/2}$ and resulting from the insertion
of a $Q_\ell$ coming from the expansion of $\sqrt{-g}$ in the
$\Lambda^{16}$ vertex. The perturbative vertex at the point $w$ is a
$Q_\ell Q_2 Q_2$ cubic interaction, where $\ell$ takes any
value allowed by the $SU(4)$ selection rules.  The amplitudes in (c),
(d) and (e) all have leading terms of order $N^{1/2}$, plus $1/N$
corrections, which correspond to a $\Lambda^{16}$ interaction with
additional perturbative vertices. We will see that in the case (c) the
dashed lines correspond to dilatini in the second Kaluza--Klein
excited mode, which transform in the {\bf 60} of $SU(4)$. In the case
(d) there are different possibilities for the bulk-to-bulk propagators
between the points $v$, $w$ and $z$ and we will analyse the
details below. Finally we will see that a potential contribution with
the structure in (e) actually vanishes.  Diagram (f) is a disconnected
amplitude. Its leading term is of order $N^{5/2}$, resulting from a
factor of $N^2$ carried by the factorised two-point function and a
factor of $N^{1/2}$ associated with the D-instanton vertex in
$z$. This contribution can be matched separately with the one of order
$N^{5/2}$ in the Yang--Mills calculation. In all the other cases it
will be crucial for our comparison with the field theory calculation
that all the interactions involved in these processes are either of
the extremal or of the next-to-extremal type
\cite{extremal,extrbk,extrwest1,extrwest2,nextextremal,nearextremal}.
In analysing the various contributions we will make use of the
techniques and results of these papers, the reader is referred, for
instance, to the appendix of \cite{nearextremal} for a detailed
discussion of the technical aspects.

To compare the results with field theory we need to normalise
the AdS amplitudes appropriately. With our definitions of
the Yang--Mills operators the normalisation of the external states in
the supergravity processes does not involve any powers of $N$ or
$\gs$. The exact numerical normalisations will not be important in our
subsequent analysis but they can be fixed by the matching of two-point
functions as, for instance, in  \cite{intri}.
We will use the standard notation $K_\D(z;x)=K_\D(z_m,z_0;x_m)$ to
denote the bulk-to-boundary propagator from the point $(z_m,z_0)$ in
$AdS_5$ to the boundary point $x_m$ for a supergravity field
corresponding to a scalar operator of dimension $\D$
\cite{gkp,wittone,freedman}. A superscript $F$ is used to distinguish
the propagators for fermionic operators.

Let us first consider the disconnected diagram \ref{Lam16Q2figAdS}(f).
This amplitude simply gives
\be
G_{\Lambda^{16}Q^2}^{\rm (f)} =
\la Q(y_1)Q(y_2)\ra\la\Lambda(x_1)\dots \Lambda(x_{16})\ra
\, ,
\label{Lam16Q2adsdisc1}
\ee
where the factorised two-point function is given by the free field theory
result and the remaining sixteen-point function coincides with the
minimal correlation function of \cite{bgkr,doreya}. Thus we get
\be
G_{\Lambda^{16}Q^2}^{\rm (f)} =
\frac{c_{\rm f}(\tau,\bar\tau,N)}{(y_1-y_2)^4} \int \frac{d^5z}{z_0^5}
\prod_{i=1}^{16} K_{7/2}^F(z;x_i) \, ,
\label{Lam16Q2adsdisc}
\ee
with
\be
c_{\rm f}(\tau,\bar\tau,N) = c_{\rm f} N^{5/2} \gs^{-12}
\er^{2\pi i\tau} \, ,
\label{ads-f-coeff}
\ee
where $c_{\rm f}$ is a non-zero numerical constant which can be
determined from the known normalisations of the two-point function and
the minimal $\la\Lambda^{16}\ra$ correlator. As already observed
the overall power of $N$ comes from a factor of $N^2$ associated with
the two-point function times a $N^{1/2}$ from the D-instanton
correction to the minimal sixteen-point function. The bulk-to-boundary
propagator for the dilatini, $K^F_{7/2}$,  corresponding to the
fermionic operator $\hat\Lambda$ of dimension $7/2$ was given in
\cite{henningsfet,bgkr} and reads
\ba
K^F_{7/2}(z_m,z_0;x_m) &=& K_{4}(z_m,z_0;x_m)
\frac{1}{\sqrt{z_0}} \left( z_0 \gamma_{\hat 5} +
(z-x)^{m}\gamma_{\hat m}\right) \nn \\ &=&
\frac{z_0^4}{[(z-x)^2+z_0^2]^4} \frac{1}{\sqrt{z_0}} \left( z_0
\gamma_{\hat 5} + (z-x)^{m}\gamma_{\hat m}\right) \, ,
\label{prop-7/2}
\ea
so that the spatial dependence in (\ref{Lam16Q2adsdisc}) agrees with
the field theory result.  With our normalisations the two
point-function is independent of $\gs$ and the minimal sixteen-point
function is proportional to $\gs^{-12}$. This contribution reproduces
the leading $N^{5/2}$ disconnected term $\hat G^{\rm
(c1)}_{\hat\Lambda^{16}\calQ^2}$  of
(\ref{Lam16Q2cfin})-(\ref{Ndepc1}) in the Yang--Mills  calculation.
Notice that in the Yang--Mills calculation the disconnected
contribution with the spatial structure matching the supergravity
result (\ref{Lam16Q2adsdisc}) arises with a coefficient
\be
C_{\rm c1}(N) = \left(N^2-1\right)N^{1/2}\left(1 + O(1/N)\right) \, .
\label{ymdisccoeff}
\ee
The above calculation reproduces the leading order term in
(\ref{ymdisccoeff}). The full two-point function in supergravity
is,
however, expected to reproduce exactly the field theory result, which
means that it actually should produce a factor of $(N-1)^2$ which
matches that in (\ref{ymdisccoeff}). Moreover the $1/N$ corrections in
the Yang--Mills result can also be explained on the gravity side: they
are produced by amplitudes involving higher $\al$ vertices that
contribute to the minimal correlator of 16 dilatini via the mechanism
described at the end of section \ref{oneinst}.

The AdS diagram in figure \ref{Lam16Q2figAdS}(a) is a contact diagram
and it is straightforward to evaluate since it has the same structure
as those contributing to the minimal correlation functions. As noted
above a contribution  with this structure is obtained from the
$(\tr\,h)^2\Lambda^{16}$ coupling produced by expanding the
$\sqrt{-g}$ factor in the $\Lambda^{16}$ vertex in the effective
action at order $\al^{-1}$ and then simply replacing all the fields
with the appropriate bulk-to-boundary  propagators. Moreover a
contribution with exactly the same form comes from the $\calR^2
\Lambda^{16}$ vertex at order $\al$. The total contribution of
diagram (a) is
\be
G_{\Lambda^{16}Q^2}^{\rm (a)} = c_{\rm a}(\tau,\bar\tau,N)
\int \frac{d^5z}{z_0^5} \, \prod_{i=1}^{16} K^F_{7/2}(z;x_i)
\prod_{j=1}^2 K_2(z;y_j) \, ,
\label{Lam16Q2adsa}
\ee
In (\ref{Lam16Q2adsa}) the coefficient $c_{\rm a}(\tau,\bar\tau,N)$ is
the sum of a term  of order $N^{1/2}$ and one of order $N^{-1/2}$. It
encodes the information about the D-instanton induced part of the
modular forms appearing in the  relevant vertices. The explicit form
of the two terms in $c_{\rm a}(\tau,\bar\tau,N)$ can be read from the
general formula (\ref{gentermstring}). For the comparison with the
result of the super Yang--Mills calculation we are interested in the
dependence on $N$ and on the coupling. For this purpose notice in
particular that the leading D-instanton term in all the modular forms
$f_l^{(0,0)}(\tau,\bar\tau)$ has no powers of $\tau_2=\gs^{-1}$. For
the coefficient in (\ref{Lam16Q2adsa}) we get
\be
c_{\rm a}(\tau,\bar\tau,N) =
c_{\rm a1} N^{1/2} \gs^{-12}  \er^{2\pi i\tau} + c_{\rm a2} N^{-1/2}
\gs^{-12}  \er^{2\pi i\tau} \, ,
\label{ads-a-coeff}
\ee
where $c_{\rm a1}$ and $c_{\rm a2}$ are numerical coefficients
independent of $N$ and $\gs$ and the powers of the coupling come from
the expansion (\ref{asymterm}) of the modular functions
$f^{(12,-12)}_1(\tau,\bar\tau)$ and $f^{(12,-12)}_2(\tau,\bar\tau)$
appearing in the $\Lambda^{16}$ and $\calR^2\Lambda^{16}$ vertices
respectively.

Diagram \ref{Lam16Q2figAdS}(b) is an exchange amplitude and we must
analyse the contribution of all the possible intermediate states. The
vertex in $z$ is a $(\tr\,h)\Lambda^{16}$ coupling of order
$\al^{-1}$, where the $\tr\,h$ insertion is obtained again by expanding
$\sqrt{-g}$. Notice that a similar diagram involving a $\tau \,
\Lambda^{16}$ coupling is not present because a cubic coupling of a
dilaton in any Kaluza--Klein mode with two $Q$'s is forbidden by the
$U(1)$ symmetry of classical supergravity. Hence the dashed line can
correspond only to a $Q$ in any Kaluza--Klein level allowed by the
$SU(4)$ selection rules. The amplitude in figure \ref{Lam16Q2figAdS}
(b) is thus
\be
G_{\Lambda^{16}Q^2}^{\rm (b)} = c_{\rm b}(\tau,\bar\tau,N)
\int \frac{d^5z}{z_0^5} \frac{d^5w}{w_0^5}\, \prod_{i=1}^{16}
K^F_{7/2}(z;x_i) G_\ell(z,w) \prod_{j=1}^2 K_2(w;y_j) \, ,
\label{Lam16Q2adsb}
\ee
where $G_\ell(z,w)$ is a bulk-to-bulk propagator for the field
corresponding to the chiral primary $\calQ_\ell$. The $SU(4)$ symmetry
only allows $\ell=2,4$. In the case $\ell=2$ the vertex in $w$ is next
to extremal and the $w$ integration can be done using the formulae
derived in \cite{zint} for the general scalar exchange. An exchange
diagram with two bulk integrals as in (\ref{Lam16Q2adsb}) can be
reduced to a sum of contact contributions. In this case
there is only one term in the sum which at leading order takes
the form
\be
\frac{c_{\rm b1}(\tau,\bar\tau,N)}{(y_1-y_2)^2} \int
\frac{d^5z}{z_0^5} \, \prod_{i=1}^{16}
K^F_{7/2}(z;x_i) \prod_{j=1}^2 K_1(z;y_j) \, ,
\label{Lam16Q2adsb1}
\ee
with
\be
c_{\rm b1}(\tau,\bar\tau,N) = c_{\rm b1} N^{1/2} \gs^{-12}
\er^{2\pi i\tau} \, .
\label{ads-b1-coeff}
\ee
In the case $\ell=4$ the coupling in $w$ is extremal and the
resulting amplitude reduces to a contact contribution.
The tree diagram in figure \ref{Lam16Q2figAdS}(b)
should be thought of as representing the ten dimensional
amplitude. Then the integration over the five-sphere selects the
intermediate states allowed by the $SO(6)$ symmetry, which are
 the scalars with $\ell=2,4$ in the present case. One can then simplify
the integral over the $AdS_5$ point $w$ by integrating by parts the
derivatives appearing in the cubic coupling. In the case of the
extremal vertex  the derivatives taken to act on the $G_\ell(z,w)$
bulk-to-bulk propagator reconstruct the wave operator, so that the
only contribution that is left is a contact diagram. Hence the
corresponding amplitude becomes
\be
c_{\rm b2}(\tau,\bar\tau,N) \int
\frac{d^5z}{z_0^5} \, \prod_{i=1}^{16} K^F_{7/2}(z;x_i) \prod_{j=1}^2
K_2(z;y_j) \, ,
\label{Lam16Q2adsb2}
\ee
where
\be
c_{\rm b2}(\tau,\bar\tau,N) = c_{\rm b2} N^{1/2} \gs^{-12}
\er^{2\pi i\tau} \, .
\label{ads-b2-coeff}
\ee
Notice that, unlike in the cases encountered in evaluating four-point
functions, in this case the new contact term obtained in $AdS_5$ is not
extremal or next-to-extremal.

In diagram \ref{Lam16Q2figAdS}(c) the two cubic vertices in $v$ and
$w$ involve a dilatino and a $Q_2$ combination, thus the dashed
lines correspond to dilatini in an allowed Kaluza--Klein mode. The
only intermediate states permitted by the $SU(4)$ symmetry are in the
representation $[1,2,0]$, \ie in the {\bf 60}. The vertex in $z$ is
hence the usual $\Lambda^{16}$ interaction where two of the fields are
taken in their second Kaluza--Klein excited level. As a consequence
the vertices in $v$ and $w$ are extremal. More precisely they are
superconformal descendants of a standard extremal coupling of chiral
primaries. Then the same mechanism discussed above reduces the
amplitude to
\be
G_{\Lambda^{16}Q^2}^{\rm (c)} = c_{\rm c}(\tau,\bar\tau,N)
\int \frac{d^5z}{z_0^5} \, \prod_{i=1}^{16} K^F_{7/2}(z;x_i)
\prod_{j=1}^2 K_2(z;y_j) \, ,
\label{Lam16Q2adsc}
\ee
where
\be
c_{\rm c}(\tau,\bar\tau,N) = c_{\rm c} N^{1/2} \gs^{-12}
\er^{2\pi i\tau} \, .
\label{ads-c-coeff}
\ee
Here again the factors of $N$ and $\gs$ are those associated with the
$\Lambda^{16}$ vertex.

The amplitude in diagram (d) is the most complicated that we have to
consider. Denoting by $\phi$ and $\phi^\prime$ the two bulk fields
exchanged between $v$ and $w$ and between $v$ and $z$
respectively we obtain
\be
G_{\Lambda^{16}Q^2}^{\rm (d)} = c_{\rm d}(\tau,\bar\tau,N)
\int \frac{d^5z}{z_0^5} \frac{d^5v}{(v_0)^5}
\frac{d^5w}{(w_0)^5} \, \prod_{i=1}^{16}
K^F_{7/2}(z;x_i) \, G_{\phi^\prime}(z,v) \, G_\phi(v,w)
\prod_{j=1}^2 K_2(w;y_j) \, .
\label{Lam16Q2adsd}
\ee
The vertex in $z$ is $\Lambda^{16}$ so that $\phi^\prime$ is a
dilatino or one of its Kaluza--Klein excited states. Starting form the
vertex in $w$ the allowed $SU(4)$ representations for the field
$\phi$ are $[0,2,0]$, $[0,4,0]$, $[2,0,2]$ and $[0,0,0]$, \ie ${\bf
20^\prime}$, ${\bf 105}$, ${\bf 84}$ and ${\bf 1}$. The first two
correspond to $Q_2$ and $Q_4$ respectively, the third one is a
scalar coming from internal components of the metric and the last one
is a graviton. When $\phi$ is in the  $[0,2,0]$ $\phi^\prime$ can only
be in the $[1,2,0]$, \ie it is a dilatino in the second Kaluza--Klein
excited level. In this case the vertex at $w$ is next-to-extremal
and the one at $v$ is a descendant of an extremal vertex. The
integration in $w$ can be carried out using the formulae in \cite{zint}
and from the resulting extremal vertex in $v$ we get a contribution of
the form
is
\be
G_{\Lambda^{16}Q^2}^{\rm (d1)} =
\frac{c_{\rm d1}(\tau,\bar\tau,N)}{(y_1-y_2)^2}
\int \frac{d^5z}{z_0^5} \, \prod_{i=1}^{16}
K^F_{7/2}(z;x_i) \, \prod_{j=1}^2 K_1(z;y_j) \, .
\label{Lam16Q2adsd1}
\ee
with
\be
c_{\rm d1}(\tau,\bar\tau,N) = c_{\rm d1} N^{1/2} \gs^{-12}
\er^{2\pi i\tau} \, .
\label{ads-d1-coeff}
\ee
When $\phi$ is in the $[0,4,0]$ the coupling in $w$ is extremal and
the integral is finite so that the amplitude vanishes. The same
happens in the case of $\phi=h_{(\a\b)}$ in the $[2,0,2]$ since this
is a superconformal descendant of $Q_4$. The case of the graviton
exchange is more involved and we will not discuss it in detail. In
this case the only allowed state for the dilatino $\phi^\prime$ is
the Kaluza--Klein ground state. The integrals in $w$ and $v$ can
be sequentially performed using the techniques of \cite{zint,freedgrav}
reducing the amplitude to a sum of terms of the same form as those in
(\ref{Lam16Q2adsa}) and (\ref{Lam16Q2adsb1}).

The amplitude corresponding to diagram \ref{Lam16Q2figAdS}(e)
involves a $\Lambda^{16}$ vertex in $z$. The bulk lines from $z$ to
$v$ and from $v$ to $w$ are dilatini in allowed Kaluza--Klein
levels. The $SU(4)$ selection rules imply in particular that the
second one is in the $[1,2,0]$. This means that the coupling in $w$
is extremal and thus the whole contribution vanishes because the
integral is finite.

There are, in principle, other possible contributions to the exchange
diagrams considered above. These involve other D-instanton induced
vertices at order $\al^{-1}$. For instance, there is a vertex of the
form \cite{gs}
\be
f_1^{(11,-11)}(\tau,\bar\tau)\Lambda^{15}\gamma^\mu\psi^*_\mu \, .
\label{lam15psi}
\ee
This could contribute to diagrams (d) and (e) in figure
\ref{Lam16Q2figAdS}. However, the possibility of such processes is
ruled out by $U(1)$ conservation at one of the cubic interactions
involved. For instance, in the case of diagram (d) if the vertex in $z$
were (\ref{lam15psi}) then the bulk state exchanged between $v$ and
$w$ would be $U(1)$ charged because the dilatino and the gravitino
have different $U(1)$ charges. But then the coupling in $w$ would
not be allowed because the $Q$'s are not charged. An analogous
argument can be applied to the case of diagram (e). In this case if
the state exchanged between $z$ and $v$ is a gravitino then $U(1)$
conservation at $v$ implies that the bulk line between $v$ and
$w$ is also a gravitino, but then the coupling in $w$ would violate
the $U(1)$ selection rule. Similar arguments can be applied to rule
out the possibility of contributions from amplitudes involving the
vertices $G\Lambda^{14}$ and $\psi^2\Lambda^{14}$.

In conclusion, combining the various AdS amplitudes we can recapitulate
the result as follows. There is an order $N^{5/2}$ disconnected amplitude
of the form of (\ref{Lam16Q2adsdisc}). This corresponds to
the Yang--Mills contribution of
(\ref{Lam16Q2cfin})-(\ref{Ndepc1}). The other terms are of order
$N^{1/2}$ and $N^{-1/2}$. These are of two different types
\be
a \int \frac{d^5z}{z_0^5} \,
\prod_{i=1}^{16} K^F_{7/2}(z;x_i) \prod_{j=1}^2 K_2(z;y_j) +
\frac{b}{(y_1-y_2)^2} \int \frac{d^5z}{z_0^5} \,
\prod_{i=1}^{16} K^F_{7/2}(z;x_i) \prod_{j=1}^2 K_1(z;y_j) \, ,
\label{Nhalffin}
\ee
where the coefficients $a$ and $b$ contain terms of order
$N^{1/2}$ and $N^{-1/2}$,
\ba
&& a = a_1 \gs^{-12} N^{1/2} + a_2 \gs^{-12} N^{-1/2} \\
&& b = b_1 \gs^{-12} N^{1/2} + b_2 \gs^{-12} N^{-1/2} \, .
\label{Nhalfcoeff}
\ea
The dependence on the parameters, $\gy^2=4\pi\gs$ and $N$, as well as
the spatial dependence in this result are exactly those found in the
Yang--Mills calculation of the previous subsection,
(\ref{lam16Q2YMsum})-(\ref{Ndepconn2}). In particular, the agreement of
the spatial structure is highly non-trivial. It is achieved after combining
many types of terms in a calculation which is significantly more
complicated than the one required in the case of minimal correlation
functions.
It would be satisfying to be able to make a detailed correspondence
between the coefficients in the Yang--Mills instanton calculation and
those of supergravity D-instanton contributions to the
$\Lambda^{16}Q^2$ interaction.  However, there are ambiguities that
our analysis has not resolved.  We can identify the amplitudes
contributing to the leading terms of order $N^{1/2}$ in
(\ref{Nhalfcoeff}). The contact amplitude --- the first term in
(\ref{Nhalffin}) --- receives contributions from the terms (a), (b2) and
(c) in the previous analysis,
\be
a_1 = c_{\rm a} + c_{\rm b2} + c_{\rm c} \, .
\label{leadcontact}
\ee
Similarly the second term in (\ref{Nhalffin}) gets contributions from
the processes (b1) and (d1), so that
\be
b_1 = c_{\rm b1} + c_{\rm d1} \, .
\label{leadexch}
\ee
These coefficients are, in principle, computable although the
calculation is rather subtle because it involves the tree level
processes described in this section. The situation is even more
complicated for the subleading terms of order $N^{-1/2}$ in
(\ref{Nhalfcoeff}). On the Yang--Mills side we have a precise
expression for the term of order $N^{-1/2}$, which is the first  $1/N$
correction  to the leading one-instanton contribution to the  $\hat
\Lambda^{16}\, \calQ^2$ correlation function (as always, in
semi-classical approximation).  This comes from expanding the
instanton measure and taking into account the subleading contributions
in the diagrams coming both from the scalar contractions and from the
$(\nub\nu)$ insertions. On the string side it should come from terms
of order $\al$.  For this specific correlation function the relevant
coupling at this order should be $\al\,\sqrt {-g}\,\er^{\phi/2}\,
\Lambda^{16}\, \calR^2$. Naively, our calculation leads to a
prediction for the coefficient of this conjectured interaction in
(\ref{actfiv}). However, another order $\alpha'$ effect is almost
certainly present, which complicates matters. This arises from the
potential presence of a $\al\, \sqrt{-g}\, \er^{9\phi/2}\,
\Lambda^{16}\, F_5^4$ term.  Since $F_5$ has a nonzero background
value this interaction makes a contribution to  the $\hat
\Lambda^{16}$ correlation function that is suppressed by a power of
$1/N$ relative to the leading instanton term. This is the effect that
produces the $1/N$ corrections to minimal correlation functions
 that were
mentioned in section \ref{oneinst}. However, the same interaction also
generates a new $\Lambda^{16} \, h$ vertex which can be joined by a
$\langle h h\rangle$ propagator to a  classical $h^3$ vertex.  This
generates a tree diagram that contributes to the $\hat \Lambda^{16}\,
\calQ^2$ correlation function at order $N^{-1/2}$, which is the same
order as the  $\al\,\sqrt{-g}\,\er^{\phi/2}\, \Lambda^{16}\,
\calR^2$. In general for all the processes considered in this section
which involved a $\Lambda^{16}$ vertex generating a contribution of
order $N^{1/2}$ there is a corresponding amplitude of order $N^{-1/2}$
in which the D-instanton induced vertex is $\Lambda^{16}\, F_5^4$, and
the five-form factors are replaced by their background value. This
explains the fact that the contribution we found at order $N^{-1/2}$
is not purely a contact amplitude as would be the case if it came
entirely from the $\al\,\sqrt{-g}\,\er^{\phi/2}\, \Lambda^{16}\,
\calR^2$ vertex.  In conclusion, although the Yang--Mills calculation
shows the presence of various order $\al$ interactions, our
calculation does not pin down their relative coefficients.

\section{Non-minimal correlation functions in the instanton
background: Kaluza--Klein excited states}
\label{kknonmincorr}

In this section we illustrate the general features of correlations
functions involving operators corresponding to Kaluza-Klein excited
modes in supergravity. As in the previous section we focus on a
specific example in order to highlight the general properties of these
processes. The new aspect that emerges is the r\^ole played by
$\bar\nu\nu$ bilinears in the {\bf 6} of $SU(4)$.

From the string theory point of view there is little distinction
between amplitudes in which the boundary fields are in their
Kaluza--Klein ground states or in excited states.  This means that for
any of the minimal processes that contribute at order $N^{1/2}$ there
should be similar contributions when the external legs are in excited
Kaluza--Klein states. The only additional selection rule is imposed by
the $SU(4)$ symmetry. On the Yang--Mills side the operators
corresponding to Kaluza--Klein excitations are higher dimensional with
respect to the ones associated with supergravity states. This means
that they soak up more fermion modes and thus their correlation
functions are non-minimal and involve the presence of $\nu$ and
$\bar\nu$ fermionic variables.

We consider here the example of the  $\Lambda^{16}$ interaction at
order $\al^{-1}$. Instead of taking all the $\Lambda$'s in their
Kaluza--Klein ground state, \ie in the ${\bf 4}$ of $SU(4)$, we might
consider the process  in which two of them are in the first excited
state, \ie in the ${\bf 20^*}$. This is a combination allowed by
$SU(4)$ selection rules. We analyse the Yang--Mills calculation first
and then compare the results with the AdS calculation.

\subsection{One-instanton contribution in $\N4$ SYM}
\label{1instcorrKK}

The process we are interested in corresponds to the correlation
function
\be
\hat G_{\hat\Lambda^{14}_{\bf 4}\hat\Lambda^2_{\bf
20^*}} (x_1,\ldots,x_{14},y_1,y_2) = \langle \hat\Lambda_{\bf 4}(x_1)
\ldots \hat\Lambda_{\bf 4}(x_{14}) \hat\Lambda_{\bf 20^*}(y_{1})
\hat\Lambda_{\bf 20^*}(y_{2}) \rangle \, .
\label{twotwent}
\ee
Here $\hat\Lambda_{\bf 4}$ is the fermionic operator dual to the
type IIB dilatino already considered in the previous section and
$\hat\Lambda_{\bf 20^*}$ denotes the operator of
(\ref{Lam20newdef}), which corresponds to the first  Kaluza--Klein
mode of the dilatino.  The correlation function (\ref{twotwent}) involves 20
fermionic modes. Analogously to the case described in the previous
section there are two types of contributions to this correlation
function at leading order in the coupling in the one-instanton
sector. The first is obtained by the standard semiclassical
approximation replacing the operators with their expressions in the
instanton  background. This means that each operator insertion soaks
up one superconformal mode and in addition the two $\hat\Lambda_{\bf
20^*}$ insertions involve a $\nsix$ bilinear each. The second type of
contribution corresponds to a leading order quantum correction to the
semiclassical approximation and arises from contracting two scalars
between the operators $\hat\Lambda_{\bf 20^*}$. Thus we have
\be
\hat G_{\hat\Lambda^{14}_{\bf 4}\hat\Lambda^2_{\bf 20^*}} =
\hat G_{\hat\Lambda^{14}_{\bf 4}\hat\Lambda^2_{\bf 20^*}}^{({\rm a})}
+ \hat G_{\hat\Lambda^{14}_{\bf 4}\hat\Lambda^2_{\bf 20^*}}^{({\rm b})}
\label{Lam20-2-dec}
\ee
and the two types of contributions are represented in figure
\ref{Lam20-2figYM}.

\FIGURE[!h]{
{\includegraphics[width=0.45\textwidth]{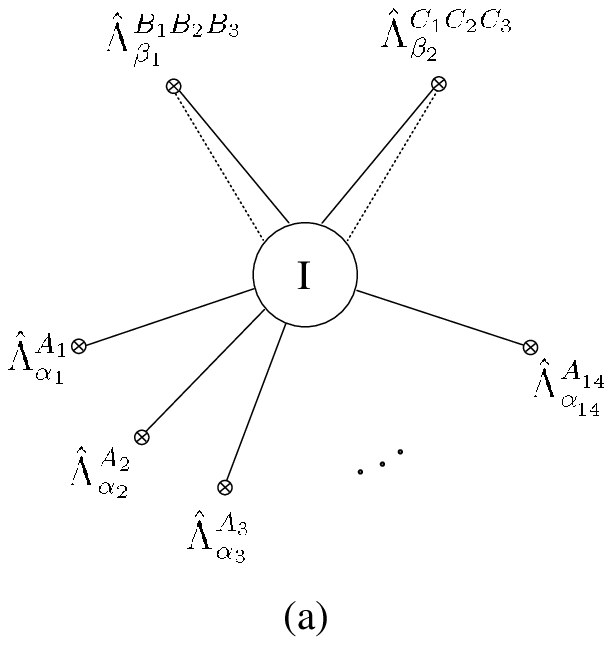}
\hspace*{0.6cm}
\includegraphics[width=0.45\textwidth]{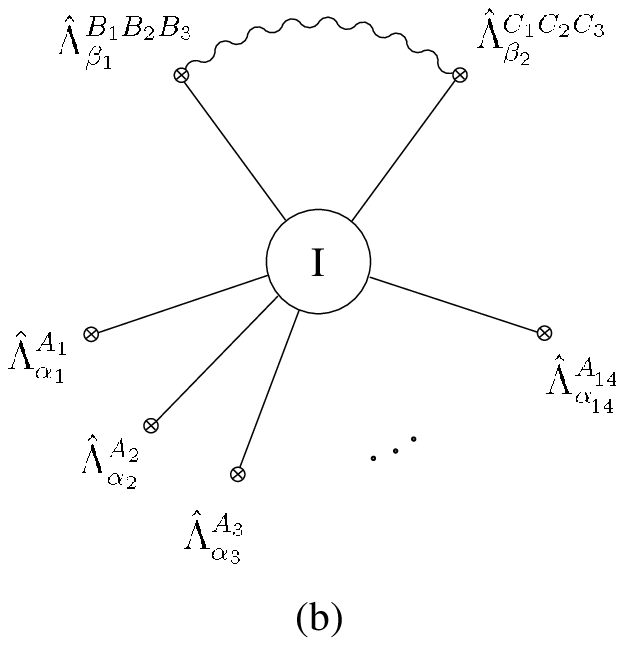}}
\caption{The two types of contributions to the correlation function
$\hat G_{\hat\Lambda^{14}_{\bf 4}\hat\Lambda^2_{\bf 20^*}}$.  As
before, plain lines correspond to superconformal modes, wiggly lines
are scalar propagators and dotted lines are used to indicate the
insertion of $\bar\nu \nu$ pairs in the {\bf 6} of $SU(4)$.}
\label{Lam20-2figYM}
}

The first term in (\ref{Lam20-2-dec}) is given by
\ba
{\hat G}_{\hat\Lambda^{14}_{\bf 4}
\hat\Lambda^2_{\bf 20^*}}^{({\rm a})}  &\!=\!&
\int d\mu_{\rm phys}\, \er^{-S_{4F}}
\left[\hat\Lambda^{A_1}_{\a_1}(x_1) \ldots
\hat\Lambda^{A_{14}}_{\a_{14}}(x_{14})
\hat\Lambda^{B_1B_2B_3}_{\b_1}(y_1)
\hat\Lambda^{C_1C_2C_3}_{\b_2}(y_2) \right] \nn \\
&\!=\!& \frac{1}{\gy^{34}N} \int d\mu_{\rm phys}\,
\er^{-S_{4F}} \, \prod_{i=1}^{14}
\frac{\rho^4\zeta^{A_i}_{\a_i}}{[(x_i-x_0)^2+\rho^2]^4}
\prod_{j=1}^{2}\frac{\rho^4}{[(y_j-x_0)^2+\rho^2]^5} \nn \\
&& \left[\left(\zeta^{B_2}_{\b_1} \bar\nu^{[B_1}\nu^{B_3]} +
\zeta^{B_3}_{\b_1} \bar\nu^{[B_1}\nu^{B_2]} \right) \left(
\zeta^{C_2}_{\b_2} \bar\nu^{[C_1}\nu^{C_3]} + \zeta^{C_3}_{\b_2}
\bar\nu^{[C_1}\nu^{C_2]} \right)\right] \, ,
\label{Lam20-2-a1}
\ea
where we have substituted the classical expressions (\ref{explLamb})
and (\ref{expl20ferm}) for the operators $\Lambda_{\bf 4}$ and
$\Lambda_{\bf 20^*}$ respectively and we have omitted numerical
coefficients which will be reinstated in the final expression. Notice
in particular the powers of $\gy$ and $N$ in the prefactor which are
dictated by the normalisation of the operators. According to the
general rule (\ref{cpoell}) for the normalisation of Yang--Mills
operators each $\hat\Lambda_{\bf 4}$ brings a factor of $\gy^{-2}$ and
each $\hat\Lambda_{\bf 20^*}$ a factor of $\gy^{-3}N^{-1/2}$. In terms
of the generating function (\ref{genfunctfin}) the above expression is
rewritten as
\ba
&& \hat G_{\hat\Lambda^{14}_{\bf 4}\hat\Lambda^2_{\bf 20^*}}^{({\rm a})}
= \frac{1}{\gy^{34} N}  \frac{\gy^{8}\er^{2\pi i\tau}}
{(N-1)!(N-2)!}  \int d\rho \, d^4x_0 \, d^5\Omega \prod_{A=1}^4
d^2\eta^A \, d^2 {\bar\xi}^A \, \rho^{4N-7} \nn \\
&& \int_0^\infty dr \, r^{4N-3} \er^{-2\rho^2 r^2} \,
\prod_{i=1}^{14} \frac{\rho^4}{[(x_i-x_0)^2+\rho^2]^4}
\zeta^{A_i}_{\a_i} \prod_{j=1}^2
\frac{\rho^4}{[(y_j-x_0)^2+\rho^2]^5}   \nn \\
&& \left\{ \left( \zeta^{B_2}_{\b_1}\zeta^{C_2}_{\b_2}
\left.  \left[ \frac{\d^4\calZ(\vt,\bvt;\Omega,r)}
{\d\vt_{u_1[B_1}\d\bvt_{B_3]}^{u_1}\d\vt_{u_2[C_1}\d\bvt_{C_3]}^{u_2}}
\right] \right\vert_{\vt=\bvt=0} + (B_2 \leftrightarrow B_3) \right) +
(C_2 \leftrightarrow C_3) \right\} \, ,
\label{Lam20-2-a2}
\ea
where the integrations over the collective coordinates have been
indicated explicitly. After performing the derivatives we can compute
the angular integrals using (\ref{spherint2}). As in the
cases described in the previous section there is a factorisation of
the angular integrals, which implies that the indices carried by the
$\nu$ and $\bar\nu$ modes are combined separately in a $SU(4)$
singlet.  Proceeding as in section \ref{1instcorr} we then carry
out the  integration over the radial variable $r$ to extract the
dependence  on $\gy$ and $N$ and get
\ba
\hat G_{\hat\Lambda^{14}_{\bf 4}
\hat\Lambda^2_{\bf 20^*}}^{({\rm a})}&\!=\!&
C_{\rm a}(N) \, \frac{\er^{2\pi i\tau}}{\gy^{24}}
\int \frac{d\rho \, d^4x_0}{\rho^5} \prod_{A=1}^4 d^2\eta^A \, d^2
{\bar\xi}^A \, \prod_{i=1}^{14}\frac{\rho^4}{[(x_i-x_0)^2+\rho^2]^4}
\zeta^{A_i}_{\a_i}  \nn  \\
&& \left( \veps^{B_1B_3C_1C_3}
\frac{\rho^5}{[(y_1-x_0)^2+\rho^2]^5}\zeta^{B_2}_{\b_1}
\frac{\rho^5}{[(y_2-x_0)^2+\rho^2]^5}\zeta^{C_2}_{\b_2} + \cdots
\right) \, ,
\label{Lam20-2-a3}
\ea
where the $\cdots$\ refers to symmetrisation in the ($B_2$,$B_3$)
and ($C_2$,$C_3$) pairs of indices. The $N$-dependence is enclosed in
the prefactor
\be
C_{\rm a}(N) = \frac{2^{44}3^{18}}{\pi^{35/2}}N^{1/2}
\left( 1 - \frac{25}{8N} + O(1/N^2) \right) \, .
\label{NdepL20-2-a}
\ee
Integration over the fermion zero modes $\eta$ and $\bar\xi$ produces
the same completely antisymmetric tensor $t_{(16)}$ that enters in the
calculation of the correlator of sixteen $\hat\Lambda$'s in the {\bf
4} \cite{bgkr}. In the next subsection we will see how this result
agrees with the AdS calculation. In particular the powers of $\gy$ and
$N$ are the same as in the minimal correlation function of sixteen
$\hat\Lambda_{\bf 4}$: the factors of $\gy\sqrt{N}$ associated with
each $\nsix$ insertion are compensated by the normalisation of the
operators. From the supergravity point of view (\ref{Lam20-2-a3}) will
be interpreted as a sum of terms involving the product of fourteen
bulk-to-boundary propagators $K^F_{7/2}$ for the $\hat\Lambda_{\bf 4}$
insertions and two propagators $K^F_{9/2}$ for the $\hat\Lambda_{\bf
20^*}$ insertions. In each term the $SU(4)$ indices are combined in a
singlet tensor of the form $\veps_{(4)} t_{(16)}$.

The second type of contribution to the correlation function we are
considering, $\hat G_{\hat\Lambda^{14}_{\bf 4}
\hat\Lambda^2_{\bf 20^*}}^{({\rm b})}$, involves a scalar propagator,
see figure \ref{Lam20-2figYM}(b). It comes from the contraction
\ba
\hat G_{\hat\Lambda^{14}_{\bf 4}
\hat\Lambda^2_{\bf 20^*}}^{({\rm b})} &\!=&
\la \Tr\!\left(F_{\a_1}{}^{\b_1}\lambda^{A_1}_{\g_1}\right)\!(x_1)
 \ldots \Tr\!\left(F_{\a_{14}}{}^{\g_{14}}\lambda^{A_{14}}_{\b_{14}}
\right)\!(x_{14}) \nn \\
&& \Tr\left(\left\{F_{\b_1}{}^{\d_1},\lambda^{B_2}_{\d_1}\right\}
\v^{B_1B_3} \right)(y_1) \,
\Tr\left(\left\{F_{\b_2}{}^{\d_2},\lambda^{C_2}_{\d_2} \right\}
\v^{C_1C_3} \right)(y_2) \ra \raisebox{-10pt}{\hspace*{-7.09cm}
\rule{0.4pt}{4pt}\rule{4.93cm}{0.4pt}\rule{0.4pt}{4pt}}
\hspace*{2cm} + \cdots \, ,
\label{Lam20-2-b1}
\ea
where we are using the notation $F_\a{}^\g=F_{mn}\s_{\:\a}^{mn\,\g}$
and the ellipsis refers again to terms obtained by symmetrisation in
($B_2$,$B_3$) and ($C_2$,$C_3$).  Notice that as in the case
examined in the previous section there is no similar contribution
involving contractions of the $\lambda$'s or the $F$'s. For this
reason in evaluating $\hat G_{\hat\Lambda^{14}_{\bf
4}\hat\Lambda^2_{\bf 20^*}}^{({\rm b})}$ we do not need to consider
the term cubic in the fermions in the operators $\hat\Lambda_{\bf
20^*}$. Using the form of the scalar propagator given in section
\ref{scalarprop} to expand (\ref{Lam20-2-b1}) we get a rather
complicated expression that is given in (\ref{Lam20-2-b2})
of appendix \ref{scalcontr}.    In particular,
unlike in the case of the $\hat G_{\hat\Lambda^{16}\calQ^2}$
correlation function, here the $1/N$ and $1/N^2$ terms in the
propagator (\ref{propfin}) do contribute because the operator
$\hat\Lambda_{\bf 20^*}$ is cubic in the elementary fields.
However,
all the contributions obtained from the expansion of
(\ref{Lam20-2-b2}) are subleading in the large $N$ limit with respect
to (\ref{Lam20-2-a3})-(\ref{NdepL20-2-a}), which had an additional
factor of $N$ coming from the colour contractions in the $\nsix$
bilinears to compensate the $1/N$ from the normalisation of the
operators. The dependence on $N$ in (\ref{Lam20-2-b2}) arises entirely
from the instanton measure and the normalisation of the Klauza--Klein
operators  and thus the dominant terms are of order $N^{-1/2}$. We
only consider these leading order contributions in the following
discussion. Expanding (\ref{Lam20-2-b2}) we obtain two types of terms
with different spatial dependence. The structure of the two
contributions is
\ba
\hat G_{\hat\Lambda^{14}_{\bf 4}
\hat\Lambda^2_{\bf 20^*}}^{({\rm b})}&\!=\!&
C_{\rm b}(N) \, \frac{\er^{2\pi i\tau}}{\gy^{24}}
\int \frac{d\rho \, d^4x_0}{\rho^5} \prod_{A=1}^4 d^2\eta^A \, d^2
{\bar\xi}^A \, \prod_{i=1}^{14}\frac{\rho^4}{[(x_i-x_0)^2+\rho^2]^4}
\zeta^{A_i}_{\a_i}  \nn  \\
&& \left[ \veps^{B_1B_3C_1C_3} \left(
c_{\rm b1}\,\frac{\rho^5}{[(y_1-x_0)^2+\rho^2]^5}\zeta^{B_2}_{\b_1}
\frac{\rho^5}{[(y_2-x_0)^2+\rho^2]^5}\zeta^{C_2}_{\b_2}
\right. \right. \nn \\
&& \left. \left. + \,c_{\rm b2} \,\frac{1}{(y_1-y_2)^2}
\frac{\rho^4}{[(y_1-x_0)^2+\rho^2]^4}\zeta^{B_2}_{\b_1}
\frac{\rho^4}{[(y_2-x_0)^2+\rho^2]^4}\zeta^{C_2}_{\b_2}
\right) + \cdots \right]\, ,
\label{Lam20-2-b3}
\ea
where, as in the previous expressions, the ellipsis stands for
symmetrisation in the pairs ($B_2$,$B_3$) and ($C_2$,$C_3$). We find
that the coefficients $c_{\rm b1}$ and $c_{\rm b2}$ in
(\ref{Lam20-2-b3}) are both non-vanishing. This is a puzzling
result. The first term has the structure of a contact amplitude. It
contributes a $1/N$ correction to (\ref{Lam20-2-a3}) which, as we
shall discuss in the next subsection, matches the supergravity
result. However, the second term has a spatial dependence that cannot
be reproduced by any supergravity amplitude and it should, therefore,
be absent.  We shall leave the presence of this subleading term as an
open question and therefore only demonstrate agreement with
supergravity at leading order in $N$.

\subsection{AdS interpretation}
\label{adsinterprKK}

As already observed the supergravity process corresponding to the
correlation function (\ref{twotwent}) is straightforward to evaluate,
presenting no new difficulties with respect to the minimal
processes. There is only one type of contribution at leading order in
the string coupling: it is a contact diagram induced by the
$\Lambda^{16}$ interaction at order $\al^{-1}$, in which upon
dimensional reduction on the five-sphere two of the fields are taken
to be in the first Kaluza--Klein excited level, see figure
\ref{Lam20-2figAdS}.
\FIGURE[!h]{
{\includegraphics[width=0.1\textwidth]{x.eps}
\includegraphics[width=0.45\textwidth]{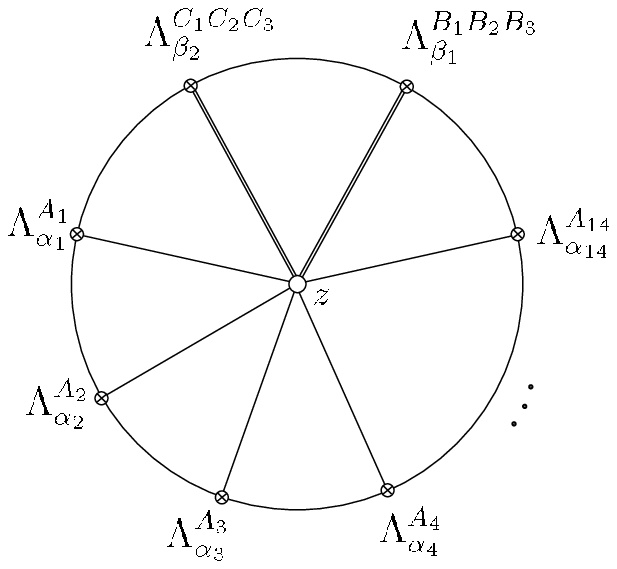}
\includegraphics[width=0.1\textwidth]{x.eps}}
\caption{The contact diagram contributing to the amplitude
$G_{\Lambda^{14}_{\bf 4}\Lambda^2_{\bf 20^*}}$.  A double line is used
to distinguish the insertions of Kaluza--Klein excited states.}
\label{Lam20-2figAdS}
}
In general integration over the five-sphere imposes the $SU(4)$
selection rules; in particular it gives a non vanishing coefficient for
the coupling we are considering of two $\Lambda$'s in the first
Kaluza--Klein level and fourteen in the ground state, in agreement
with the group theoretic  analysis on the Yang--Mills side. The five
dimensional coupling is of the form
\be
c(\tau,\bar\tau,N)\int\frac{d^5z}{z_0^5} \,\veps_{(4)} t_{(16)} \,
\Lambda_{\bf 4}^{14} \Lambda_{\bf 20^*}^2 \, ,
\label{KKcoupling}
\ee
where spinor and $SU(4)$ indices have not been indicated
explicitly. $\veps_{(4)}$ is a $SU(4)$ completely antisymmetric tensor
resulting from the angular integration on $S^5$ and $t_{(16)}$ is the
same 16-index tensor which contracts the indices in the vertex
coupling sixteen dilatini in the Kaluza--Klein ground state.
The coefficient $c(\tau,\bar\tau,N)$, apart from numerical
constants produced by the integration over the five-sphere, contains
the dependence on the coupling and on $N$, which are encoded in the
same modular form $f_1^{(12,-12)}(\tau,\bar\tau)$ which enters in the
`minimal' coupling of sixteen dilatini. At leading order in $\gs$ the
D-instanton contribution to $c(\tau,\bar\tau,N)$ is
\be
c(\tau,\bar\tau,N) \sim \gs^{-12} N^{1/2} \er^{2\pi i \tau} \, .
\label{gNdep}
\ee
The AdS amplitude is obtained as usual replacing the fields with
the appropriate bulk-to-boundary propagators
\be
G_{\Lambda_{\bf 4}^{14}\Lambda_{\bf 20^*}^2} = c(\tau,\bar\tau,N)
\int \frac{d^5z}{z_0^5} \, \veps_{(4)} t_{(16)} \,
\prod_{i=1}^{14} K_{7/2}^F(z;x_i)
\prod_{j=1}^2 K_{9/2}^F(z;y_j) \, ,
\label{Lam20-2-ads}
\ee
where $K_{7/2}^F$ is given in (\ref{prop-7/2}) and
$K_{9/2}^F$ is the bulk-to-boundary propagator for the dilatini in the
first Kaluza--Klein excited level, whose AdS mass is
$-\frac{5}{2}L^{-1}$, corresponding to an operator of dimension 9/2,
\ba
K^F_{9/2}(z,z_0;x) &=& K_{5}(z_m,z_0;x_m) \frac{1}{\sqrt{z_0}}
\left( z_0 \gamma_{\hat5} + (z-x)^{m}\gamma_{\hat m}\right)
\nn \\
&=& \frac{z_0^5}{[(z-x)^2+z_0^2]^5} \frac{1}{\sqrt{z_0}}
\left( z_0 \gamma_{\hat5} + (z-x)^{m}\gamma_{\hat m}\right) \, .
\label{prop-9/2}
\ea
In (\ref{Lam20-2-ads}) an overall constant, independent of both
$\tau$ and $N$ and including the normalisation factors for the
external fields, has been reabsorbed in $c(\tau,\bar\tau,N)$.
Although the coefficient of the $\Lambda^{16}$ coupling is known, the
exact normalisation of the supergravity amplitude requires the
detailed Kaluza--Klein reduction on the five-sphere which we
have not carried out.

The AdS result of (\ref{Lam20-2-ads}) is in exact agreement
with the Yang--Mills instanton calculation at leading order,
$N^{1/2}$, after the fermionic integrations over $\eta$ and $\bar\xi$
in (\ref{Lam20-2-a3}) have been performed. As usual to
compare the two results the coordinates in $AdS_5$, $(z_m,z_0)$, are
identified with the position and size of the instanton. In conclusion
for the class of non-minimal correlation functions involving operators
associated with Kaluza--Klein excited states in supergravity the
comparison works exactly as in the case of minimal correlators. The
angular integration over the five-sphere selects allowed processes and
then the resulting five-dimensional integrals can be matched as in the
minimal cases.

As observed at the end of the previous subsection the AdS amplitude
does not contain a term with the structure of the second one in
(\ref{Lam20-2-b3}). From the discussion of the correlation
function $\la \hat\Lambda^{16}\calQ^2_2 \ra$ in section
\ref{nonmincorr}, we would expect such a term to arise from a tree level
AdS diagram with a D-instanton induced vertex and an additional
classical supergravity interaction. It is, however, easy to convince
oneself that in this case a process of this type is not allowed by the
$U(1)$ symmetry of type IIB supergravity. Independently of what the
specific D-instanton vertex is, it is not possible to have an
additional cubic coupling involving two dilatini, which could generate
an amplitude with required structure. This is because the dilatino has
charge 3/2 and thus a coupling of the form $\Lambda\Lambda\Phi$ is not 
allowed because it can not be $U(1)$ neutral for any $\Phi$, since
there is no type IIB supergravity field with charge greater than 2.

The analysis of the example examined in this section can be
extended to other cases. In general the calculation of the AdS
amplitudes is straightforward and produces a leading contribution of
the same order in $\gs$ and $N$ as the corresponding minimal
amplitudes. On the Yang--Mills side the same powers of the coupling
and $N$ result from the combination of the normalisation factor in the
definition of the operators  (\ref{cpoell}), the factors of $\gy$ and
$N$ in the instanton measure and the factors of $\gy\sqrt{N}$
associated with each $\nsix$ insertion. Based on the general analysis
of section \ref{ginvarcorr} and the structure of the Yang--Mills
multiplets discussed in section \ref{oneinsttwo}, we conclude that all
the correlation functions of the type considered in this section have
the same dependence on the complex coupling $\tau$ and $N$ as the
minimal correlators studied in \cite{bgkr,doreya}.

\section{Other correlation functions in the one-instanton sector}
\label{othercorr}

We will now give an overview of the structure of some other
non-minimal correlation functions in which the effects analysed in the
previous two sections play an essential r\^ole.  In particular in the
examples that we consider in this section tree level processes in
$AdS_5\times S^5$ in which one of the vertices is a D-instanton
induced interaction will have crucial effects. We have already seen in
section \ref{nonmincorr} that ten-dimensional amplitudes of this type
involving extremal couplings give rise to contact terms that appear as
new effective interactions in the $AdS_5\times S^5$ background
starting at order $\al^{-1}$. The analysis of this section will suggest
that many new vertices should be expected to be appear via this
mechanism when reducing the flat ten-dimensional effective action on
$AdS_5\times S^5$.  Since the main features that enter into the
evaluation of the correlation functions studied here have already been
described in earlier sections the discussion will be briefer.

\subsection{Supercovariance}
\label{susycompl}

We saw in (\ref{ints}) that there are interactions in the IIB
effective action at order $(\al)^{-1}$ that go beyond those studied in
\cite{bgkr,doreya}. A class of such terms are those that arise from
the fact that the generalised field strengths, $\hat G$ and $\hat P$,
contain fermion bilinears, which are required to ensure that they
transform supercovariantly, \ie that their supersymmetry
transformations do not involve derivatives of the supersymmetry
parameters. For example, \cite{schw,gs},
\be
\hat{G}_{\mu\nu\rho} = G_{\mu\nu\rho}-
3{\bar\psi}_{[\mu}\g_{\nu\rho]}\Lambda -
6i{\bar\psi}^*_{[\mu}\g_\nu\psi_{\rho]} \, .
\label{Ghat}
\ee
The particular interaction that will be considered here is
\be
(\al)^{-1} \int d^{10}x \,
\sqrt{-g}\, \er^{-\phi/2}f_1^{(11,-11)}(\tau,\bar\tau) \,
\hat{G}\Lambda^{14} \, ,
\label{glamb14}
\ee
which  was also studied  in detail in \cite{gs}.  There, the
presence of the maximal number of fermions, arising from the
${\bar\psi}^*_{[\mu}\g_\nu\psi_{\rho]}$ factor in $\hat G$,
was seen to be important in determining the exact modular function
$f_1^{(11,-11)}(\tau,\bar\tau)$ in (\ref{glamb14}).

The piece of (\ref{glamb14}) involving the field strength
$G_{\mu\nu\rho}$ of (\ref{gdef}) gives rise to contact
diagrams in AdS generating amplitudes that are in correspondence with
correlation functions on the Yang--Mills side of the form
\ba
&&
\langle \hat\Lambda^{A_1}_{\a_1}(x_1) \ldots
\hat\Lambda^{A_{14}}_{\a_{14}}(x_{14}) \calE^{B_1C_1}(y)
\rangle \nn \\
&& \langle \hat\Lambda^{A_1}_{\a_1} (x_1) \ldots
\hat\Lambda^{A_{14}}_{\a_{14}}(x_{14})
\calB^{B_1C_1}_{m_1n_1}(y) \rangle \, ,
\label{Lam14Gdef}
\ea
where the scalar $\calE^{(AB)}$ and the antisymmetric tensor
$\calB^{[AB]}_{mn}$ are the operators dual to the complex combination
of the type IIB two-forms with indices respectively in $S^5$ and
$AdS_5$ directions. The correlation functions (\ref{Lam14Gdef}) are
minimal, \ie they involve only sixteen fermionic modes. Each
$\hat\Lambda^A_\a$ insertion soaks up one fermion superconformal mode
while each $\calE^{AB}$ or $\calB^{AB}_{mn}$ soaks up two, so that the
sixteen exact fermion modes are saturated. These correlation functions
agree with the supergravity results in a straightforward manner as in
the cases considered in \cite{bgkr,doreya}. However,  the interaction
(\ref{glamb14})  also gives rise to supergravity amplitudes in which
there are fourteen dilatinos and two gravitinos via the third term in
(\ref{Ghat}). If the two gravitinos have an internal vector index
these processes correspond to correlation functions of fourteen
$\hat\Lambda^A_\a$'s and two spinors $\calX^{B[CD]}_\b$'s (see
(\ref{chinewdef})) transforming in the ${\bf 20^*}$ of $SU(4)$
\be
\hat G_{\Lambda^{14}\calX^2}(x_1,x_2,\ldots,x_{14},y_1,y_2) =
\langle \hat\Lambda^{A_1}_{\a_1}(x_1) \ldots
\hat\Lambda^{A_{14}}_{\a_{14}}(x_{14})
\calX^{B_1[C_1D_1]}_{\b_1}(y_1) \calX^{B_2[C_2D_2]}_{\b_2}(y_2)
\rangle \, .
\label{Lam14chi2}
\ee
Notice that this process is allowed by the $SU(4)$ symmetry.
In the instanton background each $\hat\Lambda^{A_i}_{\a_i}$ soaks up
one superconformal mode as usual. Each $\calX^{B_i[C_iD_i]}_{\b_i}$
contains the two terms in (\ref{explChi}) and thus soaks up
three fermionic moduli. It immediately follows that the term in
(\ref{explChi}) cubic in the superconformal modes cannot contribute to
the correlation function (\ref{Lam14chi2}), since the total number of
superconformal modes would exceed sixteen.  There are, however, two
distinct kinds of possible contributions to (\ref{Lam14chi2}) that we
will denote by $\hat G^{\rm (a)}_{\hat\Lambda^{14}\calX^2}$ and $\hat
G^{\rm (b)}_{\hat\Lambda^{14}\calX^2}$,
\be
\hat G_{\hat\Lambda^{14}\calX^2}(x_1,\ldots,x_{14},y_1,y_2) =
\hat G^{\rm (a)}_{\hat\Lambda^{14}\calX^2}(x_1,\ldots,x_{14},y_1,y_2)
+ \hat G^{\rm (b)}_{\hat\Lambda^{14}\calX^2}(x_1,\ldots,x_{14},y_1,y_2)
\, .
\label{Lam14chi2dec}
\ee
$\hat G^{\rm (a)}$ is the contribution that is obtained by replacing all
the operators by their profiles in the instanton background. This
means that each $\calX$ provides one superconformal mode and a $\nten$
bilinear.  We will see that this term actually vanishes. The term
$\hat G^{\rm (b)}$ is a contribution that arises  by contracting the
$\v$'s in the two $\calX$'s with a propagator and using the remaining
$\lambda$'s,  together with the fourteen $\hat\Lambda$'s, to saturate the
fermionic integrations over the exact modes. Recall that there is no
possibility of contracting the $\lambda$'s with a propagator since
they all have the same chirality.  We shall now give the explicit
expressions for these two types of contributions that are depicted in
figure \ref{Lam14Chi2fig}.
\FIGURE[!h]{
{\includegraphics[width=0.45\textwidth]{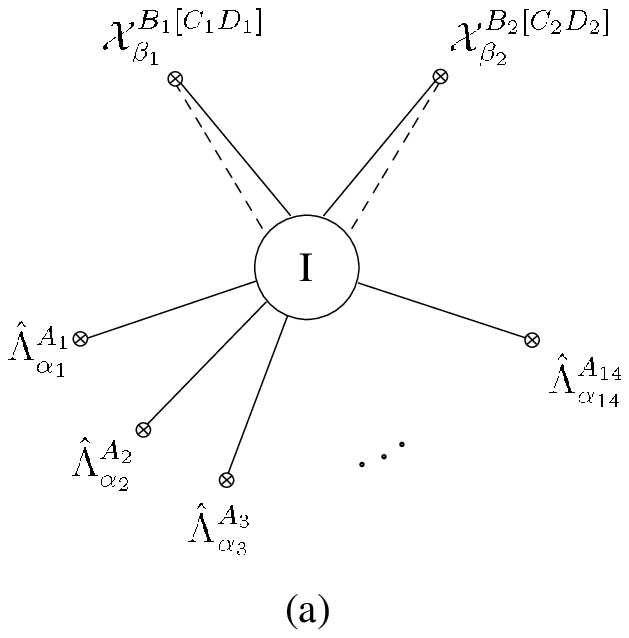}
\hspace*{0.6cm}
\includegraphics[width=0.45\textwidth]{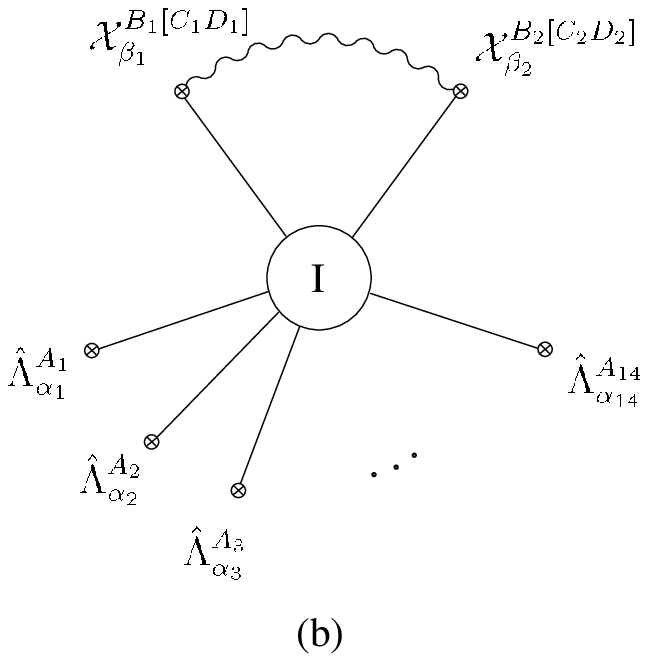}}
\caption{The two types of contributions to the correlation function
(\ref{Lam14chi2}). The notation is the same used in previous
  sections.}
\label{Lam14Chi2fig}
}

We first consider $G^{\rm (a)}_{\Lambda^{14}\calX^2}$, see figure
\ref{Lam14Chi2fig}(a), which using the formulae of appendix
\ref{appinsprofiles} is given by
\ba
\hat G^{\rm (a)}_{\hat\Lambda^{14}\calX^2} &\!=\!&
\int d\mu_{\rm phys}\, \er^{-S_{4F}}
\left[ \hat\Lambda^{A_1}_{\a_1}(x_1)
\ldots \hat\Lambda^{A_{14}}_{\a_{14}}(x_{14})
\calX^{B_1[C_1D_1]}_{\b_1}(y_1)
\calX^{B_2[C_2D_2]}_{\b_2}(y_2) \right]
\label{Lam14Chi2-a} \\
&\!=\!& \frac{1}{\gy^{32}}
\int d\mu_{\rm phys}\, \er^{-S_{4F}}
\prod_{i=1}^{14} \frac{\rho^4}{[(x_i-x_0)^2+\rho^2]^4}
\zeta^{A_i}_{\a_i} \nn \\
&& \hspace*{3.35cm} \prod_{j=1}^2 \left(
\frac{\rho^2}{[(y_j-x_0)^2+\rho^2]^3}\zeta^{D_j}
\bar\nu^{(B_j}\nu^{C_j)} - (C_j \leftrightarrow D_j)
\right) \, . \nn
\ea
It can easily be argued that this term vanishes by the following
symmetry argument. The whole expression must be a $SU(4)$ singlet. The
integration over the $\bar\nu$ and $\nu$ modes factorises and
therefore the indices carried by these variables must be combined to
form a $SU(4)$ singlet separately. This is, however, not possible
because (\ref{Lam14Chi2-a}) only contains $\nten$ bilinears and the
product ${\bf 10}\otimes{\bf 10}={\bf 20^\prime}_s\oplus{\bf 35}_s
\oplus{\bf 45}_a$ does not contain the singlet. Hence the whole
expression vanishes.

Now consider the contribution $G^{\rm (b)}_{\Lambda^{14}\calX^2}$
corresponding to the process in figure \ref{Lam14Chi2fig}(b).
We want to contract two $\v$'s in the $\calX$ operators with a
propagator instead of replacing them with their instanton profiles.
In other words, we want to consider
\ba
\hat G^{\rm (b)}_{\hat\Lambda^{14}\calX^2} &\!=\!&
\frac{1}{\gy^{32}} \langle \Tr\left(F_{m_1n_1}
\sigma^{m_1n_1}_{\:\a_1}{}^{\g_1} \lambda^{A_1}_{\g_1}
\right)(x_1) \ldots \Tr\left(F_{m_{14}n_{14}}
\sigma^{m_{14}n_{14}}_{\:\a_{14}}{}^{\g_{14}}
\lambda^{A_{14}}_{\g_{14}} \right)(x_{14}) \nn \\
&&\left[ 4 \Tr\left(\lambda^{B_1}_{\b_1}\v^{C_1D_1}\right)(y_1) \,
\Tr\left(\lambda^{B_2}_{\b_2}\v^{C_2D_2}\right)(y_2)
+\cdots \right] \rangle \; ,
\raisebox{-10pt}{\hspace*{-6.9cm}
\rule{0.6pt}{4pt}\rule{3.33cm}{0.6pt}\rule{0.6pt}{4pt}}
\label{Lam14Chi2wick}
\ea
where the dots in the last line stand for eight other terms with
the same structure obtained from the expansion of the product of the
two $\calX$'s and we will omit them in the next equations. Expanding
the last equation and using (\ref{propfin}) for the scalar propagator
we find
\ba
\hat G^{\rm (b)}_{\hat\Lambda^{14}\calX^2}
&=& \frac{1}{\gy^{32}} \langle
\Tr\left(F_{m_1n_1}\sigma^{m_1n_1}_{\a_1}{}^{\g_1}
\lambda^{A_1}_{\g_1} \right)(x_1) \ldots
\Tr\left(F_{m_{14}n_{14}}\sigma^{m_{14}n_{14}}_{\a_{14}}{}^{\g_{14}}
\lambda^{A_{14}}_{\g_{14}} \right)(x_{14}) \nn \\
&& \frac{\gy^2\veps^{C_1D_1C_2D_2}}{4\pi^2}\left[
\frac{2}{(y_1-y_2)^2} \Tr\left(\calP(y_2)\lambda^{B_1}_{\b_1}(y_1)
\calP(y_1)\lambda^{B_2}_{\b_2}(y_2)\right) \right. \nn \\
&& \left. + \frac{1}{\rho^2} \Tr\left(\calP(y_1)b\bar b
\lambda^{B_1}_{\b_1}(y_1) \right) \, \Tr\left(\calP(y_2)b\bar b
\lambda^{B_1}_{\b_1} (y_2) \right) \right] \ra \, .
\label{Lam14Chi2-b1}
\ea
Notice that the terms proportional to $1/N$ and $1/N^2$ in the scalar
propagator (\ref{propfin}) do not contribute here because of the
tracelessness of the elementary fields.
Using the explicit expressions for the instanton profiles of the
fields and for the ADHM matrices $\calP$, $b$ and $\bar b$
given in the appendices we find
\be
\Tr\left[\calP(y)b\bar b \lambda^{B_1}_{\b_1}(y) \right] = 0 \, ,
\label{vanishcontact}
\ee
so that the contact term that would arise from the last line in
(\ref{Lam14Chi2-b1}) vanishes. The evaluation of the term in the
second line is rather lengthy, but straightforward. It yields two
types of contributions. After combining the results of the various
contractions and reintroducing the numerical factors from the measure
we find
\ba
&& \hat G^{\rm (b)}_{\hat\Lambda^{14}\calX^2} =
c(N)\frac{2^{46}3^{16}\veps^{C_1D_1C_2D_2} \er^{2\pi i\tau}}
{\pi^{31/2} \gy^{22}} \int \frac{d\rho \, d^4x_0}{\rho^5}
\prod_{A=1}^4 d^2\eta^A \, d^2 {\bar\xi}^A
\prod_{i=1}^{14} \frac{\rho^4}{[(x_i-x_0)^2+\rho^2]^4}
\zeta^{A_i}_{\a_i} \nn \\
&& \left[ \prod_{j=1}^2 \left( \frac{2}{(y_1-y_2)^2}
\frac{\rho^2}{[(y_j-x_0)^2+\rho^2]^2}\zeta^{B_j}_{\b_j} +
\frac{\rho^3}{[(y_j-x_0)^2+\rho^2]^3}\zeta^{B_j}_{\b_j}
\right) \right] + \cdots \, ,
\label{Lam14Chi2-bfin}
\ea
where
\be
c(N) =  N^{1/2}\left(1 - \frac{5}{8N} + O(1/N^2) \right) \, .
\label{NdepLam14Chi2}
\ee
Therefore the correlation function $\hat G_{\hat\Lambda^{14}\calX^2}$
has a leading contribution of order $N^{1/2}$ as expected, since it is
related to AdS amplitudes involving interactions of order
$(\al)^{-1}$. The spatial dependence in (\ref{Lam14Chi2-bfin}) is,
however, different from that found in minimal correlators. In all the
cases studied in \cite{bgkr,doreya} the correlation functions took the
form of contact contributions. Here there is a term of this type (the
last one in $\hat G_{\hat\Lambda^{14}\calX^2}^{\rm (b)}$) as well as a
term with a different structure with a factorised $(y_1-y_2)^{-2}$.

Now we will show how the above result agrees with the result of the
calculation of the corresponding AdS amplitude. On the supergravity
side there are two types of contributions as well. There is an obvious
contribution from a diagram involving the D-instanton induced part of
the coupling arising from the ten-dimensional vertex
\be
\frac{1}{\al} \int d^{10}x \sqrt{-g} \, \er^{-\phi/2}
f_1^{(11,-11)}(\tau,\bar\tau) \Lambda^{14} {\bar\psi}^*\g \psi \, .
\label{Lam14Chi2-coupl}
\ee
Taking the vector indices of the gravitini in $S^5$ directions
produces a five-dimensional coupling that generates a contact diagram
contributing to the amplitude dual to the correlation function we are
considering, see figure \ref{Lam14Chi2-AdS-fig}(a).
\FIGURE[!h]{
{\includegraphics[width=0.45\textwidth]{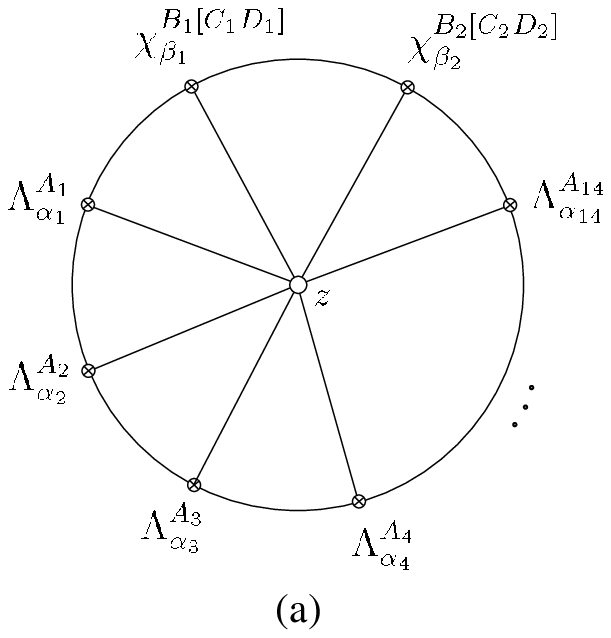}
\hspace*{0.6cm}
\includegraphics[width=0.45\textwidth]{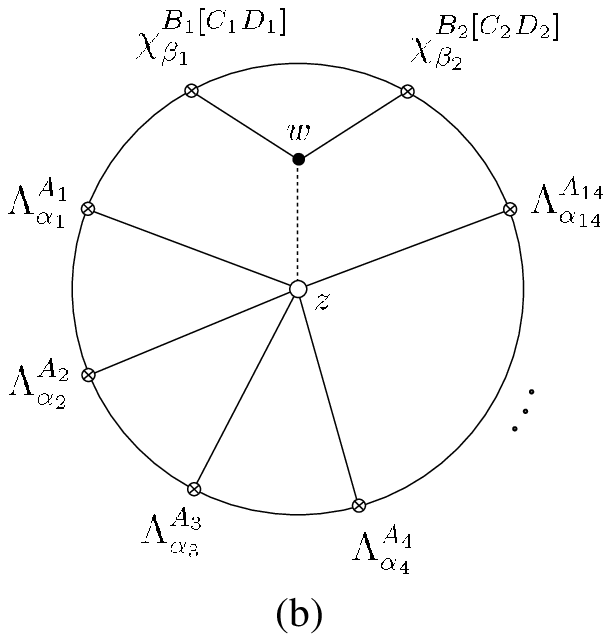}}
\caption{The two AdS amplitudes contributing to
  $G_{\Lambda^{14}\chi^2}$.}
\label{Lam14Chi2-AdS-fig}
}

The resulting amplitude is straightforward to evaluate and yields
\be
G^{\rm (a)}_{\Lambda^{14}\chi^2} = c_{\rm a}(\tau,\bar\tau,N)
\int \frac{d^5z}{z_0^5} \prod_{i=1}^{14}
K_{7/2}^F(z;x_i) \prod_{j=1}^2 K_{5/2}^F(z;y_j) \, ,
\label{Lam14Chi2-AdS1}
\ee
where, at leading order, the coefficient $c(\tau,\bar\tau,N)$ is
\be
c_{\rm a}(\tau,\bar\tau,N) = c_{\rm a} N^{1/2} \gs^{-11}
\er^{2\pi i \tau} + O(N^{-1/2}) \, .
\label{Lam14Chi2-Ads-coeff1}
\ee
In (\ref{Lam14Chi2-AdS1}) the usual notation for bulk-to-boundary
propagators has been used: $K_{7/2}^F$ was defined in
(\ref{prop-7/2}), whereas the two gravitini have been replaced by
\ba
K^F_{5/2}(z_m,z_0;x_m) &=& K_{3}(z_m,z_0;x_m) \frac{1}{\sqrt{z_0}}
\left( z_0 \gamma_{\hat5} + (z-x)^{m}\gamma_{\hat m}\right)
\nn \\
&=& \frac{z_0^3}{[(z-x)^2+z_0^2]^3} \frac{1}{\sqrt{z_0}}
\left( z_0 \gamma_{\hat5} + (z-x)^{m}\gamma_{\hat m}\right) \, .
\label{prop-5/2}
\ea

There is a second AdS amplitude that contributes to the same
process. This is an exchange amplitude of the type of those considered
in section \ref{L16Q2sugra}. The process we have to consider here
involves a D-instanton induced vertex
$f_1^{(11,-11)}(\tau,\bar\tau)\Lambda^{14}G$ with the $G$ splitting
into two $\psi$'s via a classical cubic interaction $\bar G\psi\psi$
where the spatial indices are in internal directions. This amplitude
is represented in figure \ref{Lam14Chi2-AdS-fig}(b). The dotted line
is a bulk-to-bulk propagator for a scalar associated with internal
components of $G$.  Analysing the spectrum of scalars in
\cite{kimromvan} we see that the allowed states are the Kaluza--Klein
ground state which is in the ${\bf 10}$ of $SU(4)$ and the second
excited state, in the ${\bf 50}$.  The corresponding couplings are
superconformal descendants of a next-to-extremal and an extremal
coupling respectively. Although the relevant integrals, involving
fermions, are not among the cases studied in the literature we expect
the extremal diagram to reduce to a contribution of the form of
(\ref{Lam14Chi2-AdS1})
\be
G^{\rm (b1)}_{\Lambda^{14}\chi^2} = c_{\rm b1}(\tau,\bar\tau,N)
\int \frac{d^5z}{z_0^5} \prod_{i=1}^{14}
K_{7/2}^F(z;x_i) \prod_{j=1}^2 K_{5/2}^F(z;y_j) \, ,
\label{Lam14Chi2-Adsb1}
\ee
where
\be
c_{\rm b1}(\tau,\bar\tau,N) = c_{\rm b1} N^{1/2} \gs^{-11}
\er^{2\pi i\tau} + O(N^{-1/2}) \, .
\label{coeff-Lam14Chi2-AdSb1}
\ee
Analogously the next-to-extremal diagram is expected to give
\be
G^{\rm (b2)}_{\Lambda^{14}\chi^2} =
\frac{c_{\rm b2}(\tau,\bar\tau,N)}{(y_1-y_2)^2}
\int \frac{d^5z}{z_0^5}
\prod_{i=1}^{14} K_{7/2}^F(z;x_i)
\prod_{j=1}^2 K_{3/2}^F(z;y_j) \, ,
\label{Lam14Chi2-Adsb2}
\ee
where
\ba
K^F_{3/2}(z_m,z_0;x_m) &=& K_{2}(z_m,z_0;x_m) \frac{1}{\sqrt{z_0}}
\left( z_0 \gamma_{\hat5} + (z-x)^{m}\gamma_{\hat m}\right)
\nn \\
&=& \frac{z_0^2}{[(z-x)^2+z_0^2]^2} \frac{1}{\sqrt{z_0}}
\left( z_0 \gamma_{\hat5} + (z-x)^{m}\gamma_{\hat m}\right)
\label{prop-3/2}
\ea
and
\be
c_{\rm b2}(\tau,\bar\tau,N) = c_{\rm b2} N^{1/2} \gs^{-11}
\er^{2\pi i\tau} + O(N^{-1/2}) \, .
\label{coeff-Lam14Chi2-AdSb2}
\ee
Let us now compare the AdS result with what we obtained from the
Yang--Mills calculation. The two terms (\ref{Lam14Chi2-AdS1}) and
(\ref{Lam14Chi2-Adsb1}) have the structure of contact contributions.
They reproduce the last term in the Yang--Mills result of
(\ref{Lam14Chi2-bfin}). From a five-dimensional point of view these
two contributions would not be distinguished. Proceeding as in
perturbative calculations (for example, as in  the analysis in
\cite{nearextremal}) the process in figure \ref{Lam14Chi2-AdS-fig}(b)
would not be included, but its effect would result from the correction
it induces in the five-dimensional $\Lambda^{14} \hat G$ coupling.
The AdS amplitude involving the next-to-extremal interaction
(\ref{Lam14Chi2-Adsb2})  agrees exactly with the remaining term in
(\ref{Lam14Chi2-bfin}).  The $SU(4)$ tensor in the Yang--Mills result
is obtained from the integration over the five-sphere analogously to
the cases analysed in previous sections. In conclusion we find
agreement at order $N^{1/2}$ between the two calculations. This
example illustrates how the different effects described in this paper
are relevant even for processes arising at the level of the leading
instanton corrections.

\subsection{Other higher derivative interactions in type IIB string
theory}
\label{nminhalf}

In section \ref{nonmincorr} we discussed the details of how the
instanton contribution to a  correlation function with sixteen
$\hat\Lambda$'s and two $\calQ_2$'s can be computed in the
semiclassical approximation.  This was related via the AdS/CFT
correspondence to higher derivative interactions in string theory at
order  $\al^{-1}$ and $\al$.  We will now consider other examples
of this kind that involve the correlation functions
\be
\hat G_{\hat\Lambda^{16}\calE^4}(x_1,\ldots,x_{16},y_1,\ldots,y_4)
\!=\! \langle \hat\Lambda^{A_1}_{\a_1}(x_1) \ldots
\hat\Lambda^{A_{16}}_{\a_{16}}(x_{16}) \calE^{B_1C_1}(y_1)
\ldots \calE^{B_4C_4}(y_4) \rangle
\label{Lam16E4}
\ee
and
\be
\hat G_{\hat\Lambda^{16}\calB^4}(x_1,\ldots,x_{16},y_1,\ldots,y_4)
\!=\! \langle \hat\Lambda^{A_1}_{\a_1}(x_1) \ldots
\hat\Lambda^{A_{16}}_{\a_{16}}(x_{16}) \calB^{B_1C_1}_{m_1n_1}(y_1)
\ldots \calB^{B_4C_4}_{m_4n_4}(y_4) \rangle .
\label{Lam16B4}
\ee
As observed after (\ref{Lam14Gdef}), in the AdS/CFT
correspondence $\calE^{(AB)}$ and $\calB^{[AB]}_{mn}$  are the
composite operators in the supercurrent multiplet associated with the
internal and space-time components of the complex combination of the
\NSNS\ and \RR\ two-forms in supergravity. A naive generalisation of
the analysis of \cite{bgkr,doreya} would associate the above
correlation functions with  terms in the type IIB effective action of
the form  $\al\, G^4\, \Lambda^{16}$, which were discussed in
(\ref{actfiv}) in section \ref{overii}. This specific interaction has
been recently studied in detail in \cite{aninda}.  As before, the
reason for considering the particular correlation functions in
(\ref{Lam16E4}) and (\ref{Lam16B4}) is that the analysis is simpler
for the terms with the maximal number of fermions.

In the semiclassical approximation the classical solutions of section
\ref{nusupercurr} are used to saturate the integrations over the
fermion zero-modes. Each operator $\hat\Lambda^A_\a$ soaks up one
superconformal mode and does not depend on the additional modes
$\bar\nu$ and $\nu$.  The operators $\calE^{AB}$ and $\calB^{AB}_{mn}$
each soak up two zero modes, but in the case of $\calB^{AB}_{mn}$
these are necessarily superconformal modes as follows from the
classical expression (\ref{explB}).  Thus there is a non-vanishing
contribution to the correlation function (\ref{Lam16E4}) in which the
sixteen exact fermion modes come from the $\hat\Lambda$'s and each $\calE$
gives a bilinear $\nten$ in the $\nu$'s, but no similar contribution
can exist for (\ref{Lam16B4}).  There is, however, a different type of
contribution to (\ref{Lam16B4}) which is non-zero and comes from
considering the  $\Tr(\v^{AB}F_{mn})$ term in the operator
$\calB^{AB}_{mn}$ (\ref{Bnewdef}). The new contribution is obtained
by contracting a scalar field, $\v$, in one of the $\calB^{AB}_{mn}$
operators with a propagator joining it to another $\calB^{AB}_{mn}$
operator. The remaining $F_{mn}$'s in each $\calB$ are substituted
by their non-vanishing value in the instanton background. As in
previous examples only scalar contractions are possible because of the
chirality of the fermions and the self-duality of the field
strength. This means that there is no analogous contribution involving
propagators to the correlation function (\ref{Lam16E4}) with $\calE$
insertions at leading order in the coupling $\gy$.

\FIGURE[!h]{
{\includegraphics[width=0.1\textwidth]{x.eps}
\includegraphics[width=0.45\textwidth]{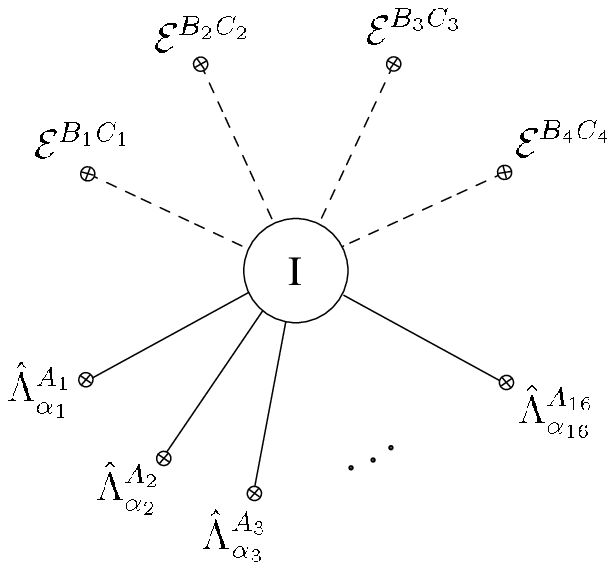}
\includegraphics[width=0.1\textwidth]{x.eps}}
\caption{Instanton contribution to the correlation function
(\ref{Lam16E4}).}
\label{Lam16E4figYM}
}

First consider the `purely instantonic' process which gives rise to
(\ref{Lam16E4}). This is represented diagrammatically in figure
\ref{Lam16E4figYM}.  In the semiclassical approximation the result is
\ba
&& \hat G_{\hat\Lambda^{16}\calE^4} = \int d\mu_{\rm phys}\,
\er^{-S_{4F}}  \left[\hat\Lambda^{A_1}_{\a_1}(x_1) \ldots
\hat\Lambda^{A_{16}}_{\a_{16}}(x_{16}) \, \calE^{B_1C_1}(y_1)\ldots
\calE^{B_2C_2}(y_4) \right] \nn \\
&& = \frac{1}{\gy^{40}}\int d\mu_{\rm phys}\,
\er^{-S_{4F}} \left[\prod_{i=1}^{16}  \frac{\rho^4
\zeta^{A_i}_{\a_i}}{[(x_i-x_0)^2+\rho^2]^4}  \prod_{j=1}^4
\frac{\rho^2({\bar\nu}^{B_ju}\nu^{C_j}_{u} +
{\bar\nu}^{C_ju}\nu^{B_j}_{u})}{[(y_j-x_0)^2+\rho^2]^3} \right] \, ,
\label{Lam16E4scl}
\ea
where we have substituted the classical expressions for the
operators $\hat\Lambda$ and $\calE$ which are given by
(\ref{explLamb}) and (\ref{explE}), respectively. The combinatorics
necessary to evaluate the $2(N-2)$ fermionic integrations over
$\bar\nu^{Au}$ and $\nu^A_u$ is simplified with the aid of the
generating function defined earlier in section \ref{ginvarcorr}.
After some algebra the result is
\ba
\hat G_{\hat\Lambda^{16}\calE^4}
&\!=\!& c(N) t^{B_1C_1\ldots B_4C_4} \frac{2^{40}3^{16}\er^{2\pi i\tau}}
{\pi^{35/2}\gy^{28}} \int \frac{d\rho \, d^4x_0}{\rho^5}
\prod_{A=1}^4 d^2\eta^A \, d^2 {\bar\xi}^A \: \nn \\
&& \left[ \prod_{i=1}^{16}
\frac{\rho^4}{[(x_i-x_0)^2+\rho^2]^4}\zeta^{A_i}_{\a_i}
\prod_{j=1}^4 \frac{\rho^3}{[(y_j-x_0)^2+\rho^2]^3} \right] \, ,
\label{Lam16E4fin}
\ea
where
\be
c(N) = N^{1/2} \left(1 - \frac{25}{8N} + O(1/N^2) \right)
\label{NdepL16E4}
\ee
and the tensor $t^{B_1C_1\ldots B_4C_4}$, determined by the
five-sphere integration, takes the form
\ba
t^{B_1C_1\ldots B_4C_4} &=& \left(\veps^{B_1B_2B_3B_4}
\veps^{C_1C_2C_3C_4} + \veps^{B_1C_2C_3C_4}\veps^{C_1B_2B_3B_4}
\right. \nn \\
&& \left. + ~ {\rm 6~ terms~ obtained~ by~ symmetrisation~ in~}
(B_i,C_i)~ \right) \, .
\label{Lam16E4tens}
\ea

We would like to compare this expression with the corresponding
supergravity process.  With the usual identification,
$\left((x_0)_m,\rho\right)\longleftrightarrow(z_m,z_0)$, and recalling
(\ref{fKrel}), we notice that the integrand in (\ref{Lam16E4fin})
contains sixteen factors that correspond to the dilatini
bulk-to-boundary propagators and four factors of the bulk-to-boundary
propagator $K_3$, which is the propagator for a scalar field of
dimension $\Delta=3$. The supergravity scalar field, $E^{AB}$,  that
couples to $\calE^{AB}$ in (\ref{Lam16E4fin}) comes from  the
second-rank potential, $B_{\alpha\beta}$, which has indices in the
five-sphere directions. The leading term (of order $N^{1/2}$) in the
expression for $\hat G_{\hat\Lambda^{16}\calE^4}$ (\ref{Lam16E4fin})
therefore appears to  correspond to a  contact interaction  of the
form $(\al)^{-1}\, \Lambda^{16}\, E^4$.  Such a term does not arise
among the interactions of order $(\al)^{-1}$ in ten dimensions
(indeed, it is not even gauge invariant). Nevertheless an amplitude
with exactly the same structure emerges from an AdS tree level process
involving a $\Lambda^{16}$ D-instanton induced interaction. A
contribution of the form of (\ref{Lam16E4fin}) results from a tree
diagram in $AdS_5\times S^5$ with a D-instanton at one vertex. The
coupling we need to consider here is of the form
\be
\int d^{10}
\sqrt{-g} \er^{-\phi/2} f_1^{(12,-12)}(\tau,\bar\tau)  \tau^2\Lambda^{16}
\label{Lam16tau2vertex}
\ee
and it arises at order $\al^{-1}$ via the mechanism described in
section \ref{nonmincorr} from the expansion of the $\er^{2\pi i\tau}$
factor. To obtain the amplitude relevant to our example each dilaton
must couple to a pair of $G$'s via a $\bar\tau G^2$ vertex in the
classical IIB supergravity.  The amplitude is represented in figure
\ref{Lam16E4figAdS}(a). From the $AdS_5\times S^5$ point of view the
allowed processes are all those compatible with the   $SU(4)\sim
SO(6)$ symmetry.  Group theoretical restrictions imply that in this
particular process the two intermediate dilatons can only be in the
${\bf 20^\prime}$ of $SU(4)$,  \ie they are in the second Kaluza-Klein
excited state which will is denoted by $\tau_{20'}$.  This means that
the cubic $\bar \tau GG$ vertices involved are extremal and as in
previously described cases the process reduces to a
contact amplitude that contributes at order $N^{1/2}$.  This can be
seen  directly from the expression of the tree diagram with all the
indices on the antisymmetric tensor potential $B_{\mu\nu}$ aligned
along the five-sphere.  Indeed, making use of the equations of motion
in \cite{kimromvan} it is easy to see that, after two integrations by
parts, each of the cubic vertices can be written as the integral of
$(\partial^2+m_{\tau_{20'}}) \bar \tau_{20'} B_{ij}B^{ij}$, where the
mass term comes from derivatives in $S^5$ directions.  In this form,
it is manifest that such a  vertex cancels the adjacent $\la\tau
\bar\tau\ra_{20'}$ propagator.

This means that from a five-dimensional point of
view this process appears to produce a new coupling at order $\al^{-1}$
of the form
\be
\gs^{-2}\int \frac{d^5z}{z_0^5} \, \er^{-\phi/2}
f_1^{(12,-12)}(\tau,\bar\tau)\Lambda^{16}E^4 \, ,
\label{effLam16E4coupl}
\ee
where the factor of $\gs^{-2}$ is obtained after canonically
normalising the kinetic  term for the dilaton.  The interaction
(\ref{effLam16E4coupl}) is not manifestly gauge invariant since it
involves the field $E$ and not the field strength $G$. This is,
however,
not a problem because gauge invariance is recovered when
(\ref{effLam16E4coupl}) is combined with the other couplings, involving
the antisymmetric tensor $B_{mn}$ and the vectors resulting from mixed
components of the ten-dimensional two-form, which also descend from the
tree level process just described.

\FIGURE[!h]{
{\includegraphics[width=0.45\textwidth]{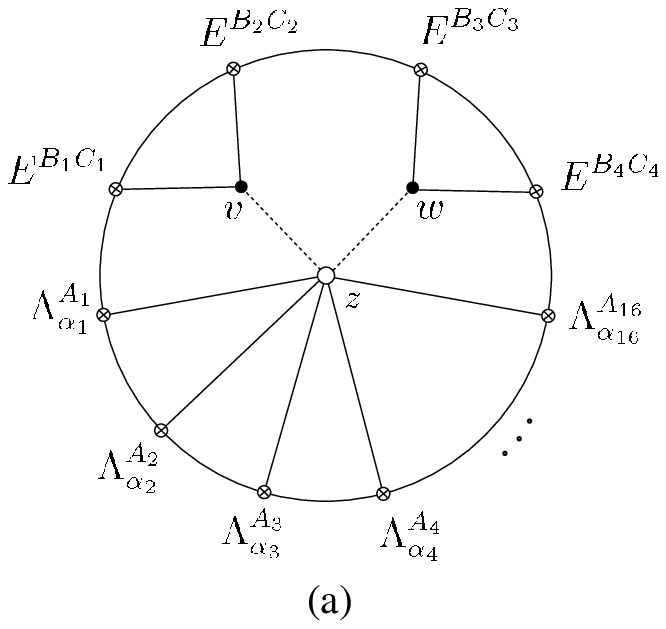}
\hspace*{0.6cm}
\includegraphics[width=0.45\textwidth]{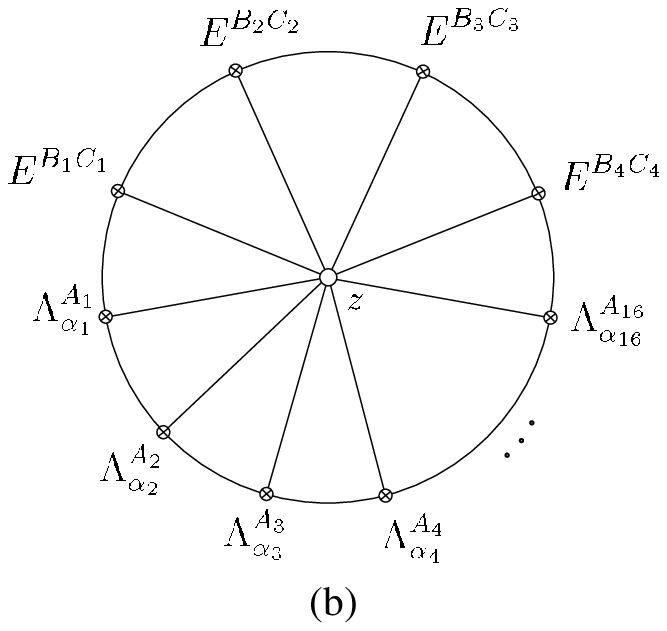}}
\caption{AdS amplitudes contributing to the process dual to the
correlation function (\ref{Lam16E4}).}
\label{Lam16E4figAdS}
}

The same amplitude gets a contribution from the contact diagram
generated by the $G^4\Lambda^{16}$ interaction that is expected to
contribute at order $\al$. This is a higher order term, $\al$ with
respect to the classical Einstein--Hilbert term, in the derivative
expansion of the string effective action that was discussed in section
\ref{overii}.  Notice that in this comparison a crucial r\^ole is
played by the self-duality property of the complex type IIB two-form
with indices in internal directions \cite{kimromvan}, which implies
that the $AdS$ diagram involves $K_3$ bulk-to-boundary propagators for
the $G$'s. This second contribution is of order $N^{-1/2}$
corresponding to a vertex of order $\al$. The situation at order
$N^{-1/2}$ is actually more complicated. There is in fact another
source of corrections at this order. At the end of section
\ref{oneinst}  we commented on how the $1/N$ corrections to the
minimal correlation functions are reproduced in supergravity. In the
case of the correlator of sixteen dilatini for example the order
$N^{-1/2}$ effects are expected to arise from amplitudes involving the
vertices
\be
\al \int d^{10}x \sqrt{-g} \, \er^{\phi/2}
f_2^{(12,-12)}(\tau,\bar\tau) \Lambda^{16} \left( \calR^2 + F_{(5)}^4
\right) \, ,
\label{Lam16-1/N}
\ee
where the $\calR^2$ and $F_{(5)}^4$ factors are set to their
non-vanishing background values. Analogously here we have to consider
similar effects associated with (\ref{Lam16tau2vertex}), \ie tree
level processes in which the D-instanton vertex is
\be
\al \int d^{10}x \sqrt{-g} \, \er^{\phi/2} f_2^{(12,-12)}(\tau,\bar\tau)
\Lambda^{16} \tau^2 \left( \calR^2 + F_{(5)}^4 \right) \, ,
\label{Lam16tau2-1/N}
\ee
with $\calR^2$ and $F_{(5)}^4$ taken to be constant. As for the
leading $N^{1/2}$ term the resulting exchange diagrams contain
extremal couplings of the dilatons with pairs of $G$'s and should
produce again a contact contribution.  Determining the correct
coefficients of all the relevant terms is extremely subtle, so we will
not go beyond a qualitative analysis. In conclusion we find that all
the different effects just described give rise to the same structure
and the complete AdS result reads
\be
G_{\Lambda^{16}E^4} = c(\tau,\bar\tau,N)
\int \frac{d^5z}{z_0^5} \, \prod_{i=1}^{16} K^F_{7/2}(z;x_i)
\prod_{j=1}^4 K_3(z;y_j) \, ,
\label{Lam16E4AdSfin}
\ee
where
\be
c(\tau,\bar\tau,N) = \gs^{-14} \er^{2\pi i\tau}
\left( c_1 N^{1/2} + c_2 N^{-1/2} \right) \, .
\label{Lam16E4AdSNdep}
\ee
Notice that the common power of the string coupling has a different
origin in the two terms in (\ref{Lam16E4AdSNdep}). In the leading order
term and in the subleading correction produced by the amplitude
induced by (\ref{Lam16tau2-1/N}) the factor of $\gs^{-14}$ is the
combination of a $\tau_2^{12}$ from the modular form
$f_1^{(12,-12)}(\tau,\bar\tau)$ and two additional factors of $\gs$
following from the proper normalisation of the kinetic term for the
dilaton. In the subleading term generated by the $G^4\Lambda^{16}$
interaction  the correct powers of the coupling constant come from the
expansion of the modular form $f_2^{(14,-14)}(\tau,\bar\tau)$
defined in
 (\ref{asymterm}) and (\ref{flw-wdef}).  Comparing this
result with the Yang--Mills expression (\ref{Lam16E4fin}) we find
agreement up to numerical constants that we have not checked.

We now turn to the case of correlation functions with insertions of
$\calB^{[AB]}_{mn}$. A natural example to consider is (\ref{Lam16B4})
which naively would be expected to correspond to an AdS amplitude
generated by the interaction $G^4\Lambda^{16}$ at order $\al$ with
the indices on the $G$'s in AdS directions. In view of the previous
discussion concerning the case of the correlation function $\hat
G_{\hat\Lambda^{16}\calE^4}$, however, we can consider the simpler
case
\be
\hat G_{\hat\Lambda^{16}\calB^2} = \la \hat\Lambda^{A_1}_{\a_1}(x_1)
\ldots \hat\Lambda^{A_{16}}_{\a_{16}}(x_{16})
\calB^{B_1C_1}_{m_1n_1}(y_1) \calB^{B_2C_2}_{m_2n_2}(y_2) \ra \, .
\label{Lam16B2def}
\ee
On the Yang--Mills side it is natural to expect a non-vanishing result
for this correlation function since it is allowed by group theory: the
indices on the sixteen $\hat\Lambda$'s are combined to form a $SU(4)$
singlet and a singlet is also contained in the product of two ${\bf
6}$'s corresponding to the $\calB$ insertions. On the supergravity
side there is no candidate vertex in the type IIB ten dimensional
effective action in Minkowski space to generate an amplitude dual to
(\ref{Lam16B2def}).  Nevertheless we expect to generate a
$\Lambda^{16} B^2$ effective interaction in $AdS_5\times S^5$ at order
$\al^{-1}$ via the same mechanism discussed in the case of
(\ref{effLam16E4coupl}).

The correlation function (\ref{Lam16B2def}) is more complicated than
previous examples because it involves antisymmetric tensors which lead
to a non-trivial tensorial structure. The calculation is
straightforward but tedious and we will discuss the general features
of this correlator without deriving a complete explicit expression. The
Yang--Mills and AdS diagrams corresponding to (\ref{Lam16B2def}) are
depicted in figure \ref{Lam16B2fig}.
\FIGURE[!h]{
{\includegraphics[width=0.45\textwidth]{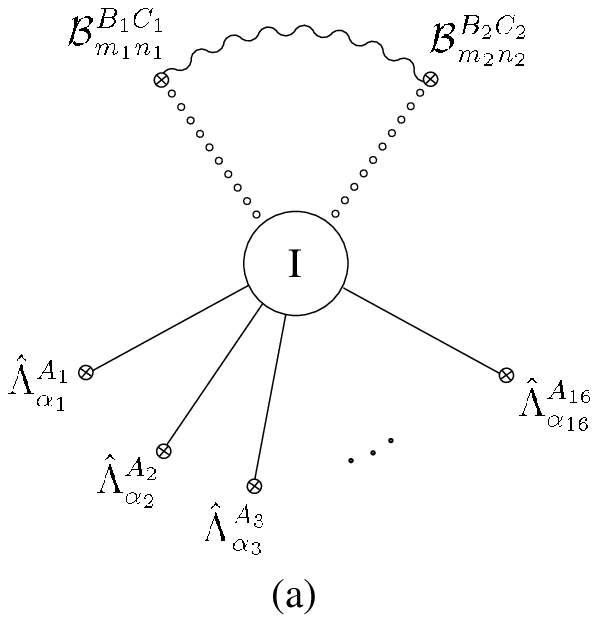}
\hspace*{0.6cm}
\includegraphics[width=0.45\textwidth]{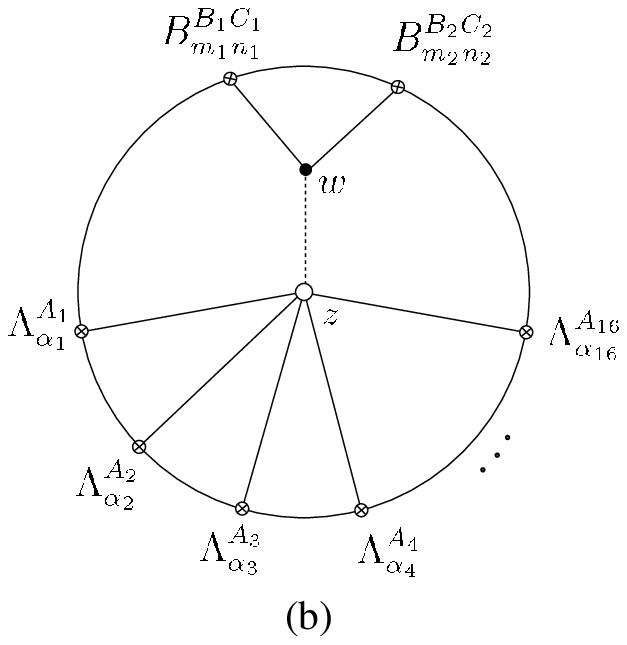}}
\caption{Diagram (a) represents the Yang--Mills contribution to the
correlation function (\ref{Lam16B2def}). The lines denoted by
small circles in (a) indicate insertions of $F_{m_in_i}$, which do
not contain fermionic modes, but have a non vanishing value in the
instanton background. Diagram (b) represents the AdS amplitude
contributing to the dual process.}
\label{Lam16B2fig}
}

Since the $\calB^{BC}_{mn}$ fields do not depend on the $\nu$ and
$\bar\nu$ modes, contractions between pairs of scalar fields in the
$\calB$ insertions must be considered to get a non-vanishing result.
This gives
\ba
&&
\hat G_{\hat\Lambda^{16}\calB^2}(x_1,\ldots,x_{16},y_1,y_2) =
\langle \hat\Lambda^{A_1}_{\a_1}(x_1) \ldots
\hat\Lambda^{A_{16}}_{\a_{16}}(x_{16}) \calB^{B_1C_1}_{m_1n_1}(y_1)
\ldots \calB^{B_4C_4}_{m_4n_4}(y_4) \rangle
\label{Lam16B2wick} \\
&& = \frac{1}{\gy^{36}} \langle
\Tr\left(F_{m_1n_1}\sigma^{m_1n_1}_{\a_1}{}^{\b_1}
\lambda^{A_1}_{\b_1} \right)(x_1) \ldots
\Tr\left(2i\v^{B_1C_1}F_{p_1q_1}\right)(y_1)
\Tr\left(2i\v^{B_2C_2}F_{p_2q_2}\right) (y_2)
\raisebox{-10pt}{\hspace*{-6.49cm}
\rule{0.6pt}{4pt}\rule{3.68cm}{0.6pt}\rule{0.6pt}{4pt}}
\hspace*{2.64cm} \ra \, .
\nn
\ea
Let us compute the contraction in (\ref{Lam16B2wick}) separately.
Using the scalar propagator (\ref{propfin}) we get
\ba
&& \Tr\left(2i\v^{B_1C_1}F_{p_1q_1}\right)(y_1)
\Tr\left(2i\v^{B_2C_2}F_{p_2q_2}\right) (y_2)
\raisebox{-10pt}{\hspace*{-6.5cm}
\rule{0.6pt}{4pt}\rule{3.68cm}{0.6pt}\rule{0.6pt}{4pt}} \nn \\
&& = \frac{\gy^2\veps^{B_1C_1B_2C_2}}{2\pi^2} \left[
\frac{1}{(y_1-y_2)^2} \Tr\left(\calP(x_2)F_{m_1n_1}(x_1)
\calP(x_1)F_{m_2n_2}(x_2) \right) \right. \nn \\
&& \left. + \frac{1}{2\rho^2} \Tr\left(\bar b
F_{m_1n_1}(x_1)\calP(x_1)b\right)
\Tr\left(\bar bF_{m_2n_2}(x_2)\calP(x_2)b\right)\right]
\label{Lam16B2-1contr} \, ,
\ea
where we have used the tracelessness of $F_{mn}$ to discard the
contribution of the $1/N$ and $1/N^2$ terms in the propagator.
It is easy to verify that the terms in the last line of
(\ref{Lam16B2-1contr}) vanish, so that for (\ref{Lam16B2wick}) we
obtain
\be
\hat G_{\hat\Lambda^{16}\calB^2} =
\frac{\gy^2\veps^{B_1C_1B_2C_2}}{2\pi^2 (y_1-y_2)^2} \,
\la \hat\Lambda^{A_1}_{\a_1}(x_1) \ldots
\hat\Lambda^{A_{16}}_{\a_{16}}(x_{16}) \Tr\left[\calP(y_2)
F_{m_1n_1}(y_1)\calP(y_1)F_{m_2n_2}(y_2) \right] \ra \, .
\label{Lam16B2YM-2}
\ee
Evaluating this expression in the instanton background leads to the
result
\ba
\hat G_{\hat\Lambda^{16}\calB^2} &=& c(N)
\frac{\veps^{B_1C_1B_2C_2} 2^{40}3^{16} \er^{2\pi i\tau}}
{\pi^{31/2}\gy^{26}}
\int \frac{d\rho \, d^4x_0}{\rho^5} \prod_{A=1}^4 d^2\eta^A \,
d^2 {\bar\xi}^A \prod_{i=1}^{16} \frac{\rho^4}{[(x_i-x_0)^2+\rho^2]^4}
\zeta^{A_i}_{\a_i} \nn \\
&& \left\{ \left[ \frac{1}{(y_1-y_2)^2}
\prod_{j=1}^2\frac{\rho^2}{[(y_j-x_0)^2+\rho^2]^2} 
\right. \right. \nn \\ 
&& \left. \left. 
-\prod_{j=1}^2\frac{\rho^3}{[(y_j-x_0)^2+\rho^2]^3}\right]
\tr(\s_{m_1n_1}\s_{m_2n_2}) + \cdots \right\} \, ,
\label{Lam16B2YMfin} 
\ea
\ie as in previous cases we obtain a term with the structure
of a contact contribution as well as a term with a factorised
$(y_1-y_2)^{-2}$. The $\cdots$ in (\ref{Lam16B2YMfin}) denotes terms
with different tensor structure. The $N$-dependence is contained in
the prefactor
\be
c(N) = N^{1/2} \left( 1 - \frac{5}{8N} + O(1/N^2) \right) \, .
\ee

The result can now be compared with the expectations from the
supergravity calculation. In order to understand the AdS side of the
correspondence for this process it is necessary to consider the
amplitude in figure \ref{Lam16B2fig}(b). This is  a tree diagram in
which the vertex in $z$ is a D-instanton induced  interaction coupling
sixteen dilatini to a dilaton. In this case the  dilaton  couples to a
pair of $B_{mn}$'s, with indices in AdS directions, which are in the
{\bf 6} of $SU(4)$. There are two kinds of allowed
contributions. These arise from the possibility of the dilaton being
in its massless Kaluza--Klein ground state, which is a singlet of
$SU(4)$ or in its second Kaluza--Klein state, which is a ${\bf
20^\prime}$ of $SU(4)$. The evaluation of the integrals entering these
amplitudes is complicated because of the presence of the antisymmetric
tensors. The vertices in the two cases (dilaton in the ${\bf 1}$ or
${\bf 20^\prime}$) are superconformal descendants of a
next-to-extremal and an extremal coupling respectively. Hence
we expect the amplitude in the case of singlet exchange to reduce to
\be
G_{\Lambda^{16}B^2}^{\rm (1)} =
\frac{c_1(\tau,\bar\tau,N)}{(y_1-y_2)^2} \int \frac{d^5z}{z_0^5}
\prod_{i=1}^{16} K_{7/2}^F(z;x_i) \prod_{j=1}^2 K_{2;m_jn_j}(z;y_j)
\, , \label{Lam16B2AdS1}
\ee
whereas for the ${\bf 20^\prime}$ exchange we expect
\be
G_{\Lambda^{16}B^2}^{\rm (2)} = c_2(\tau,\bar\tau,N)
\int \frac{d^5z}{z_0^5} \prod_{i=1}^{16} K_{7/2}^F(z;x_i)
\prod_{j=1}^2 K_{3;m_jn_j}(z;y_j)
\, . \label{Lam16B2AdS2}
\ee
In both cases the prefactors have a leading contribution of order
$N^{1/2}$ plus $1/N$ corrections
\be
c_i(\tau,\bar\tau,N) = c_i \gs^{-13} \er^{2\pi i\tau} \left(
N^{1/2} + O(N^{-1/2})\right) \, , \quad i=1,2 \, ,
\label{NdepLam16B2AdS}
\ee
where the powers of the string coupling come from the
$f_1^{(12,-12)}(\tau,\bar\tau)$ in the $\Lambda^{16}$ vertex and the
normalisation of the dilaton fluctuations. In these expressions
$K_{\Delta;mn}(z;x)$ denotes the bulk-to-boundary propagator for an
antisymmetric tensor dual to an operator of dimension $\Delta$
\cite{antisymmt}, which has AdS mass squared  $m^2L^2=\D^2-4\D+3$. The
product of the two $K_{\Delta;mn}$'s in (\ref{Lam16B2AdS1}) and
(\ref{Lam16B2AdS2}) contains several terms, in particular
\be
\prod_{j=1}^2 K_{\D;m_jn_j}(z;y_j) = \prod_{j=1}^2 K_{\D}(z;y_j)
\, \tr(\s_{m_1n_1}\s_{m_2n_2}) + \cdots \, ,
\label{Lam16B2AdStstruc}
\ee
so that the AdS result reproduces qualitatively  the correct terms in
the Yang--Mills calculation.

A comment can be added about the comparison with the five-dimensional
perspective. In a five-dimensional approach only the singlet exchange
diagram would be included, but the effective action would then contain
a genuine new interaction of the form
\be
\gs^{-1}\int \frac{d^5z}{z_0^5} \, \er^{-\phi/2}
f_1^{(12,-12)}(\tau,\bar\tau)\Lambda^{16}B^2 \, ,
\label{effLam16B2coupl}
\ee
producing the amplitude (\ref{Lam16B2AdS2}). Again, gauge invariance of
the complete effective action results from combining
(\ref{effLam16B2coupl}) with vertices containing vector fields.

The effects just described in the case of the correlator
(\ref{Lam16B2def}) are the building blocks for the evaluation of
(\ref{Lam16B4}). In this case the contact term at order $N^{-1/2}$ gets
also an additional contribution from a diagram with a
$\Lambda^{16}G^4$ vertex at order $\al$ with indices on the $G$'s
in $AdS_5$ directions.

More generally we can consider correlation functions of the form
\ba
&& \hat G_{\hat\Lambda^{16}\calE^{2k}} = \la \hat\Lambda^{A_1}_{\a_1}
(x_1)\ldots \hat\Lambda^{A_{16}}_{\a_{16}}(x_{16}) \,
\calE^{(B_1C_1)}(y_1) \ldots \calE^{(B_{2k}C_{2k})}(y_{2k}) \ra
\label{Lam16E2k} \\
&& \hat G_{\hat\Lambda^{16}\calB^{2k}} = \la \hat\Lambda^{A_1}_{\a_1}
(x_1)\ldots \hat\Lambda^{A_{16}}_{\a_{16}}(x_{16}) \,
\calB_{m_1n_1}^{[B_1C_1]}(y_1) \ldots
\calB_{m_{2k}n_{2k}}^{[B_{2k}C_{2k}]}(y_{2k}) \ra \, .
\label{Lam16B2k}
\ea
These correlation functions are naturally associated (for even $k$)
with a class of terms in the type IIB effective action which were
conjectured in \cite{berkovafa}. In that paper terms of the form
$\calR^4 G^{4g-4}$ were conjectured to arise at order $(\al)^{2g-3}$
in flat space.  On the other hand a generalisation of the Yang--Mills
calculations of this section leads us to expect all the non-vanishing
correlation functions of the form (\ref{Lam16E2k}) and
(\ref{Lam16B2k}) to have a leading contribution of order $N^{1/2}$,
\ie $(\al)^{-1}$ in terms of string parameters. It is particularly
easy to convince oneself that this is the case for (\ref{Lam16E2k})
using the general formulae of section \ref{ginvarcorr}. We have seen
that each $\calE$ contains only a bilinear $\nten$, so that
(\ref{Lam16E2k}) involves $2k$ $\nten$ and no $\nsix$
insertions. Using the general formula (\ref{genernm}) we then conclude
that
\be
\hat G_{\hat\Lambda^{16}\calE^{2k}} \sim \gy^{-24-k} \er^{2\pi i\tau}
N^{1/2} \, .
\label{Lam16E2kNdep}
\ee
To explain this mismatch in powers of $N$ we need to consider AdS
amplitudes like the ones producing the leading contribution in the
case of $\hat G_{\hat\Lambda^{16}\calE^4}$. For the general
correlation function (\ref{Lam16E2k}) the AdS diagrams to be computed
are generalised processes of the type in figure
\ref{Lam16E4figAdS}(a), involving the vertex
\be
\frac{1}{\al} \int d^{10}x \sqrt{-g} \, \er^{-\phi/2}
f_1^{(12,-12)}(\tau,\bar\tau) \Lambda^{16} \tau^k \, .
\label{Lam16E2kvertex}
\ee
and $k$ $\bar\tau G^2$ classical cubic interactions. A process of this
type will indeed produce a dependence in agreement with
(\ref{Lam16E2kNdep}).  The conclusion is that a test of the
conjectures of \cite{berkovafa} using the methods of the present paper
is impossible, because the effects of the interactions proposed in
\cite{berkovafa} are  subleading and highly suppressed in the
large-$N$ expansion in $AdS_5\times S^5$.

\subsection{Near extremal correlation functions}
\label{kkcorrel}

Among the most striking new results in $\calN=4$ SYM which were
stimulated by results in supergravity via the AdS/CFT correspondence
is the non-renormalisation of a class of $n$-point functions,
the extremal and next-to-extremal correlation functions of chiral primary
operators \cite{extremal,extrbk,extrwest1,extrwest2,nextextremal}.
These are functions of the type
\be
G_n(x_1,x_2,\ldots,x_n) = \la \calQ_{\ell_1}(x_1)\calQ_{\ell_2}(x_1)
\ldots \calQ_{\ell_n}(x_n) \ra \, ,
\label{defextremal}
\ee
where $\ell_1=\sum_{i=2}^n \ell_i$ in the extremal case and
$\ell_1+2=\sum_{i=2}^n \ell_i$ in the next-to-extremal case. For all
the $G_n$'s of this type supergravity predicts the absence of quantum
corrections to the free theory result and this prediction has been
checked by explicit calculations in $\calN=4$ SYM. In particular,
correlation functions in this class do not receive instanton corrections
\cite{extrbk,nextextremal}.

An even more interesting set of correlation functions are the
near-extremal ones \cite{nearextremal}, for which
$\ell_1+2m=\sum_{i=2}^n \ell_i$, with $m \ge 2$, in
(\ref{defextremal}). The near-extremal functions do receive quantum
corrections, but have been shown to display some `partial
non-renormalisation' properties. The best studied examples are those
with $m=2$ and the simplest case is represented by a four-point
functions of chiral primaries of dimension $\D=2$
\be
G_4^{\rm (near-extr)} = \la \calQ_2(x_1)\calQ_2(x_2)\calQ_2(x_3)
\calQ_2(x_4) \ra \, .
\label{nearextr4}
\ee
In general the spatial dependence in four- (and higher-) point
functions is not completely fixed by conformal invariance. Four-point
functions contain an {\it a priori} undetermined function of two
independent conformally invariant cross ratios. There are six
independent correlation functions of the form (\ref{nearextr4}) and
symmetry arguments suggest that they should be expressed in terms of
two distinct functions of the cross ratios.   Perturbative and
non-perturbative calculations as well as the analysis of supergravity
amplitudes have shown that, in fact, the six independent correlation
functions (\ref{nearextr4}) can all be expressed in terms of only one
function of the cross ratios \cite{partnonren}.

Another near-extremal correlation function with $m=2$ which has been found to
possess special partial non-renormalisation properties is the five-point
function
\be
G_5^{\rm (near-extr)} = \la \calQ_4(x_1)\calQ_2(x_2)\calQ_2(x_3)
\calQ_2(x_4) \calQ_2(x_5) \ra \, .
\label{nearextr5}
\ee
In \cite{nearextremal} it was shown that at the perturbative
level,
both in $\calN=4$ SYM and in type IIB supergravity in $AdS_5\times
S^5$, (\ref{nearextr4}) has a factorised structure, namely
\ba
G_5^{\rm (near-extr)} &\!=\!& N^3 \left[ \la \calQ_2(x_1)
\calQ_2(x_2)\calQ_2(x_3)\ra \la\calQ_2(x_1)\calQ_2(x_4)
\calQ(x_5)\ra \right. \label{nearextr5a} \\
&\!+\!& \left. \la\calQ_2(x_1)\calQ_2(x_2)\ra \la \calQ_2(x_1)
\calQ_2(x_3) \calQ_2(x_4)\calQ_2(x_5)\ra + \,
{\rm permutations} \: \right] \, , \nn
\ea
where only the four-point factor in the last line gets quantum
corrections.

In this section we comment on the possible generalisation of this
partial non-renormalisation result to include instanton effects.
If the factorisation (\ref{nearextr5a}) observed at the perturbative
level were also valid at the instanton level the non-perturbative
corrections should take the form of a minimal four-point function
times a free two-point function. In other words we should only get a
correction to the last line in (\ref{nearextr5a}), since the three
point functions of protected operators are not corrected by instantons.

The correlation function (\ref{nearextr5}) contains twelve scalars and thus
can,
in principle, soak up 24 fermionic modes which makes it non-minimal in
the sense discussed in this paper. The leading non-vanishing instanton
corrections involve $2r$ $(\nub\nu)$ bilinears and $s$ scalar
contractions, with $r+2s=4$, so that there are 8 scalar fields left to
saturate the superconformal fermion zero-modes. Taking into account
the way the $(\nub\nu)$ bilinears enter into the operators $\calQ_2$
and $\calQ_4$ (see  (\ref{opsnu}), (\ref{explO2}) and
(\ref{O4inst})) we find that the possible instanton contributions at
leading order in $\gy$ come from the diagrams in figure
\ref{nearextYMfig}.
\FIGURE{
{\includegraphics[width=0.275\textwidth]{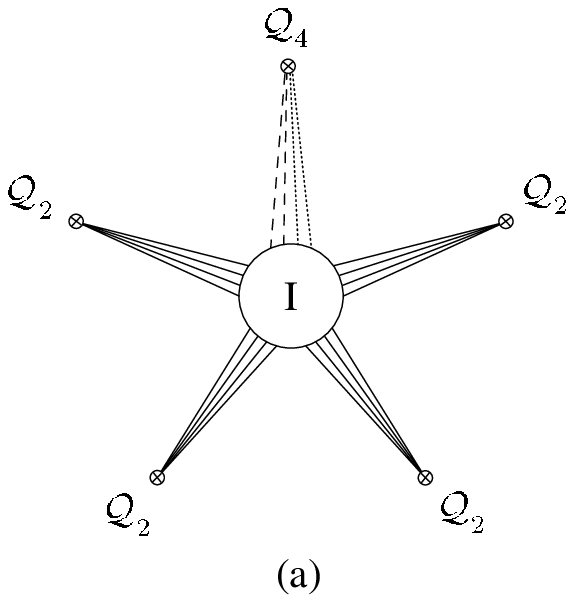}
\hspace*{0.3cm}
\includegraphics[width=0.275\textwidth]{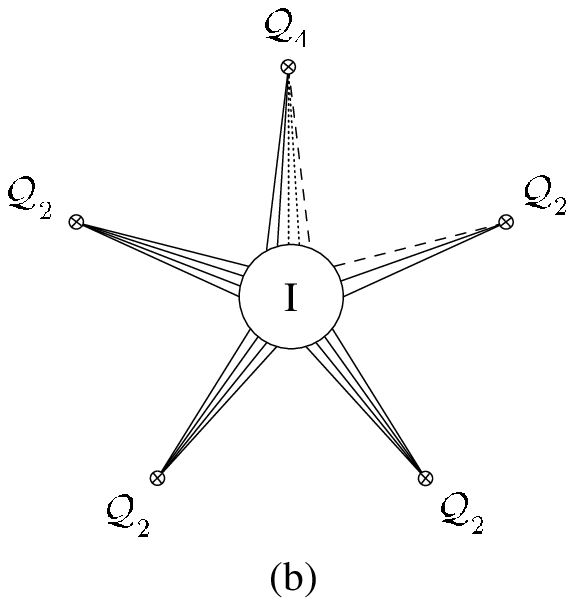}
\hspace*{0.3cm}
\includegraphics[width=0.275\textwidth]{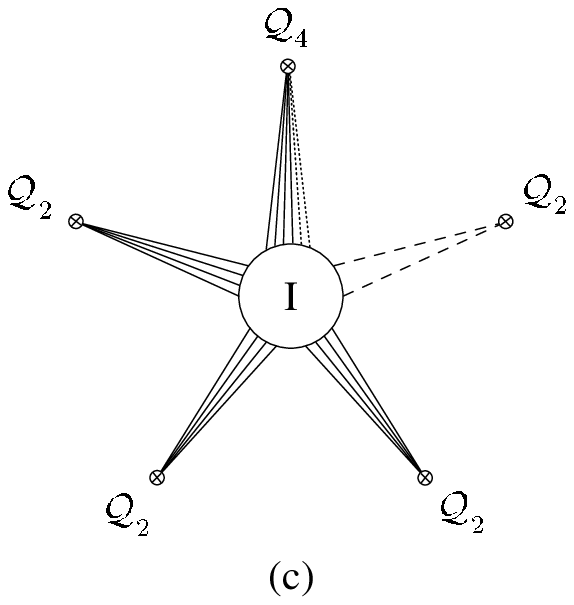}} \\
\rule{0pt}{10pt} \\
{\includegraphics[width=0.275\textwidth]{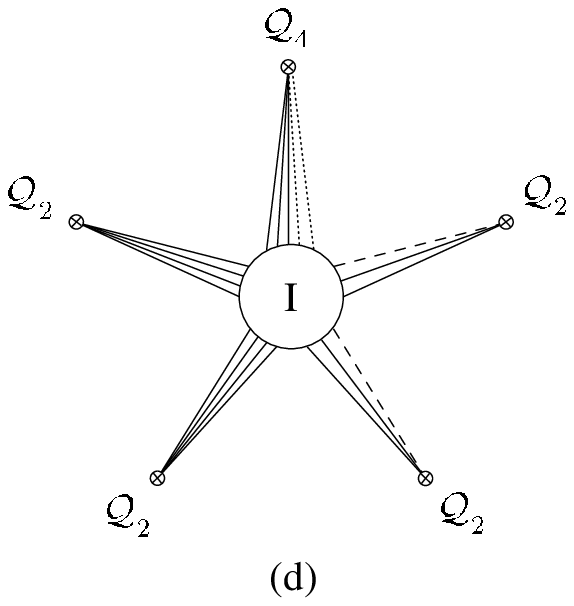}
\hspace*{0.3cm}
\includegraphics[width=0.275\textwidth]{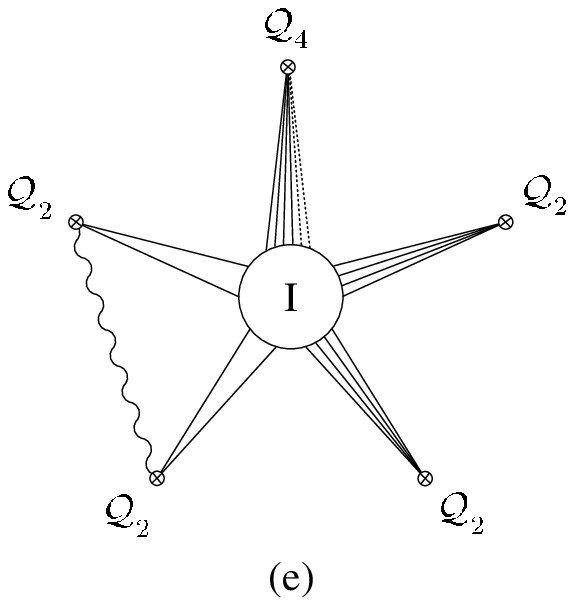}
\hspace*{0.3cm}
\includegraphics[width=0.275\textwidth]{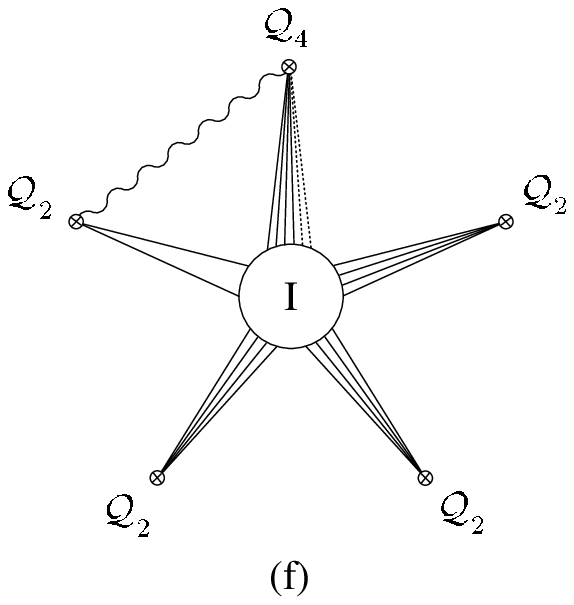}} \\
\rule{0pt}{10pt} \\
{\includegraphics[width=0.275\textwidth]{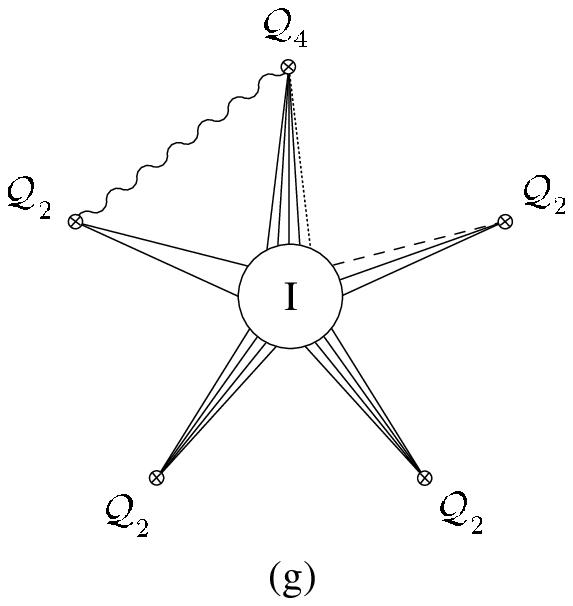}
\hspace*{0.3cm}
\includegraphics[width=0.275\textwidth]{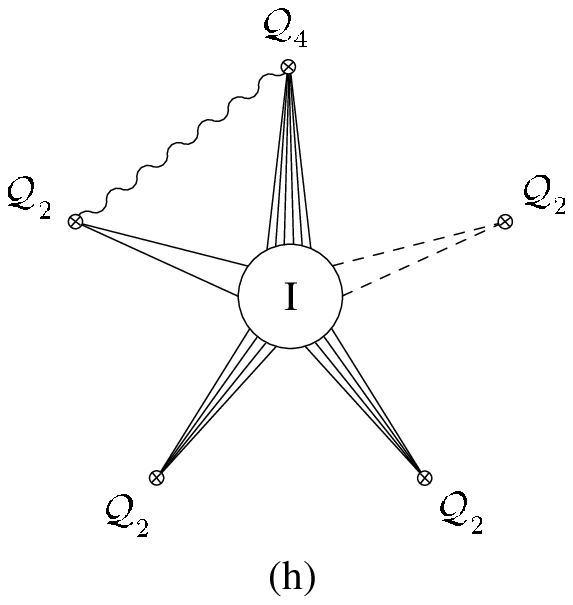}
\hspace*{0.3cm}
\includegraphics[width=0.275\textwidth]{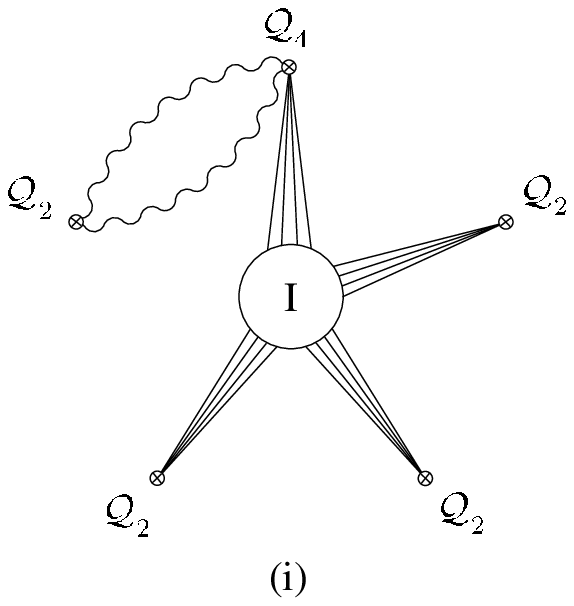}} \\
\rule{0pt}{10pt} \\
{\includegraphics[width=0.275\textwidth]{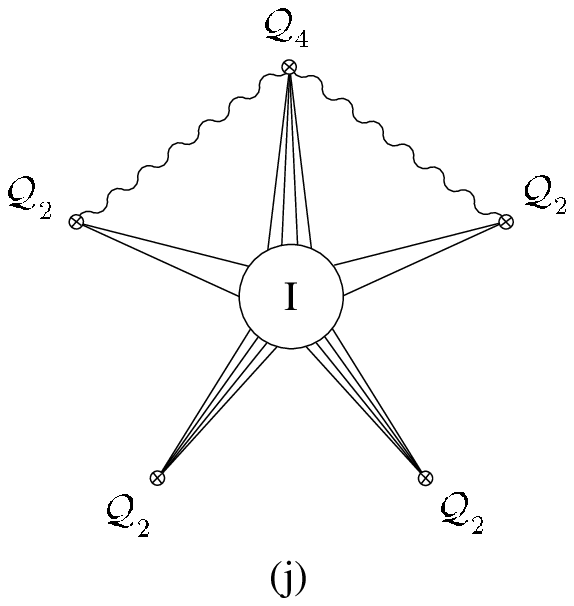}
\hspace*{0.3cm}
\includegraphics[width=0.275\textwidth]{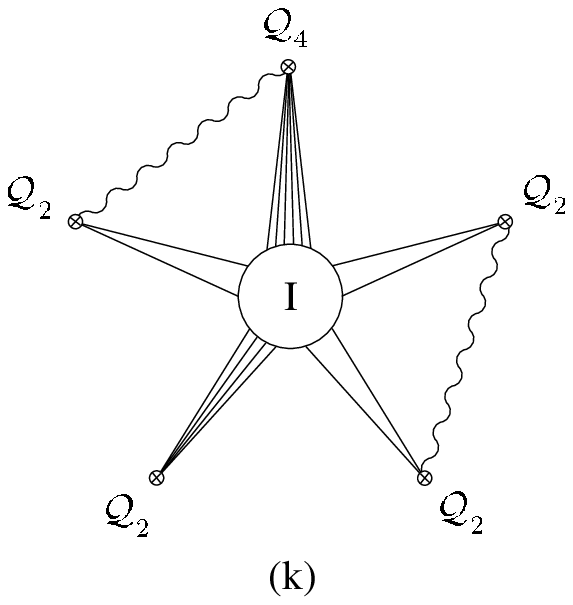}
\hspace*{0.3cm}
\includegraphics[width=0.275\textwidth]{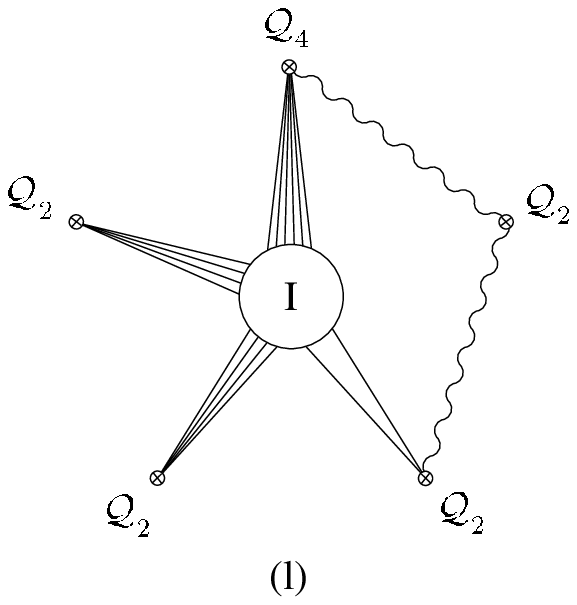}}
\caption{Diagrammatic representation of the
non-perturbative Yang--Mills contributions to the correlation function
$\langle \calQ_4 \calQ_2 \calQ_2 \calQ_2 \calQ_2\rangle$. The notation
is the same used in previous figures.}
\label{nearextYMfig}
}

Writing the $SU(4)$ indices explicitly the correlation function
(\ref{nearextr5}) becomes
\ba
\hat G_5 &\!=\!& \la
\calQ^{[A_1B_1][C_1D_1][E_1F_1][G_1H_1]}(x_1)
\calQ^{[A_2B_2][C_2D_2]}(x_2) \nn \\
&& \calQ^{[A_3B_3][C_3D_3]}(x_3) \calQ^{[A_4B_4][C_4D_4]}(x_4)
\calQ^{[A_5B_5][C_5D_5]}(x_5) \ra \, ,
\label{nearextr5b}
\ea
where as in previous sections we use a hat to distinguish the
Yang--Mills correlation function from the dual AdS amplitude to be considered
in the following.

We will not compute the individual diagrams in detail since they are rather
involved, but simply sketch how the generalisation of the perturbative
result ought to work. The factorisation (\ref{nearextr5a}) was proved in
\cite{nearextremal} using classical supergravity and is therefore a leading
order result in the $1/N$ expansion. We will also study the instanton
effects only at leading order, which is $N^{1/2}$.

The diagrams in figure \ref{nearextYMfig} can be organised in three
groups of four characterised by the number of scalar contractions,
$s=0,1,2$. Diagrams (a)-(d) form the first group ($s=0$), diagrams
(e)-(h) the second group ($s=1$) and diagrams (i)-(l) the third group
($s=3$). The first four diagrams have no contractions and thus
involve four $(\nub\nu)$ pairs. Recalling the normalisation of chiral
primary operators in (\ref{cpoell}) and the general rules for
the counting of powers of $N$ and $\gy$ derived in section
\ref{ginvarcorr}, we find that diagrams \ref{nearextYMfig}(a)-(d)
are all of order
\be
\left(\frac{N}{\gy^2N}\right)^4 \frac{N}{\gy^4 N^2} \:
\gy^{12} N \, N^{1/2} = N^{1/2} \, ,
\label{Ndepnearextr1}
\ee
where the first two factors are the normalisations of the five
operators and the remaining factors are those associated with the
instanton measure and $(\nub\nu)$ insertions according to
(\ref{Ndepnu}). Diagrams (a) and (b) can be argued to vanish after
integration over the angular variables $\Omega^{AB}$ on the
basis of group theory. In the case of (a) the operator
$\calQ_4$ soaks up all the $\nub$ and $\nu$ modes. This means that
they must be combined in the {\bf 105} of $SU(4)$ and thus the
corresponding angular integral vanishes because it selects a singlet.
In the case of diagram (b) it is easy to see that the two $\nsix$ and
one $\nten$ in the $\calQ_4$ insertion must be combined to form a {\bf
126}, which again means that the angular integral is zero because the
remaining $(\nub\nu)$ bilinear is in the {\bf 10} and the product
${\bf 126}\times{\bf 10}$ does not contain the singlet. The other two
diagrams, (c) and (d), are potentially non zero. The result for (c) is
of the form
\ba
\hat G_5^{\rm (c)} &\!=\!& c_{\rm c}(N) \, \er^{2\pi i\tau}
t_{\rm c}^{A_1B_1C_1D_1A_2B_2C_2D_2}
\int \frac{d^4x_0d\rho}{\rho^5}
\frac{\rho^6}{[(x_1-x_0)^2+\rho^2]^6}
(\zeta\zeta\zeta\zeta)^{E_1F_1G_1H_1} \nn \\
&& \frac{\rho^2}{[(x_2-x_0)^2+\rho^2]^2}
\prod_{i=3}^5 \frac{\rho^4}{[(x_i-x_0)^2+\rho^2]^4}
(\zeta\zeta\zeta\zeta)^{A_iB_iC_iD_i} + \cdots
\label{nearextr-c}
\ea
where
\be
c_{\rm c}(N) \sim N^{1/2} + O(N^{-1/2})
\label{Ndepnear-c}
\ee
and the tensor $t_{\rm c}^{A_1B_1C_1D_1A_2B_2C_2D_2}$ is determined by
the angular integration.  In (\ref{nearextr-c}) we have ordered the
insertion points clockwise starting from $x_1$ where $\calQ_4$ is
located and we have schematically indicated by
$(\zeta\zeta\zeta\zeta)^{ABCD}$ the combination of superconformal
modes entering in the operators $\calQ_4$ and $\calQ_2$ (see
(\ref{opsnu}), (\ref{explO2})  and (\ref{O4inst})). The ellipsis stands
for permutations corresponding to the other ways of distributing the
zero modes with the fixed structure in figure \ref{nearextYMfig}(c).

The contribution of (d) takes the form
\ba
\hspace*{-0.5cm}
\hat G_5^{\rm (d)} &\!=\!& c_{\rm d}(N) \, \er^{2\pi i\tau}
t_{\rm d}^{A_1B_1C_1D_1A_2B_2A_3B_3}
\int \frac{d^4x_0d\rho}{\rho^5}
\frac{\rho^6}{[(x_1-x_0)^2+\rho^2]^6}
(\zeta\zeta\zeta\zeta)^{E_1F_1G_1H_1} \nn \\
&& \hspace*{-0.3cm}
\prod_{i=2}^3 \frac{\rho^3}{[(x_i-x_0)^2+\rho^2]^3}
(\zeta\zeta)^{C_iD_i}
\prod_{i=4}^5 \frac{\rho^4}{[(x_i-x_0)^2+\rho^2]^4}
(\zeta\zeta\zeta\zeta)^{A_iB_iC_iD_i} + \cdots \, ,
\label{nearextr-d}
\ea
where the dots again stand for permutations and as in case (c)
we have
\be
c_{\rm d}(N) \sim N^{1/2} + O(N^{-1/2}) \, .
\label{Ndepnear-d}
\ee

Of the second set of four diagrams, which have one scalar contraction,
only the first two are of the same order, $N^{1/2}$, as the previous
ones and need to be taken into account. Diagrams (g) and (h) would be
subleading, but actually they vanish after integration over the five
sphere since the $\nub$ and $\nu$ modes can not be combined to form a
singlet. Diagrams (e) and (f) have two $\nsix$ insertions and thus
the counting of powers of $\gy$ and $N$ for these contributions goes
exactly as in (\ref{Ndepnearextr1}). The evaluation of these two
diagrams is rather lengthy and the result we get is of the form
\ba
&& \hspace*{-0.2cm}
\hat G_5^{\rm (e)} = c_{\rm e}(N) \, \er^{2\pi i\tau}
\veps^{A_1B_1C_1D_1}
\int \frac{d^4x_0d\rho}{\rho^5}
\frac{\rho^6}{[(x_1 -x_0)^2+\rho^2]^6}
(\zeta\zeta\zeta\zeta)^{E_1F_1G_1H_1} \nn \\
&& \hspace*{-0.2cm} \veps^{C_2D_2C_3D_3} \left[
\frac{(\zeta\zeta\zeta\zeta)^{A_2B_2A_3B_3}}{(x_2-x_3)^2}
\prod_{i=2}^3\frac{\rho^2}{[x_i-x_0)^2+\rho^2]^2} +
(\zeta\zeta\zeta\zeta)^{A_2B_2A_3B_3} \prod_{i=2}^3
\frac{\rho^3}{[x_i-x_0)^2+\rho^2]^3} \right] \nn \\
&& \hspace*{-0.2cm} \prod_{i=4}^5
\frac{\rho^4}{[(x_i-x_0)^2+\rho^2]^4}
(\zeta\zeta\zeta\zeta)^{A_iB_iC_iD_i} + \cdots \, ,
\label{nearextr-e}
\ea
for (e) and
\ba
\hspace*{-0.2cm}
\hat G_5^{\rm (f)} &\!=\!& c_{\rm f}(N) \, \er^{2\pi i\tau}
\veps^{A_1B_1C_1D_1}
\int \frac{d^4x_0d\rho}{\rho^5} \,
\veps^{E_1F_1A_2B_2} \left[ \frac{\rho^5}{[(x_1 -x_0)^2+\rho^2]^5}
\frac{\rho^3\:(\zeta\zeta\zeta\zeta)^{G_1H_1C_2D_2}}
{[(x_2-x_0)^2+\rho^2]^3} \right. \nn \\
&& \hspace*{-0.2cm} \left.+ \frac{1}{(x_1-x_2)^2} \frac{\rho^4}
{[(x_1-x_0)^2+\rho^2]^4} \frac{\rho^2}{[(x_2-x_0)^2+\rho^2]^2}
(\zeta\zeta\zeta\zeta)^{G_1H_1C_2D_2} \right] \nn \\
&& \hspace*{-0.2cm} \prod_{i=4}^5
\frac{\rho^4}{[(x_i-x_0)^2+\rho^2]^4}
(\zeta\zeta\zeta\zeta)^{A_iB_iC_iD_i} + \cdots \, ,
\label{nearextr-f}
\ea
for (f). In both (\ref{nearextr-e}) and (\ref{nearextr-f}) the dots
stand for permutations and other contributions with the same structure
corresponding to contractions between different pairs of scalars.
As already observed both the coefficients $c_{\rm e}(N)$ and
$c_{\rm f}(N)$ start at order $N^{1/2}$.

The remaining four diagrams have no $\nu$ and $\nub$ insertions
and so have no additional powers of $N$ associated with colour
contractions in $(\nub\nu)$ bilinears. A counting of powers of $\gy$
an $N$ for these diagrams gives
\be
\left(\frac{N}{\gy^2N}\right)^4 \frac{N}{\gy^4 N^2} \:
\gy^{12} \, N^{1/2} = N^{-1/2} \, .
\label{Ndepnearextr2}
\ee
In diagram (i) we get an additional power of $N$ from the double
scalar contraction, hence this is the only contribution that we need
to take into account at leading order, $N^{1/2}$. The remaining
diagrams (j), (k) and (l) are subleading and we will not examine them
here. In the previous diagrams the only term in the propagator which
gave a non-vanishing contribution was the one in the first line of
(\ref{propfin}), this is also true for (i) at leading
order.  The other terms only give rise to subleading contributions. 
The order $N^{1/2}$ part of (i) reads
\ba
\hspace*{-0.2cm}
\hat G_5^{\rm (i)} &\!=\!& c_{\rm i}(N) \, \er^{2\pi i\tau}
\veps^{E_1F_1A_2B_2}\veps^{G_1H_1C_2D_2}
\int \frac{d^4x_0d\rho}{\rho^5}\, \left[ \frac{1}{(x_1-x_2)^4}
\frac{\rho^4\:(\zeta\zeta\zeta\zeta)^{A_1B_1C_1D_1}}
{[x_1-x_0)^2+\rho^2]^4}  \right. \nn \\
&& \hspace*{-0.2cm} \left. + \frac{1}{(x_1-x_2)^2}
\frac{\rho^5}{[x_1-x_0)^2+\rho^2]^5}
\frac{\rho}{[(x_1 -x_0)^2+\rho^2]}
(\zeta\zeta\zeta\zeta)^{E_1F_1G_1H_1} \right] \nn \\
&& \hspace*{-0.2cm} \prod_{i=2}^5\frac{\rho^4}{[x_i-x_0)^2+\rho^2]^4}
(\zeta\zeta\zeta\zeta)^{A_iB_iC_iD_i} + \cdots \, .
\label{nearextr-i}
\ea
Notice that this expression indeed contains a term with the factorised
structure $\la\calQ_2(x_1)\calQ_2(x_2)\ra
\la\calQ_2(x_1)\calQ_2(x_1)\calQ_2(x_1)\calQ_2(x_1)\ra_{\rm inst}$,
where the second factor is the instanton contribution to the minimal
four-point function of chiral primaries of dimension 2, \ie
\be
\frac{\veps^{E_1F_1A_2B_2}\veps^{G_1H_1C_2D_2}}{(x_1-x_2)^4}
\: N^{1/2} \, \er^{2\pi i\tau} \int \frac{d^4x_0d\rho}{\rho^5}
\prod_{ \begin{array}{c}
\begin{scriptstyle}i=1 \end{scriptstyle} \\
\raisebox{15pt}{$\begin{scriptstyle}i\ne 2 \end{scriptstyle}$}
\end{array}}^5
\frac{\rho^4}{[x_i-x_0)^2+\rho^2]^4}
(\zeta\zeta\zeta\zeta)^{A_iB_iC_iD_i} \, .
\label{factorinst} \vspace*{-20pt}
\ee

In conclusion,  our results are consistent with the possibility that
the factorisation property (\ref{nearextr5a}) of the near-extremal
five-point function remains valid when instanton corrections are
included. However, for this to happen several contributions would have
to cancel.  The contributions of (\ref{nearextr-c}),
(\ref{nearextr-d}), (\ref{nearextr-e}), (\ref{nearextr-f}) and the
second term in (\ref{nearextr-i}), which do not individually vanish,
do not possess the factorised  structure of (\ref{nearextr5a}) but
their sum might vanish when the precise coefficients are evaluated.

We now  discuss the AdS processes contributing to the dual amplitude
at leading order. All the relevant $AdS_5$ Witten diagrams involve a
$\calR^4$ D-instanton induced vertex and can be represented as in
figure \ref{nearextAdSfig}. All the lines in figures represent scalar
fields dual to the chiral primary operators $\calQ_\ell$. As in
previous sections we denote this field and its Kaluza--Klein excited
modes by $Q$. In the figure we have indicated the dimension of the
associated field next to each line.
\FIGURE{
{\includegraphics[width=0.3\textwidth]{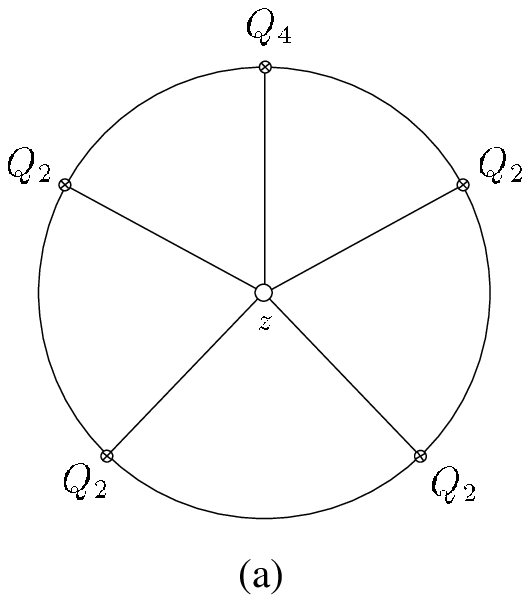}
\hspace*{0.3cm}
\includegraphics[width=0.3\textwidth]{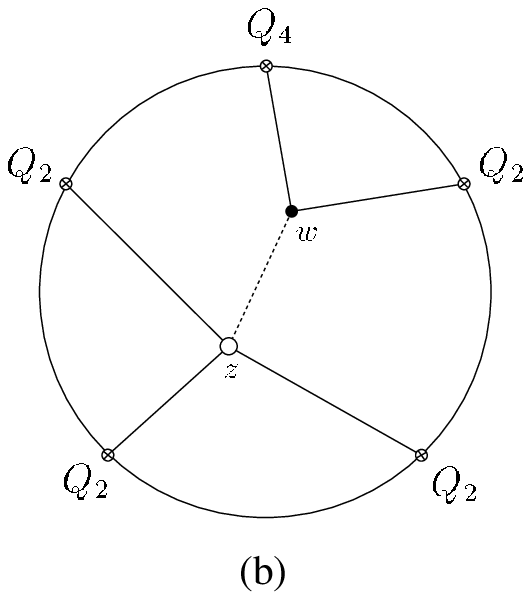}
\hspace*{0.3cm}
\includegraphics[width=0.3\textwidth]{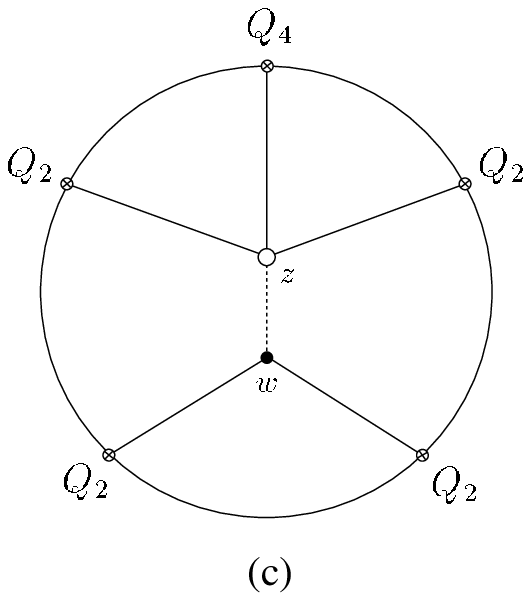}}
\caption{Diagrammatic representation of the AdS amplitudes
contributing to the process dual to the correlation function
$\langle \calQ_4 \calQ_2 \calQ_2 \calQ_2 \calQ_2\rangle$.}
\label{nearextAdSfig}
}

There is an obvious contribution which comes from a contact diagram
obtained expanding the $\calR^4$ coupling. This is represented in
figure \ref{nearextAdSfig}(a) and the resulting amplitude takes the
form
\be
G^5_{\rm a} = c_{\rm a}(\tau,\bar\tau,N) \int \frac{d^5z}{z_0^5} \,
\frac{\del^8}{\del z^8}\left[K_4(z;x_1)\prod_{i=2}^5
K_2(z;x_i) \right] \, ,
\label{nearextrAdS-a}
\ee
where we have indicated schematically the eight derivatives entering
the $\calR^4$ coupling that act on the bulk-to-boundary propagators.
The coefficient $c_{\rm a}(\tau,\bar\tau,N)$ encodes the dependence on
$N$ and the coupling which are those obtained from the $\calR^4$
interaction at order $(\al)^{-1}$
\be
c_{\rm a}(\tau,\bar\tau,N) = c_{\rm a} \, \er^{2\pi i\tau} \left(
N^{1/2} + O(N^{-1/2}) \right) \, .
\label{NdepnearexAdSa}
\ee
The second type of contribution corresponds to exchange diagrams
with four lines coming out of a $\calR^4$ vertex and an additional
cubic interaction, see figure \ref{nearextAdSfig}(b) and (c). There
are four possible contributions of this type corresponding to the fact
that the bulk-to-bulk lines in these figures can each be
dual to an operator of dimension 2
or 4. In the case of figure \ref{nearextAdSfig}(b)
when the intermediate line is a
dimension 2 field, the vertex in $w$ is extremal and the result reduces
to
\be
G^5_{\rm b} = c_{\rm b}(\tau,\bar\tau,N) \frac{1}{(x_1-x_2)^4}
\int \frac{d^5z}{z_0^5} \,
\frac{\del^8}{\del z^8}
\prod_{ \begin{array}{c}
\begin{scriptstyle}i=1 \end{scriptstyle} \\
\raisebox{15pt}{$\begin{scriptstyle}i\ne 2 \end{scriptstyle}$}
\end{array}}^5 K_2(z;x_i)  \, .
\label{nearextrAdS-b}  \vspace*{-20pt}
\ee
The contribution in which the intermediate state has dimension 4 can
be computed using the formulae derived in \cite{zint}. However, it is
expected to vanish, because the D-instanton vertex in $z$ was argued
in \cite{nextextremal} to be zero to reproduce the non-renormalisation
of next-to-extremal correlation functions. For the same reason the process of
figure \ref{nearextAdSfig}(c) in which the intermediate state is dual
to a dimension 2 operator also vanishes. The remaining possibility is
the exchange of a dimension 4 operator. In this case the vertex in $w$
is extremal and the result is of the form
\be
G^5_{\rm c} = c_{\rm c}(\tau,\bar\tau,N) \int \frac{d^5z}{z_0^5} \,
\frac{\del^8}{\del z^8}\left[K_4(z;x_1)\prod_{i=2}^5
K_2(z;x_i) \right] \, ,
\label{nearextrAdS-c}
\ee
\ie the same as (\ref{nearextrAdS-a}). For the two contributions in
(\ref{nearextrAdS-b}) and (\ref{nearextrAdS-c}) we find again
\be
c_{\rm b,c}(\tau,\bar\tau,N) = c_{\rm b,c} \, \er^{2\pi i\tau} \left(
N^{1/2} + O(N^{-1/2}) \right) \, .
\label{NdepnearexAdSbc}
\ee
The AdS result therefore has potentially two contribution. One is a
contact term resulting from the sum of (\ref{nearextrAdS-a}) and
(\ref{nearextrAdS-c}) and the other, (\ref{nearextrAdS-b}), has the
factorised structure found in perturbation theory. It is striking
to conjecture that the first two contributions should cancel leaving
no contact amplitude and a total amplitude with the factorised form.

To summarise, our calculations are compatible with the possibility that
the partial non-renormalisation of the near-extremal five-point
function (\ref{nearextr5b}) is valid beyond perturbation theory.  Although
we have only carried out a qualitative analysis which does not allow us
to check all the required cancellations and give a complete proof,
it is impressive that it is a possibility.

\subsection{Anti-instanton contributions}
\label{antiinst}

We conclude this section with a comment about anti-instanton
contributions to correlation functions like those considered in
\cite{bgkr,doreya}, which are minimal in the instanton ($K=+1$)
background. Equivalently, we can study instanton contributions to
correlation functions with insertions of the fields conjugate to those
in (\ref{opsnu}) that we have considered up to this point. These fields
contain more fermionic zero-modes in the background of an instanton
and thus give rise to non-minimal correlation functions. The simplest
cases to analyse are again those with the maximal number of fermionic
insertions. We can consider for instance
\be
\hat G_{\hat{\bar\Lambda}^{16}} = \la
\hat{\bar\Lambda}_{A_1}^{\adot_1}(x_1) \ldots
\hat{\bar\Lambda}_{A_{16}}^{\adot_{16}}(x_{16}) \ra \, ,
\label{barLam16}
\ee
where $\hat{\bar\Lambda}_{A}^{\adot}$ is the operator
\be
\hat{\bar\Lambda}_{A}^{\adot} = \Tr \left(\sbar^{\adot}_{mn\:\bdot}
F^{mn}\bar\lambda^\bdot_A \right) \, ,
\label{barLamdef}
\ee
transforming in the representation ${\bf 4^*}$ of $SU(4)$.

In the instanton background $F^+_{mn}$, which appears in
(\ref{barLamdef}), contains at least four zero-modes and
$\bar\lambda^\bdot_A$ three, so there are seven fermionic modes for
each insertion in (\ref{barLam16}), for a total of 112. Of these,
sixteen must be superconformal modes that produce a non-vanishing
result, which leaves 48 $\bar\nu\nu$ bilinears, which we expect to be
only of the $\nten$ type. According to the general formulae of section
\ref{ginvarcorr} we therefore expect a leading
$\gy$ and $N$ dependence of the type
\be
\hat G_{\hat{\bar\Lambda}^{16}} \sim N^{1/2} \gy^{24} \er^{2\pi i\tau}
\, ,
\label{barLam16gNdep}
\ee
where the powers of $\gy$ are obtained by combining a factor of
$\gy^{-32}$ from the normalisation of the operators (see
(\ref{cpoell})) and factors of $\gy^{48}$ and $\gy^8$ from the $\nten$
insertions and the instanton measure respectively.

The dependence on the parameters in (\ref{barLam16gNdep}) agrees with
the prediction of supergravity. Indeed the vertex $\bar\Lambda^{16}$,
which generates the amplitude dual to (\ref{barLam16}), appears in the
type IIB effective action at order $\al^{-1}$ multiplied by the modular
form $f_1^{(-12,12)}(\tau,\bar\tau)$, which has a leading instanton
contribution starting as $\tau_2^{-12}=\gs^{12}$.

In general, correlation functions such as (\ref{barLam16}) with
insertions of operators conjugate to those entering minimal
correlation functions are much more involved, because they are
suppressed by many powers of the coupling $\gy$, which allows for the
possibility of complicated contributions with scalar contractions at
leading non-vanishing order. A detailed analysis of these processes is
outside the scope of this paper.

\section{Discussion}
\label{discuu}

In this paper we have studied the structure of a variety of
correlation  functions of gauge invariant composite operators in $\N4$
$SU(N)$ Yang--Mills theory in a one-instanton background  at large $N$
and at leading order in the coupling constant, $\gy$.  One of the main
motivations was to understand to what extent the $1/N$ expansion of
instanton-induced contributions to correlation functions in the
$\calN=4$ supersymmetric Yang--Mills theory encodes information about
higher-derivative terms in the type IIB effective action via the
AdS/CFT correspondence.  Another motivation was to understand
instanton effects for processes involving composite Yang--Mills
operators that are dual to Kaluza--Klein excitations of the
supergravity fields.

The calculation of Yang--Mills correlation functions that we have
considered includes the contributions of the fermionic moduli, $\nu^A$
and $\bar\nu^A$, that are superpartners of the gauge rotations that
describe the embedding of $SU(2)$ in $SU(N)$.  Within the context of
the AdS/CFT conjecture such modes are essential in accounting for the
Kaluza--Klein excitations of the type IIB supergravity fields on the
five-sphere in the presence of a single D-instanton.  They are also
essential in accounting for higher-order interactions in the effective
type IIB string theory action in the \AdS5s5\ background.  A prime
example of this is the $\al G^4\Lambda^{16}$ interaction. This
corresponds to a contribution to the $\langle \hat \Lambda^{16}
\calE^4\rangle$ of order $N^{-1/2}$, which is suppressed by a power of
$1/N$ relative to the dominant instanton contributions.  However, we
found that the explicit Yang--Mills calculation gives a leading term
of order $N^{1/2}$. In order to explain this it was important to
appreciate that the \AdS5s5\ contributions are not entirely given by a
single instanton-induced interaction.  It is essential to include
certain tree diagrams which have  one induced D-instanton vertex and
classical interactions at the other vertices.  For example, the tree
diagram for the $\langle \Lambda^{16} G^4\rangle$ amplitude has two
propagators for a $\tau_{20'}$, which is a Kaluza--Klein excitation of
the dilaton.  The coupling of these fields to the external $G$'s
arises from the $\bar\tau\,G^2$ coupling of classical type IIB
supergravity.  This generates a  new kind of instantonic contact
interaction, thereby explaining  the $N^{1/2}$ contribution to the
$\langle \hat \Lambda^{16} \calE^4\rangle$ correlation function.
Since the Kaluza--Klein excitations are absent from the consistently
truncated five-dimensional gauged $N=8$ supergravity, we would expect
such contact terms to arise directly as D-instanton contact terms in
the dimensionally reduced theory.

In the simplest instanton-induced correlation functions all of the
fields in the composite operators are replaced by their `classical'
profiles, which include the fermionic zero modes.  More generally, in
the examples that we studied, even at leading order in $\gy$, the
effect of  quantum fluctuations which give rise to propagators in the
instanton background needs to be taken into account.  In practice, the
only propagator we needed to consider is the scalar propagator
$\langle \v^a(x_1) \v^b(x_2) \rangle_{\rm inst}$.  Although the
complete expression for this propagator (\ref{propfin}) looks
complicated, its insertion in correlation functions of gauge-invariant
composite operators gives rise to simple spatial structures.  The
combination of the terms in (\ref{propfin}) produces contributions to
Yang--Mills correlation functions which are found to reproduce the
supergravity amplitudes with D-instanton induced vertices and
additional tree-level interactions described above.  Contributions of
this type arose, for example, in determining the form of the $\la
\hat\Lambda^{16} \calQ_2^2 \ra$ correlation function. In this example,
that we discussed in detail, the combination of different effects on
both sides of the correspondence is required to reconcile the results
with the AdS/CFT duality.

These considerations constitute a small step towards unraveling the
systematics of the $N$-dependence of instanton effects. We have
restricted our analysis to the one-instanton sector. On the other
hand, as shown in \cite{doreya}, the large-$N$ limit simplifies
drastically the study of multi-instantons and their comparison
with D-instanton effects in type IIB string theory. It would be
interesting to extend the results of this paper to the generic
$K$-instanton sector at large $N$. This would generalise the work of
\cite{doreya} to the class of non-minimal correlation functions
and allow to investigate the large-$N$ limit of the $K$-instanton
contributions beyond leading order. This program involves systematically
expanding the instanton measure as well as the expressions for the
composite operators.  Both of these tasks are formidable and still
incomplete.

There are other obvious areas in which the fermionic moduli $\nu^A$
and $\bar \nu^A$ and the other effects discussed in this paper should
play an essential r\^ole.  One of these is the instanton contribution
to the BPS Wilson loop \cite{maldloop} in $\calN =4$ supersymmetric
$SU(N)$  Yang--Mills theory.  The generalization of the arguments for
the $SU(2)$ case of \cite{bgk} to the $SU(N)$ case requires the
inclusion of the additional $\nu$ and $\nub$ modes as well as the
contribution of contractions. In particular, in the $N\to\infty$ limit
the sum of these additional contributions becomes an infinite series
which could exponentiate and, combined with
the perturbative result, give rise to an expression that transforms
covariantly under
S-duality. Likewise, the BMN limit \cite{bmn} of  superconformal
Yang--Mills theory involves an infinite boost along a great circle of
the five-sphere which is dominated by highly excited angular momentum
states.  This will involve the $\nu^A$ and $\bar \nu^A$ moduli in an
essential manner.  D-instanton configurations in a plane-wave
background have been constructed \cite{ppdinst}, suggesting that
instanton effects should also play a r\^ole in this limit.  An
interesting class of observables in the $\calN=4$ supersymmetric
Yang--Mills theory comprises correlation functions which are not
protected by symmetry from receiving quantum corrections, but
display `partial non-renormalisation properties'. Examples of this
type of quantity are the near-extremal correlation functions, such as the
one discussed in section \ref{kkcorrel}. The partial
non-renormalisation properties of four-point functions of chiral
primary operators have been recently studied in the case of operators
of dimension $\D=3$ \cite{ados,hh} and it was found that all these
four-point correlation functions, as in the case of dimension two operators,
can be expressed in terms of a unique function of two conformally
invariant cross ratios. The proof that this property is still valid
when instanton corrections are included would require the application
of the techniques discussed in this paper. Finally, the Konishi
multiplet in $\calN=4$ super Yang--Mills is known to possess some
unexpected non-perturbative  non-renormalisation properties
\cite{konishi}, namely its two-, three- and four-point functions do
not receive instanton corrections. Higher-point functions with
insertions of operators in the Konishi multiplet are non-minimal in
the sense discussed here and appear to have non-vanishing instanton
contributions \cite{instkoni}.

\acknowledgments
We are grateful to Massimo Bianchi for numerous conversations. SK is
grateful for the support of a Marie Curie Fellowship. This work was
partially supported by  EEC contract HPRN-2000-00122.

\appendix

\section{Conventions and useful relations}
\label{conventions}

In this appendix we will summarize a few useful identities
associated with $SO(4)$ and $SO(6)$ spinors.

The four-dimensional sigma matrices are defined by
\be
\s^m = (\one,i\tau^i) \, , \qquad
\bar\s^m = (\one,-i\tau^i) \, ,
\label{defsig4}
\ee
where $\tau^i$ are the Pauli matrices.
The self-dual tensor $\s^{mn}{}_{\b}{}^{\g}$ is defined by
\be
\s^{mn}{}_{\b}{}^{\g} =
\frac{1}{4}(\s^m{}_{\b\adot}\sbar^{n\adot\g} -
\s^n{}_{\b\adot}\sbar^{m\adot\g}) \, .
\label{defsigmn}
\ee

The $SO(6)$ sigma matrices are defined by
\be
\S^a_{AB} = (\eta^i_{AB},i\bar\eta^i_{AB}) \, , \qquad
\bar\S_a^{AB} = (-\eta_i^{AB},i\bar\eta^{AB}_i) \, ~~ {\rm with} ~
a=1,\dots,6
\label{defsig6}
\ee
and the 't Hooft symbols, for $i=1,2,3$, are
\ba
&& \eta^i_{AB} = \bar\eta^i_{AB} = \veps_{iAB}
\qquad A,B=1,2,3 \, , \nn \\
&& \eta^i_{A4} = \bar\eta^i_{4A} = \d^i_A \, , \nn \\
&& \eta^i_{AB} = - \eta^i_{BA} \, , \qquad
\bar\eta^i_{AB} = - \bar\eta^i_{BA} \, .
\label{etadef}
\ea

The following properties of the 't Hooft symbols are of use,
\be
\eta^i_{mn} \s^{mn}_{\,\a}{}^\b = 2i \tau^i_{\,\a}{}^\b \, , \qquad
\bar\eta^i_{mn} \bar\s^{mn}_{\,\adot}{}^\bdot =
2i \tau^i_{\,\adot}{}^\bdot \, .
\label{etasig}
\ee
\be
\eta^i_{mn} \tau^i_{\,\a}{}^\b = - 2i \s^{mn}_{\,\a}{}^{\b} \, ,
\qquad \bar\eta^i_{mn} \tau^i_{\,\adot}{}^\bdot =
-2i \bar\s^{mn}_{\,\adot}{}^{\bdot} \, .
\label{etatau}
\ee
\be
\tr\left(\s_{mn}\tau^i\right) = i \eta^i_{mn} \, , \qquad
\tr\left(\bar\s_{mn}\tau^i\right) = i \bar\eta^i_{mn} \, .
\label{trsigtau}
\ee

The $SO(6)$ sigma matrices satisfy the following relations
\be
\veps^{ABCD} \S^a_{CD} = -2 {\bar\S}^{a\,AB} \: , \qquad
\veps_{ABCD} {\bar\S}^{a\,CD} = -2 \S^a_{AB} \, .
\label{raiseindex}
\ee
\be
\frac{1}{2} \left(\S^a_{AC}
\Sbar^{b\,CB} + \S^b_{AC}\Sbar^{a\,CB}\right)
= \delta^{ab} \delta_A^B  \, ,
\label{siga}
\ee
\be
\S^a_{AC}\Sbar^{b\,CB} = \delta^{ab} \delta_A^B + \S_A^{ab\,B}
\ee
where $\S_A^{ab\,B}=\frac{1}{2}\left(\S^{a\, AC} \Sbar^b_{CB} -
\S^{b\, AC}\Sbar^a_{CB}\right)$.
The following identities are also of use,
\be
\label{sigb}
\Sbar^{a\, AB} \S^a_{CD} = 2(\delta^A_D \delta^B_C - \delta^A_C
\delta^B_D) \, .
\ee
\be
\label{sigthree}
(\S^a\Sbar^a)^A_B = 6 \delta^A_B \, .
\ee
\be
\Sigma^a_{AB}\Sigma^a_{CD} = 2 \veps_{ABCD} \, .
\label{sigd}
\ee
\be
\S_A^{ab\,B}\S_B^{cd\,A} = -4 \left( \d^{ac}\d^{bd} -
\d^{ad}\d^{bc} \right) \, .
\label{sige}
\ee
\be
\label{xtrans}
\chi_{AB} = {1\over \sqrt 8} \S^a_{AB}\chi_a \, , \qquad
\chi_a = {1\over \sqrt 2} \Sbar_a^{AB} \chi_{BA} \, .
\ee
\be
\label{multrace}
\tr(\Sbar^a\S^b\Sbar^c\S^d) = 4 \delta^{ab}\delta^{cd} - 4 \delta^{ac}
\delta^{bd} + 4\delta^{ad}\delta^{bc} \, .
\ee

\section{Summary of ADHM for $\N4$ theory}
\label{adhmsum}

In this appendix we include for completeness a few formulae used in
the description of multi-instanton configurations in the ADHM
formalism. The discussion is by no means complete, we refer the reader
for instance to \cite{dhkm-rep} for a very comprehensive review.

The classical $K$-instanton solution is defined in terms of
a $[N+2K]\times [2K]$ dimensional matrix $\Delta_{ai}{}^{\a;j\adot}$
which is a linear function of the space-time coordinate $x_m$,
\be
\label{deltadef}
\Delta_{ui\a;}{}^j{}_{\adot} = a_{ui\a;}{}^j{}_{\adot} +
b_{u i\a;}{}^{j\b}\, x_{\b\adot} \, ,
\ee
and its conjugate,
\be
\label{deltadefc}
\bar\Delta_{i}{}^{\adot;j\a} = \bar a_{i}{}^{\adot;j\a}
+ x^{\adot \b}\, \bar b_{i\b;}{}^{uj\a}
\ee
(we use latin letters from the mid of the alphabet, $i,j,k,...=1,...K$,
to denote instanton indices).
The ADHM gauge field is written in the form
\be
(A_m)_{u;}{}^v = {\bar U}_{u;}{}^{ri\a} \partial_m U_{ri\a;}{}^v \,,
\label{adhmgaugef}
\ee
where the complex $[N]\times[N+2K]$ matrix $U(x)$ and its hermitian
conjugate ${\bar U}(x)$ satisfy
\ba
&& {\bar U}_{u;}{}^{ri\a} U_{ri\a;}{}^v = \d_u{}^v \, , \\
&& {\bar \D}_{i}{}^{\adot;rj\b} U_{rj\b;}{}^v = 0 \, , \qquad
{\bar U}_{u;}{}^{ri\a} \D_{ri\a;}{}^j{}_\adot = 0 \, .
\label{adhmdata1}
\ea
Equation (\ref{adhmgaugef}) gives a gauge configuration with self-dual
field strength provided the matrices $\D$ and ${\bar \D}$ satisfy
\be
{\bar\D}_i{}^{\adot;rk\b} \D_{rk\b;}{}^j{}_\bdot =
\d^\adot{}_\bdot \left(f^{-1}\right)_i{}^j \, ,
\label{fintro}
\ee
where $f(x)$ is an arbitrary ($x$-dependent) $[K]\times[K]$ hermitian matrix.
From this relation and the definitions (\ref{deltadef}) and
(\ref{deltadefc}) it follows that the coefficients $a$ and $b$ satisfy
the bosonic ADHM constraints
\be
\label{consabar}
\bar a_i{}^{\adot}{}^{;u k\a}\, a_{uk\a;}{}^j{}_{\bdot} =
{\half}(\bar a a)_{i}{}^k\, \delta^\adot_\bdot \, ,
\ee
\be
\label{consabarb}
\bar a_{ui}{}^{\adot;j\a}\, b_{uj\a;k}{}^{\b} =
\epsilon^{\b\g}\, \epsilon^{\adot\bdot}\, \bar b_{i\g;}{}^{uj\a}
\, a_{uj\a;}{}^{k}{}_{\bdot} \, ,
\ee
\be
\label{consbarbb}
\bar b_{i\b;}^{uk\a} \, b_{uk\a;}{}^{j\g} = \half (\bar b b)_{i}{}^{j}
\, \delta_\b^\g \, .
\ee

A choice of special frame allows to put the bosonic parameters in the
form
\be
\label{bspec}
b_{ui\a;}{}^{j\b} = \left(
\begin{array}{c} 0_{u;}{}^{j\b} \\
\delta_{i}{}^{j}\,\delta_\a{}^{\b} \end{array} \right) \,,
\qquad \bar b_{i\a;}{}^{uj\b} =  \left(
\begin{array}{c} 0_{i\a;}{}^{u} \\
\delta_{i}{}^{j} \, \delta_\a{}^{\b} \end{array} \right) \, ,
\ee
\be
\label{aspec}
a_{ui\a;}{}^{j}{}_{\adot} =  \left(
\begin{array}{c} w_{u;}{}^{j}{}_{\adot} \\
(\aap_{\a\adot})_{i}{}^{j} \end{array} \right) \, ,
\qquad \bar a_i{}^{\adot;u j \a} =
\left(\bar w_i{}^{\adot; u} \, , (\bar\aap^{\adot\a})_{i}{}^{j} \right) \, ,
\ee
where $\aap_{\a\adot} = \sigma_{\a\adot}^m \, \aap_m$ and the components
satisfy the matrix constraints
\be
\label{bosconss}
\tr_2(\tc\, \bar a a )= 0 \, , \qquad \aap_n^\dagger = \aap_n \, .
\ee

Similarly, the fermionic variables enter as $[N+2K] \times [K]$
grassmann valued matrices, $\calH$ and $\bar \calH$, that satisfy the
ADHM constraints
\be
\label{calmdef}
\bar \calH_{i;}^{A\, uj\a}\, a_{uj\a;}{}^{k}{}_{\adot} =
- \epsilon_{\adot\bdot} \, \bar a_i{}^{\bdot;uj\a} \,
\calH_{uj\a;}{}^{k}, \qquad \bar\calH^A_{i;}{}^{uj\a} \,
b_{uj\a;}{}^{k\b} = \epsilon^{\b\g} \,
\bar b_{i \g;}{}^{uj\a}\, \calH_{uj\a;}{}^{k} \, ,
\ee
Parametrising the fermionic matrices by
\be
\calH^A_{ui\a;}{}^{j} =  \left(
\begin{array}{c} \nu^{A\,}_{u;}{}^{j}+w_{u;}{}^{k\adot}\,
\xib^A_{\adot k;}{}^{j} \\
\Mp^A_{i\a;}{}^{j} \end{array} \right)  \equiv \left(
\begin{array}{c} \mu^A_{u;}{}^{j} \\
\Mp^A_{i\a;}{}^{j} \end{array} \right)  \, ,
\ee
\be
\label{fermmatb}
\bar \calH^{A\,}_{i;}{}^{u j \a} = \left(
\bar\nu^A_{i;}{}^{u} + \bar{\tilde\xi}^A_{\adot i;}{}^{k} \,
\bar w^{k\adot;u}\, ,\: \Mp^{A\a}_{\:i;}{}^{j} \right)
\equiv \left( \bar \mu^A_{i;}{}^{u} \, , \:
\bar{\calH}^{\prime A}_{i;}{}^{\,j\a} \right) \, .
\ee

The ADHM conditions imply that
\be
\label{wdefs}
\bar w^{\adot}_{k;}{}^{u}\, \nu^A_{u;}{}^{i}=0, \qquad
\bar \nu^A_{i;}{}^{u}\, w_{u;}{}^{k\adot} = 0, \qquad
\bar w^\adot_{i;}{}^{u}\, w_{u;\bdot}{}^{j} =
W^\adot{}_{\bdot i ;}{}^{j} \, ,
\ee
\be
\label{mbarlo}
\bar \calH^{\prime A\ \a}_{i;}{}^{j} \, \delta_{j}{}^{k}\, \delta_\a{}^{\b}
= \epsilon^{\b\g} \, \delta_{i}{}^{j}\, \delta_\g{}^{\a} \,
\Mp^A_{j\a;}{}^{k} \, .
\ee
The latter equation implies
\be
\label{mbarag}
\bar \calH^{\prime A\ \b}_{i;}{}^{k } =\epsilon^{\b\a} \,
\Mp^A_{i\a;}{}^{k} \, .
\ee

The gauge invariant bosonic variables are contained in
\be
\label{gaginvb}
W^\bdot{}_{\adot \, i;}{}^{j} = \bar w^\bdot{}_{i;}{}^{u}\,
w_{u;\adot}{}^{j} \, ,
\ee
which can be written as
\be
\label{gaugdef}
(W^\bdot{}_{\adot})_{i}{}^{j} = \half \, (W^0)_{i}{}^{j} \,
\delta_\adot{}^{\bdot} + \half (W^c)_{i}{}^{j}\, \tc^{\bdot}{}_\adot \, .
\ee
Inverting this expression we have
\be
\label{wodef}
(W^0)_{i}{}^{j} = \tr_2\, W_{i}{}^{j} = (W^\adot{}_{\adot})_{i}{}^{j} \, ,
\qquad (W^c)_{i}{}^{j} = \tr_2\, (\tc W_{i}{}^{j}) = \tau^{c\,\bdot}{}_\adot
\, (W^\adot{}_{\bdot})_i{}^{j} \, .
\ee

The ADHM constraints can then be expressed as
\be
\label{bosadhm}
W^c_{i}{}^{j} + [\aap_m,\aap_n]_{i}{}^{j} \, \tc_\adot^{\ \bdot}\,
\bar\sigma^{mn\ \adot}_{\ \ \bdot} = W^c_{i}{}^{j} - i
[\aap_m,\aap_n]_{i}{}^{j}\, \bar \eta^c_{mn} = 0
\ee
(where $\bar\eta$ is the conjugate 't Hooft symbol) and
\be
\label{fermadhm}
\bar{\tilde \xi}^A_{\adot i}{}^{j} \, W^\adot_{\ \bdot \,j}{}^{k} +
\epsilon_{\bdot\gdot} \, W^\gdot_{\ \adot \,i}{}^{j} \,
\xib^{\adot A}_{j}{}^{k} + \left[\Mp^{A\a},
\aap_{\a\bdot} \right]_{i}{}^{k} = 0 \, .
\ee

\section{Component fields in the one-instanton background}
\label{compfields}

In this appendix we summarise the expressions for the elementary
fields in the $\N4$ SYM multiplet in terms of the ADHM variables, in
particular we give explicit formulae in the one-instanton sector.  All
the fields in the multiplet are in the adjoint representation of the
gauge group $SU(N)$ and can be represented as $[N]\times[N]$
matrices.

The (self-dual part of the) gauge field strength $F^-_{mn}$ in the
instanton background follows from the construction of the
gauge field $A_m$ in the previous section. The result is
\be
\left(F^{mn}\right)_{u;}{}^{v} = {\bar U}_{u;}{}^{ri\a}
b_{ri\a;}{}^{j\b} \s^{mn}{}_{\b}{}^{\g} f_{j;}{}^{k}
{\bar b}_{k\g;}{}^{sl\d}U_{sl\d;}{}^{v} \, .
\label{fieldstre}
\ee

In the semiclassical approximation the Weyl fermions $\l^A_\a$ are
replaced by the solutions to the zero-eigenvalue Dirac equation in the
instanton background
\be
\Dsm \l^A_\a = 0 \, .
\label{direq}
\ee
In terms of the ADHM variables  the classical field $\l^A_\a$ is
\be
\left(\l^A_\a\right)_{u;}{}^{v} = {\bar U}_{u;}{}^{ri\b}
\left(\calH^A_{ri\b;}{}^{j} f_{j;}{}^{k} {\bar b}_{k\a;}{}^{sl\g}
- \veps_{\a\d} b_{ri\b;}{}^{j\d} f_{j;}{}^{k}
{\bar\calH}^A_{k;}{}^{sl\g}\right) U_{sl\g;}{}^{v}
\label{gaugino}
\ee
and it can be checked that it satisfies (\ref{direq}).

Analogously the scalar fields $\v^{AB}$ are replaced by the solution
to their equation of motion in the instanton background at leading
order in the Yang--Mills coupling
\be
D^2 \v^{AB} = \frac{i}{\sqrt{2}} [\l^A,\l^B] \, .
\label{scaleq}
\ee
The solution was constructed in \cite{dkm}. In the $\N4$ SYM
theory in the superconformal phase of interest here the solution can
be written in the form
\ba
i\v^{AB} &=& \half {\bar U}_{u;}{}^{ri\a}
\left(\calH^B_{ri\a;}{}^{j} f_{j;}{}^{k}
{\bar\calH}^A_{k;}{}^{sl\b} - \calH^A_{ri\a;}{}^{j} f_{j;}{}^{k}
{\bar\calH}^B_{k;}{}^{sl\b} \right) U_{sl\g;}{}^{v}  \nn \\
&+& \half{\bar U}_{u;}^{ri\a} \left(
\begin{array}{cc} 0_{r;}{}^{s} & 0_{r;}{}^{j\b} \\
0_{i\a;}{}^{s} & \calA^{AB}_{i;}{}^{j} \d_\a^\b
\end{array} \right) U_{sj\b;}{}^{v} \, ,
\label{scalar}
\ea
where the $[K]\times[K]$ matrix $\calA^{AB}_{i;}{}^{j}$ is defined by
\be
{\bf L} \calA^{AB} = \Lambda^{AB} \, .
\label{eqcalA}
\ee
In this equation $\Lambda^{AB}$ is a bilinear in the fermionic
collective coordinates, namely
\be
\Lambda^{\!AB}_{\ i;}{}^{j} =
\frac{1}{\sqrt{2}} \left({\bar\calH}^A_{i;}{}^{rk\a} \calH^B_{rk\a;}{}^{j}
- {\bar\calH}^B_{i;}{}^{rk\a} \calH^A_{rk\a;}{}^{j} \right)
\label{defLambda}
\ee
and the linear operator ${\bf L}$ acting in the space of
$[K]\times[K]$ Hermitian matrices is given by
\be
{\bf L} X =
\frac{1}{2} \{X,W^0\} + [a^\prime_m,[a^\prime_m,X]] \, ,
\label{defopL}
\ee
with $W^0$ defined in (\ref{wodef}).

The gauge invariant composite operators we are interested in are
traces over colour indices of products of elementary fields. In
evaluating correlation functions of such operators in the
semiclassical approximation in the instanton background one must then
compute expressions of the form
\be
{\cal O} = \Tr_N \left({\bar U}{\tilde F}U
\ldots {\bar U}{\tilde\l} U \ldots {\bar U}{\tilde\v} U
\ldots\right) \, .
\ee
It is then convenient to rewrite such
expressions as traces over $[N+2K]\times[N+2K]$ matrices in the
following way
\ba
{\cal O} &=& \Tr_N \left({\bar U}{\tilde F}U \ldots
{\bar U}{\tilde\l} U \ldots {\bar U}{\tilde\v} U \ldots\right)  \nn \\
&=& \Tr_{N+2K}\left[ ({\cal P} {\tilde F}) \ldots ({\cal
P}{\tilde\l}) \ldots ({\cal P}{\tilde\v}) \ldots \right] \, ,
\label{projtrace}
\ea
where we have defined the projection operator
\be
{\cal P}_{ui\a;}{}^{vj\b} = U_{ui\a;}{}^{r} {\bar U}_{r;}{}^{vj\b} =
\d_{u;}{}^{v}\d_{i;}{}^{j}\d_{\a}^{\b} - \D_{ui\a;}{}^{k\g} f_{k;}{}^{l}
\Db_{l\g;}{}^{vj\b} \, ,
\label{proj}
\ee
where $\D$ and $\Db$ are the matrices of bosonic ADHM variables
defined in the previous section.

Now consider the one-instanton ($K=1$) sector in more detail. In this
case the only collective coordinates on which the classical
expressions for the fields can depend are the position, size and gauge
orientation of the instanton and the corresponding superpartners, that
is the 16 exact zero-modes $\zeta^A_\a = (\rho\eta^A_\a - y_{\a\adot}
\xi^{\adot A})/\sqrt{\rho}$ and the $\nu^A_u$ and ${\bar\nu}^{Au}$
modes. All the previous formulae simplify drastically.  The bosonic
collective coordinates are related to the ADHM variables as
follows. The components of the  vector $a^\prime_m$ correspond to the
coordinates of the instanton position
\be
a^\prime_m = -x_{0\,m} \, .
\ee
The gauge invariant combination
\be
{\bar w}^{\adot;u} w_{u;\bdot} = \rho^2 \d^\adot_\bdot
\ee
gives the instanton size $\rho$.

The $8N$ fermionic zero-modes are collected in the ADHM matrices
$\calH$ and ${\bar\calH}$
\be
\calH^A_{\a u} = \left( \begin{array}{c}
4w_{u;\adot} {\bar\xi}^{\adot A} + \nu^A_u \\
4\eta^A_\a \end{array} \right) \; ,
\qquad {\bar\calH}^{A\,u\a} = \left( -4{\bar\xi}^A_\adot
{\bar w}^{\adot;u}+{\bar\nu}^{A\,u} \; , \; 4\eta^{\a A} \right) \, .
\label{oneinstfermi}
\ee
Here $\eta^A_\a$ and $\bar\xi^A_\adot$ denote the 16 exact
zero-modes.

The matrix $f_{i;}{}^{j}$ reduces to the function
\be
f = f(x;x_0,\rho) =
\frac{1}{(x-x_0)^2 + \rho^2} = \frac{1}{y^2 + \rho^2} \, .
\label{foneinst}
\ee
The operator ${\bf L}$ in (\ref{eqcalA}) is simply
\be
{\bf L} = 2 \rho^2 \, .
\ee
Finally the projector $\calP$ becomes
\be
\calP_{u\a;}{}^{v\b} = \d_{u;}{}^{v}\d_\a^\b -
\frac{1}{y^2+\rho^2} \left(
\begin{array}{cc} w_{u;\adot} {\bar w}^{\adot;v}
& \quad w_{u;\adot} y^{\adot \b} \\
y_{\a\adot}{\bar w}^{\adot;v} & \quad
y^2 \d_\a^\b \end{array} \right) \, .
\label{projoneinst}
\ee
Notice in particular that it satisfies
\be
\Tr_{N+2}\left[\calP(x)\right] = N \, .
\label{trproj}
\ee
The calculations of correlation functions in which pairs of scalar
fields are contracted also involve the traces
\ba
&& \Tr_{N+2}\left[\calP(x_1)\calP(x_2)\calP(x_1)\calP(x_2)\right]
= N + \frac{2\rho^4(x_1-x_2)^4}{(y_1^2+\rho^2)^2(y_2^2+\rho^2)^2}
- \frac{4\rho^2(x_1-x_2)^2}{(y_1^2+\rho^2)(y_2^2+\rho^2)}
\nn \\
&& \Tr_{N+2}\left[\calP(x_1)\calP(x_2)\right] =
N - \frac{2\rho^2(x_1-x_2)^2}{(y_1^2+\rho^2)(y_2^2+\rho^2)}
\nn \\
&& \Tr_2 \left[ \bar b\calP(x_1)\calP(x_2)b\bar b\calP(x_2)
\calP(x_1)b\right] = \frac{2\rho^4}{[(x_1-x_0)^2+\rho^2]
[(x_2-x_0)^2+\rho^2]} \label{tr24proj} \\
&& \hspace*{5.7cm} -\frac{2\rho^6(x_1-x_2)^2}{[(x_1-x_0)^2+\rho^2]^2
[(x_2-x_0)^2+\rho^2]^2} \nn \\
&& \Tr_2 \left[ \bar b\calP(x_1)\calP(x_2)\calP(x_1)b\right] =
\frac{2\rho^2}{[(x_1-x_0)^2+\rho^2]}
-\frac{2\rho^4(x_1-x_2)^2}{[(x_1-x_0)^2+\rho^2]^2
[(x_2-x_0)^2+\rho^2]} \nn \\
&& \Tr_2 \left[ \bar b\calP(x)b\bar b\calP(x)b\right]=
\frac{2\rho^4}{[(x-x_0)^2+\rho^2]^2} \nn \\
&& \Tr_2 \left[\bar b\calP(x)b\right] =
\frac{\rho^2}{[(x-x_0)^2+\rho^2]} \nn \, .
\ea
As observed above the gauge invariant composite operators can be
expressed as traces over $[N+2]\times[N+2]$ matrices, we give then the
formulae for the `projected' $[N+2]$-dimensional matrices of the
elementary fields. For this purpose we introduce the following
notation for a generic Yang--Mills field $A$
\ba
A_{u;}{}^{v} &=& {\bar U}_{u;}{}^{r\b} \tilde{A}_{r\b;}{}^{s\g}
U_{s\g;}{}^{v} \nn \\
{\hat A}_{u\a;}{}^{v\b} &=& \calP_{u\a;}{}^{r\g}
\tilde{A}_{r\g;}{}^{v\b} \, .
\label{hatfield}
\ea
For the field strength $F_{mn}$ we get
\ba
({\hat F}_{mn})_{u\a;}{}^{v\b} = \left(
\begin{array}{cc} 0 & \quad
\left({\hat F}^{(1)}_{mn}\right)_{u;}{}^{\b} \\
0 & \quad \left({\hat F}^{(2)}_{mn}\right)_{\a;}{}^{\b}
\end{array} \right) \, ,
\label{hatF}
\ea
where
\be
\left({\hat F}^{(1)}_{mn}\right)_{u;}{}^{\b} = -
\frac{4}{(y^2+\rho^2)^2} w_{u;\adot} y^{\adot\g} \s_{mn\g}{}^{\b} \: ,
\qquad \left({\hat F}^{(2)}_{mn}\right)_{\a;}{}^{\b} =
\frac{4}{(y^2+\rho^2)^2} \rho^2 \s_{mn\a}{}^{\b} \: .
\label{hatFcomp}
\ee
The classical value of the fermions $\l^A_\a$ is
\be
({\hat\l}^A_{\a})_{u\b;}{}^{v\g} = \left( \begin{array}{cc} \left(
{\hat\l}^{(1)A}_\a \right)_{u;}{}^{v} & \quad
\left({\hat\l}^{(2)A}_{\a}\right)_{u;}{}^{\g} \\
\left({\hat\l}^{(3)A}_\a \right)_{\b;}{}^{v} & \quad
\left({\hat\l}^{(4)A}_{\a}\right)_{\b;}{}^{\g} \end{array} \right) \, ,
\label{hatlambda}
\ee
where
\ba
\left({\hat\l}^{(1)A}_\a \right)_{u;}{}^{v} &=&
\frac{1}{(y^2+\rho^2)^2} \veps_{\a\d} w_{u;\adot} y^{\adot \d}
(-4{\bar\xi}^A_\bdot {\bar w}^{\bdot;v}+{\bar\nu}^{A\,v}) \nn \\
\left({\hat\l}^{(2)A}_\a \right)_{u;}{}^{\g} &=&
\frac{1}{(y^2+\rho^2)^2} \left[4\left(y^2
w_{u;\adot}{\bar\xi}^{\adot A} \d_\a^\g -
w_{u;\adot}y^{\adot\d}\eta^A_\d \d_\a^\g +
\veps_{\a\d}w_{u;\adot}y^{\adot\d}
\eta^{\g A} \right) \right. \nn \\
&+& \left. (y^2+\rho^2)\nu^A_u\d_\a^\g \right] \nn \\
\left({\hat\l}^{(3)A}_\a \right)_{\b;}{}^{v} &=&
\frac{\rho^2}{(y^2+\rho^2)^2} \veps_{\a\b}
(4{\bar\xi}^A_\adot {\bar w}^{\adot;v} -
{\bar\nu}^{A\,v}) \nn \\
\left({\hat\l}^{(4)A}_\a \right)_{\b;}{}^{\g} &=&
\frac{4\rho^2}{(y^2+\rho^2)^2} \left(-y_{\b\adot} {\bar\xi}^{\adot A}
\d_\a^\g + \eta^A_\b \d_\a^\g - \veps_{\a\b}\eta^{\g A} \right) \, .
\label{hatlambcomp}
\ea
Finally the scalar field $\v^{AB}$ reads
\be
(i{\hat\v}^{AB})_{u\b;}{}^{v\g} = \left( \begin{array}{cc} \left(
{\hat\v}^{(1)AB} \right)_{u;}{}^{v} & \quad
\left({\hat\v}^{(2)AB}\right)_{u;}{}^{\g} \\
\left({\hat\v}^{(3)AB}\right)_{\b;}{}^{v} & \quad
\left({\hat\v}^{(4)AB}\right)_{\b;}{}^{\g} \end{array} \right) \, ,
\label{hatscalar}
\ee
where
\ba
\left({\hat\v}^{(1)AB}\right)_{u;}{}^{v} &\!\!=\!\!&
\frac{1}{4(y^2+\rho^2)^2} \left\{ y^2 \left[ -16
({\bar\xi}^{\adot B}{\bar\xi}^A_\bdot - {\bar\xi}^{\adot A}
{\bar\xi}^B_\bdot)w_{u;\adot}{\bar w}^{\bdot;v}
\right. \right. \nn \\
&\!\!+\!\!& \left. 4 w_{u;\adot}
({\bar\xi}^{\adot B} {\bar\nu}^{A\,v}-{\bar\xi}^{\adot A}
{\bar\nu}^{B\,v}) \right] \nn \\
&\!\!+\!\!&(y^2+\rho^2) \left[ -4({\bar\xi}^B_\adot
\nu^A_u-{\bar\xi}^A_\adot \nu^B_u){\bar w}^{\adot;v} +
(\nu^B_u{\bar\nu}^{A\,v} - \nu^A_u{\bar\nu}^{B\,v})\right]  \nn \\
&\!\!+\!\!& \left. y^{\adot\d}\left[ 16(\eta^B_\d{\bar\xi}^A_\bdot -
\eta^A_\d{\bar\xi}^B_\bdot)w_{u;\adot}{\bar w}^{\bdot;v} -
4w_{u;\adot} (\eta^B_\d {\bar\nu}^{Av} - \eta^A_\d {\bar\nu}^{Bv})
\right]\right\} \nn \\
\left({\hat\v}^{(2)AB}\right)_{u;}{}^{\g} &\!\!=\!\!&
\frac{1}{4(y^2+\rho^2)^2} \left\{ 16y^2
w_{u;\adot}({\bar\xi}^{\adot B}\eta^{\g A} - {\bar\xi}^{\adot A}
\eta^{\g B}) + 4 (y^2+\rho^2)(\nu^B_u \eta^{\g A} - \nu^A_u
\eta^{\g B})  \rule{0pt}{18pt} \right. \nn \\
&\!\!-\!\!& \left. w_{u;\adot} \left[
16y^{\adot\d} (\eta^B_\d \eta^{\g A} - \eta^A_\d \eta^{\g B}) +
\frac{1}{2} \frac{y^2+\rho^2}{\rho^2} y^{\adot\g} ({\bar\nu}^{Au}
\nu^B_u - {\bar\nu}^{Br} \nu^A_r) \right] \right\} \nn \\
\left({\hat\v}^{(3)AB}\right)_{\b;}{}^{v} &=&
\frac{1}{4(y^2+\rho^2)^2}\left\{ \rho^2
\left[ 16y_{\b\adot}({\bar\xi}^{\adot B}{\bar\xi}^A_\bdot
- {\bar\xi}^{\adot A}{\bar\xi}^B_\bdot) {\bar w}^{\bdot;v}
- 4y_{\b\adot} ({\bar\xi}^{\adot B} {\bar\nu}^{Av} -
{\bar\xi}^{\adot A} {\bar\nu}^{Bv} )  \right. \right. \nn \\
&\!\!-\!\!& \left.\left. 16 (\eta^B_\b{\bar\xi}^A_\adot -
\eta^A_\b{\bar\xi}^B_\adot) {\bar w}^{\adot;v} + 4(\eta^B_\b
{\bar\nu}^{Av} - \eta^A_\b {\bar\nu}^{Bv}) \right]\right\} \nn \\
\left({\hat\v}^{(4)AB}\right)_{\b;}{}^{\g} &=&
\frac{\rho^2}{4(y^2+\rho^2)^2} \left[
-16y_{\b\adot}({\bar\xi}^{\adot B}\eta^{\g A} - {\bar\xi}^{\adot B}
\eta^{\g A}) +16 (\eta^B_\b \eta^{\g A} - \eta^B_\b \eta^{\g A})
\right.  \nn \\
&\!\!+\!\!& \left. \frac{1}{2}\frac{y^2+\rho^2}{\rho^2}
\delta_\b^\g ({\bar\nu}^{Ar} \nu^B_r - {\bar\nu}^{Br} \nu^A_r)
\right] \, .
\label{hatscalcomp}
\ea

\section{Instanton profiles of composite operators}
\label{appinsprofiles}

Here we will summarise the classical expressions (i.e., leading
order in $\gy$) for some of the gauge invariant composite operators
of the Yang--Mills theory in an instanton background. Terms which
potentially contain additional fermionic modes of the $\bar\nu$ and
$\nu$ type are omitted.

\vspace*{0.3cm}
\noindent
{\bf The supercurrent multiplet}

\vspace*{0.2cm}
\noindent
In this subsection we will give the explicit instanton profiles of some
of the composite operators that form the supermultiplet that contains
the energy-momentum tensor and the other currents.

Using the results of appendix \ref{compfields} the top component of
the supercurrent multiplet is given by
\be
\calC = \Tr_{N+2}
\left({\hat F}_{mn} {\hat F}^{mn} \right) = \Tr_2 \left({\hat
F}^{(2)}_{mn} {\hat F}^{(2)mn} \right) = - \frac{96
\rho^4}{[(x-x_0)^2+\rho^2]^4} \: .
\label{explcalC}
\ee
The fermionic composite operator $\hat\Lambda^A_\a$  is given by
\ba
&& \hat\Lambda^A_\a =
\Tr_{N+2} \left( {\hat F}_{mn} \sigma^{mn}{}_{\a}{}^{\b}
{\hat\lambda}^A_\b \right) = \sigma^{mn}{}_{\a}{}^{\b} \left\{
\Tr_N\left({\hat F}^{(1)}_{mn} {\hat \lambda}^{(3)A}_\b \right) +
\Tr_2\left({\hat F}^{(2)}_{mn} {\hat \lambda}^{(4)A}_\b \right)
\right\} \nn \\
&& = - \frac{96 \rho^4}{[(x-x_0)^2+\rho^2]^4} \, \zeta^A_\a \: .
\label{explLamb}
\ea
The two composite operators associated with the internal and external
components of the complex type IIB two-form are respectively
$\calE^{AB}$ and $\calB^{AB}_{mn}$.  Their classical expressions
in the one-instanton sector are
\ba
&& \calE^{(AB)} = \Tr_{N+2}
\left({\hat\lambda}^{\a A}{\hat\lambda}^B_\a \right) \nn \\
&& = \Tr_N\left( {\hat\lambda}^{(1)\a A}
{\hat\lambda}^{(1)B}_\a + {\hat\lambda}^{(2)\a
A}{\hat\lambda}^{(3)B}_\a \right) + \Tr_2 \left( {\hat\lambda}^{(3)\a
A} {\hat\lambda}^{(2)B}_\a + {\hat\lambda}^{(4)\a A}
{\hat\lambda}^{(4)B}_\a \right) \nn \\
&& = -\frac{96
\rho^4}{[(x-x_0)^2+\rho^2]^4} \, \zeta^{\a A} \zeta^B_\a -
\frac{2\rho^2}{[(x-x_0)^2+\rho^2]^3} \left( {\bar\nu}^{Au} \nu^B_u +
{\bar\nu}^{Bu} \nu^A_u \right) \:
\label{explE}
\ea
and
\ba && \calB^{[AB]}_{mn} = \Tr_{N+2}\left( {\hat\lambda}^{\a A}
\sigma_{mn\a}{}^{\b} {\hat\lambda}^B_\b + 2i {\hat\v}^{AB} {\hat
F}_{mn} \right) \nn \\
&& = \sigma_{mn\a}{}^{\b} \left\{ \Tr_N\left(
{\hat\lambda}^{(1)\a A} {\hat\lambda}^{(1)B}_\b + {\hat\lambda}^{(2)\a
A}{\hat\lambda}^{(3)B}_\b \right) + \Tr_2 \left( {\hat\lambda}^{(3)\a
A} {\hat\lambda}^{(2)B}_\b + {\hat\lambda}^{(4)\a A}
{\hat\lambda}^{(4)B}_\b \right) \right\} \nn \\
&& + 2i \, \Tr_2
\left( {\hat\v}^{(3)AB} {\hat F}^{(1)}_{mn} + {\hat\v}^{(4)AB} {\hat
F}^{(2)}_{mn} \right) \nn \\
&& = - \frac{96 \rho^4}{[(x-x_0)^2+\rho^2]^4} \,
\zeta^{\a A} \sigma_{mn\a}{}^{\b} \zeta^B_\b \: ,
\label{explB}
\ea
As already pointed out we have kept only terms that are relevant for our
calculations at leading order in the semiclassical approximation. In
particular we have neglected the term cubic in the scalar fields in
$\calE^{AB}$ \cite{bgkr}, which plays a crucial r\^ole in perturbative
calculations, \eg in the proof of the non-renormalisation of three-point
functions \cite{dfs}.

The other spinor in the multiplet that we consider is
$\calX^{A_1[A_2A_3]}_{\a}$ and its classical expression is
\ba
&& \calX^{A_1[A_2A_3]}_{\a} =
\Tr_{N+2} \left(2\hat\lambda^{A_1}_\a {\hat\v}^{A_2A_3}
+ \hat\lambda^{A_2}_\a {\hat\v}^{A_1A_3} - \hat\lambda^{A_3}_\a
{\hat\v}^{A_1A_2}  \right) \nn \\
&& = 2 \left\{ \Tr_N\left(\hat\lambda^{(1)A_1}_\a {\hat\v}^{(1)A_2A_3}
+ {\hat\lambda}^{(2)A_1}_\a {\hat\v}^{(3)A_2A_3} \right)
+ \Tr_2\left( {\hat\lambda}^{(3)A_1}_\a {\hat\v}^{(2)A_2A_3}
+ {\hat\lambda}^{(4)A_1}_\a {\hat\v}^{(4)A_2A_3} \right) \right.
\nn \\
&& \left. + \half (A_1 \longleftrightarrow A_2) + \half
 (A_1 \longleftrightarrow A_3) \right\} \nn \\
&& = \frac{96\rho^4}{[(x-x_0)^2+\rho^2]^4}
\left(\zeta^{A_3}_\a \zeta^{\b A_1}\zeta^{A_2}_\b
- \zeta^{A_2}_\a \zeta^{\b A_1}\zeta^{A_2}_\b \right)
\label{explChi} \\
&& + \frac{3\rho^2}{[(x-x_0)^2+\rho^2]^3}
\left[ \zeta^{A_2}_\a ({\bar\nu}^{A_1u} \nu^{A_3}_u +
{\bar\nu}^{A_3u} \nu^{A_1}_u) - \zeta^{A_3}_\a ({\bar\nu}^{A_1u}
\nu^{A_2}_u + {\bar\nu}^{A_2u} \nu^{A_1}_u) \right] \, . \nn
\ea
The lowest component of the multiplet is the chiral
primary operator $\calQ_2$. In the one-instanton background
this is given by
\ba
&&
\left(\calQ_2\right)^{[A_1B_1][A_2B_2]} = \Tr_{N+2} \left(
2{\hat\v}^{A_1B_1} {\hat\v}^{A_2B_2} +{\hat\v}^{A_1A_2}
{\hat\v}^{B_1B_2} + {\hat\v}^{A_1B_2} {\hat\v}^{A_2B_1} \right) \nn \\
&& = 2\left\{ \Tr_N\left(
{\hat\v}^{(1)A_1B_1} {\hat\v}^{(1)A_2B_2} + {\hat\v}^{(2)A_1B_1}
{\hat\v}^{(3)A_2B_2}
\right) + \Tr_2 \left( {\hat\v}^{(3)A_1B_1} {\hat\v}^{(2)A_2B_2}
\right. \right. \nn \\
&& \left. \left. + {\hat\v}^{(4)A_1B_1} {\hat\v}^{(4)A_2B_2} \right)
+ \half (B_1 \longleftrightarrow A_2) + \half
(B_1 \longleftrightarrow B_2) \right\} \nn \\
&& = \frac{96 \rho^4}{[(x-x_0)^2+\rho^2]^4} \left(
\zeta^{\a A_1}\zeta^{A_2}_\a \zeta^{\b B_1}\zeta^{B_2}_\b -
\zeta^{\a A_1}\zeta^{B_2}_\a \zeta^{\b B_1}\zeta^{A_2}_\b
\right) \nn \\
&& + \frac{3\rho^2}{[(x-x_0)^2+\rho^2]^3} \left[
\zeta^{\a A_1}\zeta_\a^{A_2} ({\bar\nu}^{B_1u}\nu^{B_2}_u +
{\bar\nu}^{B_2u}\nu^{B_1}_u) - \zeta^{\a A_1}\zeta_\a^{B_2}
({\bar\nu}^{B_1u}\nu^{A_2}_u + {\bar\nu}^{A_2u}\nu^{B_1}_u)
\right. \nn \\
&&\left.  - \zeta^{\a B_1}\zeta^{A_2}_\a ({\bar\nu}^{A_1u}\nu^{B_2}_u
+ {\bar\nu}^{B_2 u}\nu^{A_1}_u) + \zeta^{\a B_1}\zeta^{B_2}_\a
({\bar\nu}^{A_1u}\nu^{A_2}_u + {\bar\nu}^{A_2u}\nu^{A_1}_u)
\right. \rule{0pt}{18pt} \nn \\
&& + \frac{3}{32[(x-x_0)^2+\rho^2]^2}
\left[({\bar\nu}^{A_1u}\nu^{A_2}_u + {\bar\nu}^{A_2u}\nu^{A_1}_u)
({\bar\nu}^{B_1v}\nu^{B_2}_v + {\bar\nu}^{B_2v}\nu^{B_1}_v)
\right. \nn \\
&& \left. - ({\bar\nu}^{A_1u}\nu^{B_2}_u + {\bar\nu}^{B_2u}\nu^{A_1}_u)
({\bar\nu}^{B_1v}\nu^{A_2}_v + {\bar\nu}^{A_2v}\nu^{B_1}_v) \right]
\rule{0pt}{18pt}  \, .
\label{explO2}
\ea

\vspace*{0.3cm}
\noindent
{\bf Kaluza--Klein excitations}

\vspace*{0.2cm}
\noindent
Two specific examples will now be given of the composite Yang--Mills
theory operators that correspond to Kaluza--Klein excited states on
the $AdS$ side. These examples illustrate the general structure of
these operators and in particular the way they depend on the
$\bar\nu^{Au}$ and $\nu^A_u$ modes.  The simplest case is the operator
corresponding to the first excited  Kaluza--Klein mode of the dilaton,
$\calC_{\bf 6}$, which is in the ${\bf 6}$ of $SU(4)$.  Using the
formulae in appendix~\ref{compfields} for the elementary fields we
obtain
\ba
&& (\calC^{[AB]})_{\bf 6} = \frac{1}{(\gy^2N)^{3/2}}\Tr_{N+2}
\left({\hat F}^{-\,2} {\hat\v}^{AB}\right) \nn \\
&& = \frac{1}{(\gy^2N)^{3/2}} \left\{
\Tr_N\left({\hat F}^{(1)}_{mn} {\hat F}^{(2)mn}
{\hat\v}^{(3)AB} \right) + \Tr_2\left( {\hat F}^{(2)}_{mn}
{\hat F}^{(2)mn} {\hat\v}^{(4)AB} \right)\right\} \nn \\
&& = -\frac{12}{(\gy^2N)^{3/2}}\frac{\rho^4}{[(x-x_0)^2+\rho^2]^5}
({\bar\nu}^{Au} \nu^B_u - {\bar\nu}^{Bu} \nu^A_u) \, .
\label{expl6dil}
\ea

The second explicit example is the the fermionic operator
$(\hat \Lambda)_{\bf 20^*}$ in   (\ref{Lam20newdef}).  This soaks up
three fermionic zero modes in the instanton background and, unlike the
corresponding operator in the supercurrent multiplet,  depends on the
additional modes $\bar\nu$ and $\nu$ in the combination $\nsix$.
The explicit expression is
\ba
&& \!\! \hat\Lambda^{A_1(A_2A_3)}_\a =
\frac{1}{(\gy^2N)^{3/2}} \left\{ \Tr \left[ 2 \lambda^{A_1}_\a
\left( \lambda^{\b\,A_2}\lambda_\b^{A_3} + \lambda^{\b\,A_3}
\lambda_\b^{A_2} \right) + \lambda^{A_2}_\a \left(
\lambda^{\b\,A_1}\lambda_\b^{A_3} + \lambda^{\b\,A_3}
\lambda_\b^{A_1} \right) \right. \right. \nn \\
&& \!\! \left. + \lambda^{A_3}_\a \left(
\lambda^{\b\,A_1}\lambda_\b^{A_2} + \lambda^{\b\,A_2}
\lambda_\b^{A_1} \right) \right] \nn \\
&& \!\! \left. + \Tr \left[ F_{mn}
\s^{mn \, \b}_{\,\a} \left( \left\{ \lambda^{A_2}_\b, \v^{A_1A_3}
\right\} + \left\{ \lambda^{A_3}_\b, \v^{A_1A_2} \right\} \right)
\right] + \cdots \right\} \label{expl20ferm} \\
&& \!\! = \frac{24}{(\gy^2N)^{3/2}} \frac{\rho^4}{[(x-x_0)^2+\rho^2]^5}
\left[\zeta^{A_2}_\a(\nub^{A_1u}\nu^{A_3}_u-\nub^{A_3u}\nu^{A_1}_u)
+\zeta^{A_3}_\a(\nub^{A_1u}\nu^{A_2}_u-\nub^{A_2u}\nu^{A_1}_u)
\right] \, . \nn
\ea

The higher dimensional chiral primary operators, $\calQ_3$ and
$\calQ_4$, are more complicated, but they are expected to
have the schematic form described in section \ref{nukaluza}.

\section{Scalar contractions in the instanton background}
\label{scalcontr}

In this appendix we calculate some of the single and double
contractions between pairs of scalar fields which enter in the
calculations of sections \ref{1instcorr}, \ref{1instcorrKK} and
\ref{othercorr}.

Let us consider first the single contraction in the contribution $\hat
G^{\rm (b)}_{\hat\Lambda^{16}\calQ_2^2}$ to the correlation function
studied in section \ref{1instcorr}. Using the notation of appendix
\ref{compfields} we have
\ba
&& \Tr\left(\v^{B_1C_1}\v^{D_1E_1}\right)(y_1) \,
\Tr\left(\v^{B_2C_2}\v^{D_2E_2}\right)(y_2)
\raisebox{-10pt}{\hspace*{-6.6cm}
\rule{0.4pt}{4pt}\rule{3.65cm}{0.4pt}\rule{0.4pt}{4pt}}
\hspace*{2.85cm} \nn \\
&& = \calP_{u_1\a_1}{}^{v_1\b_1}(y_1)
\tilde{\v}_{v_1\b_1}^{B_1C_1 \, r_1\g_1}(y_1) \hat{\v}_{r_1\g_1}^{D_1E_1
\,u_1\a_1}(y_1) \calP_{u_2\a_2}{}^{v_2\b_2}(y_2)
\tilde{\v}_{v_2\b_2}^{B_2C_2 \, r_2\g_2}(y_2) \hat{\v}_{r_2\g_2}^{D_2E_2
\,u_2\a_2}(y_2) \raisebox{-10pt}{\hspace*{-11.6cm}
\rule{0.4pt}{4pt}\rule{6.9cm}{0.4pt}\rule{0.4pt}{4pt}}
\hspace*{4.7cm}  \nn \\
&& = \frac{\gy^2 \veps^{B_1C_1B_2C_2}}{2\pi^2} \left\{
\frac{1}{(y_1-y_2)^2} \Tr_{N+2} \left[ \calP(y_2) \hat\v^{D_1E_1}(y_1)
\calP(y_1)\hat\v^{D_2E_2}(y_2) \right] \right. \nn \\
&& \left. + \frac{1}{2\rho^2} \Tr_{N+2}\left[ \calP(y_1) \bar b b
\hat\v^{D_1E_1}(y_1) \right] \Tr_{N+2}\left[ \calP(y_2) \bar b b
\hat\v^{D_2E_2}(y_2) \right] \right\} \, .
\label{onecontr}
\ea
In the second line of (\ref{onecontr}) the original $[N]\times[N]$
traces have been rewritten in terms of $[N+2]$ dimensional traces and
in the third line the expression for the propagator given in
(\ref{propfin}) has been used together with the fact that $\calP$ is a
projector, \ie $\calP^2=\calP$. The first of the last two terms in
(\ref{onecontr}) comes from the term ${\tilde
B}[_{r\a,s\b}\hspace*{-21pt}^{u\g,v\d}] (y_1,y_2)$ in the propagator
(\ref{propfin}) while the second one arises from the non singular
term ${\tilde C} [_{r\a,s\b}\hspace*{-21pt}^{u\g,v\d}](y_1,y_2)$. Only
the terms in the first and fourth lines in the expression for the
propagator (\ref{propfin}) contribute, because the other terms produce
traces over single $\v$ fields, which vanish since the fields are in
$SU(N)$.  Computing the traces in (\ref{onecontr}) we obtain
\ba
&& \Tr\left(\v^{B_1C_1}\v^{D_1E_1}\right)(y_1) \,
\Tr\left(\v^{B_2C_2}\v^{D_2E_2}\right)(y_2)
\raisebox{-10pt}{\hspace*{-6.6cm}
\rule{0.4pt}{4pt}\rule{3.65cm}{0.4pt}\rule{0.4pt}{4pt}}
\hspace*{2.85cm} \nn \\
&& = \frac{\gy^2\veps^{B_1C_1B_2C_2}}{32\pi^2(y_1-y_2)^2}
\frac{1}{[(y_1-x_0)^2+\rho^2][(y_2-x_0)^2+\rho^2]} \nn \\
&& \left[ (\bar\nu^{[D_1}\nu^{D_2]})(\bar\nu^{[E_2}\nu^{E_1]})
-(\bar\nu^{[D_1}\nu^{E_2]})(\bar\nu^{[D_2}\nu^{E_1]})
-(\bar\nu^{[D_1}\nu^{E_1]})(\bar\nu^{[D_2}\nu^{E_2]}) \right. \nn \\
&& \left.
+(\bar\nu^{(D_1}\nu^{D_2)})(\bar\nu^{(E_1}\nu^{E_2)}) -
(\bar\nu^{(D_1}\nu^{E_2)})(\bar\nu^{(D_2}\nu^{E_1)}) \right]
\, . \label{onecontr2}
\ea
Notice that in this expression a contact term of the form
\be
\frac{\gy^2\veps^{B_1C_1B_2C_2}\rho^2}{32\pi^2
[(y_1-x_0)^2+\rho^2]^2[(y_2-x_0)^2+\rho^2]^2}
(\nub^{[D_1}\nu^{E_1]})(\nub^{[D_1}\nu^{E_1]})
\label{contcanc}
\ee
is cancelled when the contributions of the terms ${\tilde
B}[_{r\a,s\b}\hspace*{-21pt}^{u\g,v\d}] (y_1,y_2)$ and ${\tilde C}
[_{r\a,s\b}\hspace*{-21pt}^{u\g,v\d}](y_1,y_2)$ in the propagator are
combined.

The double contraction that enters in $\hat  G^{\rm
(c)}_{\hat\Lambda^{16}\calQ_2^2}$ is more complicated because all the
terms in the propagator (\ref{propfin}) give a non-vanishing
contribution. The result of the double contraction in terms of ADHM
matrices reads
\ba
&& \Tr\left(\v^{B_1C_1}\v^{D_1E_1}\right)(y_1) \,
\Tr\left(\v^{B_2C_2}\v^{D_2E_2}\right)(y_2)
\raisebox{-15pt}{\hspace*{-6.6cm}
\rule{0.4pt}{10pt}\rule{4.6cm}{0.4pt}\rule{0.4pt}{10pt}}
\raisebox{-10pt}{\hspace*{-3.75cm}
\rule{0.4pt}{4pt}\rule{2.6cm}{0.4pt}\rule{0.4pt}{4pt}}
\hspace*{3cm} \nn \\
&& = \frac{\gy^4\veps^{B_{1}C_{1}D_{2}E_{2}}\veps^{D_{1}E_{1}B_{2}C_{2}}}
{4\pi^{4}(y_{1}-y_{2})^4} \left\{ \left(1+\frac{1}{N^2}\right)
\left(\Tr_{N+2}\left[\calP(y_1)\calP(y_2)\right]\right)^2
\right.\nn \\
&&\left. -\frac{2}{N} \Tr_{N+2} \left[\calP(y_{1})\calP(y_{2})
\calP(y_{1})\calP(y_{2})\right] \right\} \nn \\
&& + \frac{\gy^4\veps^{B_{1}C_{1}D_{2}E_{2}}\veps^{D_{1}E_{1}B_{2}C_{2}}}
{8\pi^{4}\rho^2(y_{1}-y_{2})^2} \left\{ \Tr_2\left[\bar b \calP(y_1)
\calP(y_2)b\bar b \calP(y_2)\calP(y_1)b \right] \right. \nn \\
&& -\frac{1}{N} \left( \Tr_2\left[\bar b\calP(y_1)b\right]
\Tr_2\left[\bar b\calP(y_2)\calP(y_1)\calP(y_2) b\right]
+ \Tr_2\left[\bar b\calP(y_2)b\right]
\Tr_2\left[\bar b\calP(y_1)\calP(y_2)\calP(y_1) b\right] \right)
\nn \\
&& \left. + \frac{1}{N^2} \Tr_{N+2}\left[\calP(y_1)\calP(y_2)\right]
\Tr_2\left[\bar b\calP(y_1)b\right]
\Tr_2\left[\bar b\calP(y_1)b\right] \right\} \nn \\
&& + \frac{\gy^4\veps^{B_{1}C_{1}D_{2}E_{2}}\veps^{D_{1}E_{1}B_{2}C_{2}}}
{16\pi^{4}\rho^4} \left\{ \rule{0pt}{16pt}
\Tr_2\left[ \bar b\calP(y_1)b\bar b\calP(y_1)b\right]
\Tr_2\left[ \bar b\calP(y_2)b\bar b\calP(y_2)b\right] \right. \nn \\
&& -\frac{1}{N}\left(\left(\Tr_2\left[\bar b\calP(y_1)b\right]\right)^2
\Tr_2\left[\bar b\calP(y_2)b\bar b\calP(y_2)b\right]
+ \left(\Tr_2\left[\bar b\calP(y_2)b\right]\right)^2
\Tr_2\left[\bar b\calP(y_1)b\bar b\calP(y_1)b\right] \right) \nn \\
&& \left. + \frac{1}{N^2}\left(\Tr_2\left[\bar
b\calP(y_1)b\right]\right)^2 \left(\Tr_2\left[\bar
b\calP(y_2)b\right]\right)^2 \right\} \, .
\label{doublecontr}
\ea
Evaluating the traces we obtain
\ba
&& \Tr\left(\v^{B_1C_1}\v^{D_1E_1}\right)(y_1) \,
\Tr\left(\v^{B_2C_2}\v^{D_2E_2}\right)(y_2)
\raisebox{-15pt}{\hspace*{-6.6cm}
\rule{0.4pt}{10pt}\rule{4.6cm}{0.4pt}\rule{0.4pt}{10pt}}
\raisebox{-10pt}{\hspace*{-3.75cm}
\rule{0.4pt}{4pt}\rule{2.6cm}{0.4pt}\rule{0.4pt}{4pt}}
\hspace*{3cm} \nn \\
&& = \frac{\gy^4\veps^{B_{1}C_{1}D_{2}E_{2}}\veps^{D_{1}E_{1}B_{2}C_{2}}}
{4\pi^{4}(y_1-y_2)^4} \left\{ \left(N^2-1\right)^2 -
\frac{4N\rho^2(y_1-y_2)^2}{[(y_1-x_0)^2+\rho^2][(y_2-x_0)^2+\rho^2]}
\right. \nn \\
&& + \frac{4\rho^4(y_1-y_2)^4}{[(y_1-x_0)^2+\rho^2]^2
[(y_2-x_0)^2+\rho^2]^2} +\frac{1}{N} \left(
\frac{4\rho^2(y_1-y_2)^2}{[(y_1-x_0)^2+\rho^2][(y_2-x_0)^2+\rho^2]}
\right. \nn \\
&&\left.\left. - \frac{4\rho^4(y_1-y_2)^4}{[(y_1-x_0)^2+\rho^2]^2
[(y_2-x_0)^2+\rho^2]^2} \right) + \frac{1}{N^2}
\frac{4\rho^4(y_1-y_2)^4}{[(y_1-x_0)^2+\rho^2]^2
[(y_2-x_0)^2+\rho^2]^2} \right\} \nn \\
&& + \frac{\gy^4\veps^{B_{1}C_{1}D_{2}E_{2}}
\veps^{D_{1}E_{1}B_{2}C_{2}}\rho^2}
{4\pi^{4}(y_1-y_2)^2[(y_1-x_0)^2+\rho^2]^2[(y_2-x_0)^2+\rho^2]^2}
\left\{ -\left(1-\frac{4}{N}+\frac{4}{N^2}\right) \rho^2(y_1-y_2)^2
\right. \nn \\
&& \left. + \left(1-\frac{2}{N}\right)[(y_1-x_0)^2+\rho^2]
[(y_2-x_0)^2+\rho^2] \right\}  \nn \\
&& + \frac{\gy^4\veps^{B_{1}C_{1}D_{2}E_{2}}\veps^{D_{1}E_{1}B_{2}C_{2}}}
{2\pi^{4}} \frac{\rho^4}{[(y_1-x_0)^2+\rho^2]^2[(y_2-x_0)^2+\rho^2]^2}
\left(1-\frac{4}{N}+\frac{4}{N^2}\right) \, .
\label{doublecontr2}
\ea
In this expression we can recognise the contributions of the terms
${\tilde B}[_{r\a,s\b}\hspace*{-21pt}^{u\g,v\d}] (y_1,y_2)$ and
${\tilde C}[_{r\a,s\b}\hspace*{-21pt}^{u\g,v\d}](y_1,y_2)$ in the
propagator: the first three lines are $\tilde B$-$\tilde B$ terms, the
last line comes from $\tilde C$-$\tilde C$ terms and the remaining
block of two lines is the result of mixed $\tilde B$-$\tilde C$
terms. As a check notice that the latter two types of terms vanish for
$N=2$, as they should, because in that case the non-singular part of
the propagator is zero.

In computing the correlation function (\ref{twotwent}) we have to
evaluate the single scalar contraction in (\ref{Lam20-2-b1}).
This gives the following expression in terms of ADHM matrices
\ba
&& \hat G_{\hat\Lambda^{14}_{\bf 4}
\hat\Lambda^2_{\bf 20^*}}^{({\rm b})}
 = \frac{\veps^{B_1B_3C_1C_3}}{2\pi^2\gy^{32}N} \la
\Tr\!\left(\hat F_{\a_1}{}^{\b_1}
\hat\lambda^{A_1}_{\g_1}\right)\!(x_1) \ldots
\Tr\!\left(\hat F_{\a_{14}}{}^{\g_{14}}
\hat\lambda^{A_{14}}_{\b_{14}} \right)\!(x_{14}) \nn \\
&& \left\{ \frac{1}{(y_1-y_2)^2}\left[ \Tr \left( \calP(y_2)
\left\{\hat F_{\b_1}{}^{\d_1}(y_1), \hat\lambda^{B_2}_{\d_1}(y_1)
\right\} \,\calP(y_1) \left\{\hat F_{\b_2}{}^{\d_2}(y_2),
\hat\lambda^{C_2}_{\d_2}(y_2) \right\} \right) \right. \right. \nn \\
&& -\frac{1}{N} \left( \Tr \left(\calP(y_1)\calP(y_2)
\left\{\hat F_{\b_1}{}^{\d_1}(y_1),\hat\lambda^{B_2}_{\d_1}(y_1)
\right\} \right) \Tr\left(\left\{\hat F_{\b_2}{}^{\d_2}(y_2),
\hat\lambda^{C_2}_{\d_2}(y_2) \right\} \right) \right. \nn \\
&& \left. + \Tr \left(\left\{\hat F_{\b_1}{}^{\d_1}(y_1),
\hat\lambda^{B_2}_{\d_1}(y_1)\right\} \right) \Tr\left(\calP(y_2)
\calP(y_1)\left\{\hat F_{\b_2}{}^{\d_2}(y_2),
\hat\lambda^{C_2}_{\d_2}(y_2) \right\} \right) \right)
\label{Lam20-2-b2} \\
&& + \left.\frac{1}{N^2} \Tr\left(\left\{
\hat F_{\b_1}{}^{\d_1}(y_1),\hat\lambda^{B_2}_{\d_1}(y_1)\right\}
\right) \Tr\left(\left\{ \hat F_{\b_2}{}^{\d_2}(y_2),
\hat\lambda^{C_2}_{\d_2}(y_2) \right\} \right)
\Tr\left(\calP(y_1)\calP(y_2)\right) \right] \nn \\
&& + \frac{1}{2\rho^2} \left[ \Tr\left(\bar b
\left\{\hat F_{\b_1}{}^{\d_1}(y_1),\hat\lambda^{B_2}_{\d_1}(y_1)
\right\} \calP(y_1) b\right) \Tr\left(\bar b
\left\{\hat F_{\b_2}{}^{\d_2}(y_2),\hat\lambda^{C_2}_{\d_2}(y_2)
\right\} \calP(y_2) b\right) \right. \nn \\
&& -\frac{1}{N} \left( \Tr \left(\left\{\hat F_{\b_1}{}^{\d_1}(y_1),
\hat\lambda^{B_2}_{\d_1}(y_1)\right\} \right)
\Tr\left(\bar b
\left\{\hat F_{\b_2}{}^{\d_2}(y_2),\hat\lambda^{C_2}_{\d_2}(y_2)
\right\} \calP(y_2) b\right) \Tr\left(\bar b\calP(y_1)b\right)
\right. \nn \\
&& \left. + \Tr\left(\bar b
\left\{\hat F_{\b_1}{}^{\d_1}(y_1),\hat\lambda^{B_2}_{\d_1}(y_1)
\right\} \calP(y_1) b\right)
\Tr \left(\left\{\hat F_{\b_2}{}^{\d_2}(y_2),
\hat\lambda^{C_2}_{\d_2}(y_2)\right\} \right)
\Tr\left(\bar b\calP(y_2)b\right) \right) \nn \\
&& \left. \left.
+ \frac{1}{N^2} \Tr \left(\left\{\hat F_{\b_1}{}^{\d_1}(y_1),
\hat\lambda^{B_2}_{\d_1}(y_1)\right\} \right)
\Tr \left(\left\{\hat F_{\b_2}{}^{\d_2}(y_2),
\hat\lambda^{C_2}_{\d_2}(y_2)\right\} \right)
\Tr\left(\bar b\calP(y_1)b\right) \Tr\left(\bar b\calP(y_2)b\right)
\right] \right\} \nn \\
&& + \cdots \ra  \, . \nn
\ea
This result is much more complicated than the one we obtained for the
single contraction in (\ref{onecontr}). This is because the operators
involved, $\Lambda_{\bf 20^*}$, are cubic and thus the terms in
(\ref{propfin}) which make the propagator traceless also contribute.

\end{document}